\documentclass[aps,prb,twocolumn,longbibliography,superscriptaddress,floatfix]{revtex4-2}

\usepackage{graphicx}
\usepackage{bm}
\usepackage{color}
\usepackage[table]{xcolor}
\usepackage{makecell}
\usepackage{subfigure}
\usepackage{amsmath}
\usepackage{contour}
\usepackage[normalem]{ulem}
\usepackage{comment}
\usepackage{amssymb}
\usepackage[percent]{overpic}
\usepackage{bbold}
\usepackage{enumerate}
\usepackage{float}
\usepackage{xspace}
\usepackage{amsmath,amssymb,amsfonts,mathtools,bm}
\usepackage{mathrsfs}
\usepackage[colorlinks=true,linkcolor=blue,citecolor=blue,urlcolor=blue]{hyperref}
\usepackage{orcidlink}
\usepackage{tikz}
\usepackage{siunitx}
\usepackage[utf8]{inputenc}
\usepackage{nicefrac}
\usepackage[capitalise]{cleveref}
\usepackage{bm}
\usepackage{float}

\definecolor{color1}{rgb}{0,0.25,0.70}
\hypersetup{colorlinks=true,
    linkcolor={color1},
    citecolor={color1},
    urlcolor={color1}
}

\newcommand{\Tr}{\operatorname{Tr}}

\begin{document}

\title{Atomistic theory of the phonon angular momentum Hall effect}

\author{Daniel~A.~Bustamante~Lopez}\email{dabl@bu.edu}
\thanks{D.~A.~Bustamante~Lopez and V.~Brehm contributed equally to this work.}
\affiliation{Department of Physics, Boston University, Boston, Massachusetts 02215, USA}
\affiliation{Department of Applied Physics and Science Education,
Eindhoven University of Technology, 5612 AP Eindhoven, Netherlands}
\author{Verena Brehm}
\email{v.j.brehm@tue.nl}
\affiliation{Department of Applied Physics and Science Education,
Eindhoven University of Technology, 5612 AP Eindhoven, Netherlands}
\author{Dominik M. Juraschek}
\email{d.m.juraschek@tue.nl}
\affiliation{Department of Applied Physics and Science Education,
Eindhoven University of Technology, 5612 AP Eindhoven, Netherlands}

\date{\today}

\begin{abstract}
The spin and orbital Hall effects convert longitudinal charge currents into transverse flows of electronic angular momentum. Here we develop an atomistic theory of the recently proposed lattice-vibrational analogue, in which a longitudinal heat current driven by a thermal gradient is converted into a transverse current of phonon angular momentum. We derive a microscopic real-space expression for this current and show that it originates from thermally induced mixing of polarized vibrational motion, leading to a characteristic edge accumulation of phonon angular momentum. We demonstrate the effect in minimal square- and honeycomb-lattice models and compute the resulting phonon angular momentum accumulations for a range of example materials using input from first-principles calculations. Our results confirm that the phonon angular momentum Hall effect is a universal response of crystalline solids and our framework is generically applicable to all materials.
\end{abstract}

\maketitle

\section{Introduction}

\begin{figure*}[t]
  \centering
  \includegraphics[width=0.99\textwidth]{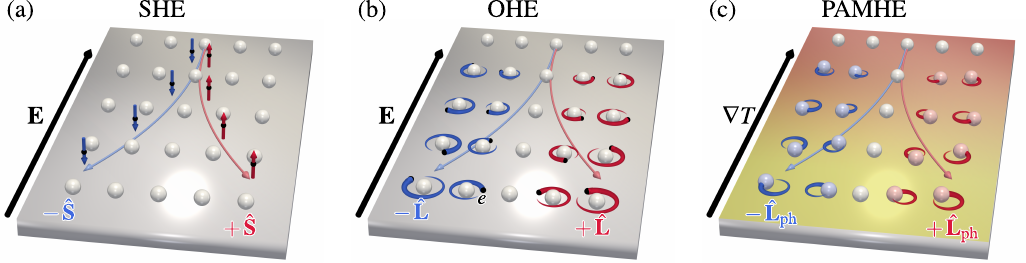}
  \caption{Angular momentum Hall effects. (a) Spin Hall effect: an electric field generates a transverse current and edge accumulation of spin angular momentum \cite{SinovaSpinHallEffect}. (b) Orbital Hall effect: an electric field generates a transverse current and edge accumulation of orbital angular momentum \cite{Go2018OHE}. (c) Phonon angular momentum Hall effect: a temperature gradient generates a transverse current and edge accumulation of phonon angular momentum \cite{ParkPAMHE}.}
  \label{fig:introHall}
\end{figure*}

In the spin Hall effect, a longitudinal charge current driven by an electric field generates a transverse spin current, leading to an edge accumulation of spin angular momentum, as illustrated in Fig.~\ref{fig:introHall}(a) \cite{SinovaSpinHallEffect}. More recently, the orbital Hall effect has been predicted and observed, in which a longitudinal charge current produces a transverse flow of orbital angular momentum, shown schematically in Fig.~\ref{fig:introHall}(b) \cite{Bernevig2005Orbitronics,Kontani2009OHE,Go2018OHE,Lee2021OHT,Choi2023OHE}. The spin and orbital Hall effects have become central mechanisms for angular-momentum injection and torque generation in spintronic and orbitronic devices \cite{Jungwirth2016,Baltz2018,Fukami2025}.

Related transverse responses also arise for charge-neutral excitations. Magnons, for instance, exhibit both thermal Hall~\cite{Onose2010MagnonHall,MookMertigMagnHall,IdeueTokuraMagnHall,MurakamiOkamotoMagnThermalHall,Katsura2010ThermalHall} and spin Nernst effects \cite{ChengXiaoAFMSpinNernst,ZyuzinKovalevAFMSpinNernst,MarkusAltermagnetNernstEff,brehm2024intrinsicspinnernsteffect}. Phonons can likewise exhibit a thermal Hall effect when time-reversal symmetry is broken, for example by an intrinsic magnetization or an external magnetic field \cite{Strohm2005,Sheng2006,Kagan2008,Zhang2010PHE,Qin2012,Grissonnanche2019GiantPhonThermHAll,JinLiUniversalPhonThermHallEff,ChenPyonPlanarPhonThermHallEff,LiBehniaPhonThermHallBlackPhosphorus,ChenTailleferPlanarPhonThermHallEff,FlebusMacDonaldPhHallViscosity,LiBehniaPhThermHallEffStrontiumTitanate,OhNagosaThermHallKitaev,OhNagosaPhonThermHallEffSkewScattering}. Beyond heat transport, lattice vibrations can carry phonon angular momentum when they are circularly or elliptically polarized \cite{Zhang2015ChiralPhonons,Zhu2018ChiralPhonons,Ishito2023,Ueda2023}. Such axial or chiral phonons have been linked to the phonon Hall effect \cite{Grissonnanche2020} and to a broad range of other magnetic responses, including the phonon Zeeman effect \cite{juraschek2:2017,Juraschek2019,Cheng2020,Baydin2022,Hernandez2023,Mustafa2025,Wu2023_phononmagneticmoments,Lujan2024,Mai2025,Che2025}, the phonon Barnett and Einstein--de Haas effects \cite{Zhang2014,Garanin2015,Mentink2019,Dornes2019,Tauchert2022,Davies2024}, and the generation of phono-magnetic fields \cite{Juraschek2020_3,Juraschek2022_giantphonomag,Luo2023,Basini2024,Shabala2024,biggs2025ultrafastfaradayrotationprobe}.

A natural question is therefore whether a temperature gradient can generate a transverse flow of phonon angular momentum (PAM). Early work showed helicity-dependent deflection of transverse elastic waves in inhomogeneous media, reminiscent of a spin Hall effect of phonons \cite{Bliokh2006}. More recently, Park and Yang proposed a phonon angular momentum Hall effect from a semiclassical treatment based on Luttinger's gravitational-field approach \cite{ParkPAMHE}, which motivates the present work. A general theoretical framework for this effect in finite lattices driven out of equilibrium has thus far remained elusive.

Here, we develop an atomistic theory of the phonon angular momentum Hall effect (PAMHE) for finite lattices away from thermal equilibrium. This framework requires only the harmonic approximation and coupling to local thermal reservoirs. Our central result contains three analytical formulae for the Hall formalism, including  the PAM current, PAM density, and PAM conductivities, expressed in terms of nonequilibrium correlation functions. This allows us to obtain a real-space-resolved transverse current response and an associated PAM edge accumulation generated by a longitudinal temperature gradient. We apply the theory to minimal models of two-dimensional square and honeycomb lattices and show that a transverse PAM current emerges in both cases. The effect originates from thermally induced mixing of polarized vibrational modes and therefore does not require lattice chirality or broken inversion symmetry. We then compute the resulting edge accumulation in representative materials, using input from first-principles calculations. These results confirm that the PAMHE is a universal nonequilibrium response of the crystal lattice and provide a practical framework for quantitative predictions in real materials.

\section{Theoretical formalism}
\label{sec:formalism}

We derive the theoretical formalism for the nonequilibrium steady state of a finite harmonic crystal coupled to thermal baths through Langevin dynamics. Details of the derivation are given in Supplemental Material~\cite{supplementary}.

\subsection{Lattice model and nonequilibrium steady state}

We model a finite crystal in the harmonic approximation as a lattice of $N$ sites coupled by harmonic forces and attached to local thermal baths. Each site $s\in\{1,\dots,N\}$ is characterized by a mass $m_s$, an equilibrium position $\bm r_s$, and a displacement $\bm u_s(t)\in\mathbb{R}^3$ from equilibrium. Collecting all displacements into the vector $\bm u=(\bm u_1,\dots,\bm u_N)^\top\in\mathbb{R}^{3N}$, the equations of motion of the full system are
\begin{equation}
\mathbf{M}\ddot{\bm u}+\kappa\,\mathbf{M}\dot{\bm u}+\mathbf{K}\bm u=\bm\eta(t),
\label{eq:EOM_realspace}
\end{equation}
where $\mathbf{M}=\mathrm{diag}(m_1,\dots,m_N)\otimes \mathbf{I}_3$ is the mass matrix, $\mathbf{K}\in\mathbb{R}^{3N\times 3N}$ is the force-constant matrix, and $\kappa$ is the damping rate. The right-hand side of the equation contains the stochastic forces exerted by the thermal baths, modeled as Langevin noise with correlations
\begin{equation}
\big\langle \eta_{si}(t)\,\eta_{tj}(t')\big\rangle
=
2\kappa\,m_s k_B T_s\,\delta_{st}\delta_{ij}\delta(t-t'),
\label{eq:noise_covariance}
\end{equation}
where indices $i,j\in\{x,y,z\}$ label Cartesian components, indices $s,t$ label lattice sites, and $T_s$ is the bath temperature at site $s$. For a static nonuniform temperature profile, the system adopts a nonequilibrium steady state. Our goal is to determine this state and use it to describe the transport and accumulation of phonon observables in real space.

To obtain the steady state, we transform the equations of motion to a basis in which the deterministic part is diagonal. This is achieved by diagonalizing the dynamical matrix
\(
\mathbf{D}=\mathbf{M}^{-1/2}\mathbf{K}\mathbf{M}^{-1/2},
\)
which, being real and symmetric, yields
\begin{equation}
\mathbf{D}\mathbf{U}=\mathbf{U}\bm\Omega^2,
\qquad
\mathbf{U}^\top\mathbf{U}=\mathbf{I},
\label{eq:modal_diag}
\end{equation}
where $\bm\Omega^2=\mathrm{diag}(\Omega_1^2,\dots,\Omega_{3N}^2)$ contains the eigenfrequencies of the normal modes and the columns of $\mathbf{U}$ are the corresponding mass-normalized polarization vectors.

We then define $\mathbf{R}=\mathbf{M}^{-1/2}\mathbf{U}$ and $\bm u=\mathbf{R}\bm Q$, so that $\bm Q$ contains the modal amplitudes. In this basis, Eq.~\eqref{eq:EOM_realspace} becomes
\begin{equation}
\ddot{\bm Q}+\kappa\,\dot{\bm Q}+\bm\Omega^2\bm Q=\bm\xi(t),
\label{eq:EOM_modal}
\end{equation}
where $\bm\xi=\mathbf{U}^\top\mathbf{M}^{-1/2}\bm\eta$. The modal noise correlations are
\begin{equation}
\langle \bm\xi(t)\bm\xi^\top(t')\rangle
=
\mathbf{W}\,\delta(t-t'),
\qquad
\mathbf{W}=2\kappa k_B\,\mathbf{U}^\top \mathbf{T}\,\mathbf{U},
\label{eq:W_def}
\end{equation}
where $\mathbf{T}=\mathrm{diag}(T_1,\dots,T_N)\otimes \mathbf{I}_3$ encodes the temperature profile applied to the system. For a uniform temperature profile, $\mathbf{W}\propto \mathbf{I}$ and the normal modes remain decoupled. A nonuniform temperature profile produces off-diagonal elements of $\mathbf{W}$, which mix different modes in the steady state.

Equation~\eqref{eq:EOM_modal} shows that in modal space the deterministic dynamics is decoupled, so the steady-state solution can be obtained analytically. In frequency space, each mode $\mu$ satisfies
\begin{equation}
Q_\mu(\omega)=\frac{\xi_\mu(\omega)}{\Omega_\mu^2-\omega^2+i\kappa\omega},
\label{eq:Qomega}
\end{equation}
which allows us to compute the equal-time correlations of the modal displacements and velocities. These correlations fully characterize the steady state and provide the quantities from which we later construct local phonon angular momentum, its current, and kinetic-energy transport in real space.

The equal-time steady-state covariances factorize into the temperature-dependent matrix $\mathbf{W}$ and kernel matrices determined only by the lattice dynamics, through the mode eigenfrequencies and damping,
\begin{subequations}
\label{eq:Sigma_modal_factor}
\begin{align}
\langle \bm Q\bm Q^\top\rangle &= \mathbf{W}\circ \mathbf{C}_{QQ},
&
\langle \bm Q\dot{\bm Q}^\top\rangle &= \mathbf{W}\circ \mathbf{C}_{Q\dot Q},
\label{eq:Sigma_modal_factor_a}
\\
\langle \dot{\bm Q}\dot{\bm Q}^\top\rangle &= \mathbf{W}\circ \mathbf{C}_{\dot Q\dot Q}.
\label{eq:Sigma_modal_factor_b}
\end{align}
\end{subequations}
Here $\circ$ denotes the Hadamard product. The scalar kernels for displacement, displacement--velocity, and velocity correlations are
\begin{subequations}
\label{eq:kernels_closed}
\begin{align}
C_{QQ}(\mu,\nu)
&=
\frac{2\kappa}{
(\Omega_\mu^2-\Omega_\nu^2)^2+2\kappa^2(\Omega_\mu^2+\Omega_\nu^2)},
\\[3pt]
C_{Q\dot Q}(\mu,\nu)
&=
\frac{\Omega_\mu^2-\Omega_\nu^2}{
(\Omega_\mu^2-\Omega_\nu^2)^2+2\kappa^2(\Omega_\mu^2+\Omega_\nu^2)},
\\[3pt]
C_{\dot Q\dot Q}(\mu,\nu)
&=
\frac{\kappa(\Omega_\mu^2+\Omega_\nu^2)}{
(\Omega_\mu^2-\Omega_\nu^2)^2+2\kappa^2(\Omega_\mu^2+\Omega_\nu^2)}.
\end{align}
\end{subequations}
where $\mu,\nu$ label normal modes. The kernel matrices $\mathbf{C}_{QQ}$ and $\mathbf{C}_{\dot Q\dot Q}$ are symmetric, whereas $\mathbf{C}_{Q\dot Q}$ is antisymmetric. It follows that $\langle \bm Q\dot{\bm Q}^\top\rangle=0$ for a uniform temperature profile, since in that case $\mathbf{W}$ is diagonal.

The previous calculations were carried out in the modal basis. To study transport in real space, we transform the correlations back to the lattice coordinates. Denoting by $\mathbf{R}_s$ the $3\times 3N$ block of $\mathbf{R}$ associated with site $s$, we obtain
\begin{subequations}
\label{eq:block_proj}
\begin{align}
\langle \bm u_s\bm u_t^\top\rangle
&=
\mathbf{R}_s\langle \bm Q\bm Q^\top\rangle\mathbf{R}_t^\top,
&
\langle \bm u_s\dot{\bm u}_t^\top\rangle
&=
\mathbf{R}_s\langle \bm Q\dot{\bm Q}^\top\rangle\mathbf{R}_t^\top,
\label{eq:block_proj_a}
\\
\langle \dot{\bm u}_s\dot{\bm u}_t^\top\rangle
&=
\mathbf{R}_s\langle \dot{\bm Q}\dot{\bm Q}^\top\rangle\mathbf{R}_t^\top.
\label{eq:block_proj_b}
\end{align}
\end{subequations}
These real-space correlations are the basic quantities used below to evaluate the local accumulation and transport of phonon angular momentum and kinetic energy at the microscopic level. Details of the derivation are given in Supplemental Material~\cite{supplementary}.

\subsection{Phonon angular momentum density and current}

We now use the steady-state correlations derived above to describe how phonon angular momentum (PAM) arises and is transported in real space. The local phonon angular momentum at site $s$ is
\begin{equation}
\bm L_s(t)=m_s\,\bm u_s(t)\times \dot{\bm u}_s(t),
\end{equation}
so that its $i$-th component in the steady state can be written as
\begin{equation}
L_i(s)\equiv -m_s\,\Tr\!\left[\mathbf{E}_i\,\langle \bm u_s\dot{\bm u}_s^\top\rangle\right],
\label{eq:L_def}
\end{equation}
where $\mathbf{E}_i$ is the antisymmetric generator of rotations about axis $i\in\{x,y,z\}$. Explicit forms are given in Supplemental Material~\cite{supplementary}. Equation~\eqref{eq:L_def} shows that the local PAM density is determined by the displacement--velocity correlations at the same site. This also makes clear why the relation
\(
\langle \bm Q\dot{\bm Q}^\top\rangle=\mathbf{W}\circ \mathbf{C}_{Q\dot Q}
\)
is central: for a uniform temperature profile, $\mathbf{W}$ is diagonal, while $\mathbf{C}_{Q\dot Q}$ vanishes on the diagonal, so the PAM density is zero. PAM accumulation therefore requires a nonuniform temperature profile, for example created by a temperature gradient.

To describe transport between lattice sites, we resolve the harmonic forces into contributions from individual bonds. Let $\mathcal N(s)$ denote the set of neighbors of site $s$, and let $\bm{\Phi}^{(st)}=\bm{\Phi}^{(ts)}$ be the symmetric $3\times 3$ force-constant tensor associated with bond $(s,t)$. The elastic force exerted by site $t$ on site $s$ is then given by
\(
-\bm{\Phi}^{(st)}(\bm u_s-\bm u_t).
\)
A direct manipulation of the equations of motion leads to the local balance equation
\begin{equation}
\frac{d}{dt}L_i(s)+\sum_{t\in\mathcal N(s)} j^{(L_i)}_{s\to t}=-\kappa L_i(s),
\label{eq:PAM_continuity_main}
\end{equation}
which shows that local PAM is transported through the lattice and relaxed by damping. Here, the bond-resolved PAM current is identified as
\begin{equation}
j^{(L_i)}_{s\to t}
=
\Tr\!\left[
\mathbf{E}_i\bm{\Phi}^{(st)}
\left(
\langle \bm u_s\bm u_s^\top\rangle
-
\langle \bm u_t\bm u_s^\top\rangle
\right)
\right].
\label{eq:jL_def}
\end{equation}
This expression shows that PAM transport is governed by nonlocal displacement correlations across a bond. The tensor $\bm{\Phi}^{(st)}$ converts the bond stretch $\bm u_s-\bm u_t$ into the corresponding elastic force, while $\mathbf{E}_i$ extracts the angular-momentum component about axis $i$. Eqs.~\eqref{eq:L_def} and \eqref{eq:jL_def} can therefore be used to distinguish between local PAM accumulation and bond-resolved PAM transport.

Equation~\eqref{eq:jL_def} further identifies the structural requirement for a transverse PAM response. If the force constants do not mix different Cartesian components of motion, then the force tensors and displacement-correlation blocks are diagonal in Cartesian space. The matrix multiplying $\mathbf{E}_i$ in Eq.~\eqref{eq:jL_def} is then also diagonal, and therefore symmetric, so the PAM current vanishes because $\Tr[\mathbf{E}_i\mathbf{S}]=0$ for any symmetric matrix $\mathbf{S}$. A finite transverse PAM current therefore requires polarization mixing generated by the force-constant matrix. In real crystals, this mixing is generally present due to oblique bond angles, long-ranged force constants, anisotropic interactions, or multi-atom unit cells.

Equation~\eqref{eq:jL_def} describes the flow of angular momentum along a single bond. To characterize net transport resolved by site, we define the PAM current vector at site $s$ by summing the bond currents from neighboring sites weighted by their bond directions,
\begin{equation}
\bm j^{(L_i)}(s)
=
\sum_{t\in\mathcal N(s)}
j^{(L_i)}_{s\to t}\,\hat{\bm d}_{st}.
\label{eq:jsite_main}
\end{equation}
So far, this current and the PAM density are defined for a generic temperature profile. We now turn to the response to a weak temperature bias, which allows us to define a PAM conductivity. For this purpose, it is useful to decompose the modal noise matrix into contributions from individual bath sites. Let $\mathbf{U}_r$ be the $3\times 3N$ block of $\mathbf{U}$ associated with site $r$, and define
\(
\bm{\Pi}_r\equiv \mathbf{U}_r^\top\mathbf{U}_r,
\)
which satisfies
\(
\sum_{r=1}^{N}\bm{\Pi}_r=\mathbf{I}.
\)
The matrix $\bm{\Pi}_r$ measures the overlap of the normal modes on site $r$ and determines how a temperature bias applied at that site populates the modal steady state. It is also convenient to introduce the bond-difference matrix
\(
\Delta \mathbf{R}_{st}\equiv \mathbf{R}_s-\mathbf{R}_t.
\)
Using the decomposition
\(
\mathbf{W}=2\kappa k_B\sum_{r=1}^{N} T_r\,\bm{\Pi}_r,
\)
the PAM current can be written as
\begin{equation}
\bm j^{(L_i)}(s)=\sum_{r=1}^{N}\bm{\sigma}^{(L_i)}(s|r)\,T_r,
\label{eq:j_sigma_decomp}
\end{equation}
where the site-resolved response kernel is
\begin{align}
\bm{\sigma}^{(L_i)}(s|r) & 
= 
2\kappa k_B
\sum_{t\in\mathcal N(s)}
\hat{\bm d}_{st} \nonumber\\
& \times \Tr\!\left[
\mathbf{E}_i\bm{\Phi}^{(st)}
\,\Delta \mathbf{R}_{st}
\left(\bm{\Pi}_r\circ \mathbf{C}_{QQ}\right)
\mathbf{R}_s^\top
\right].
\label{eq:sigma_kernel_main}
\end{align}
This kernel describes the current of the $i$-th component of PAM flowing through site $s$ per unit temperature bias applied at bath site $r$. It is determined by the force-constant matrix, the lattice geometry, and the vibrational correlations of the system.

\begin{figure*}[t]
  \centering
  \includegraphics[width=0.95\textwidth]{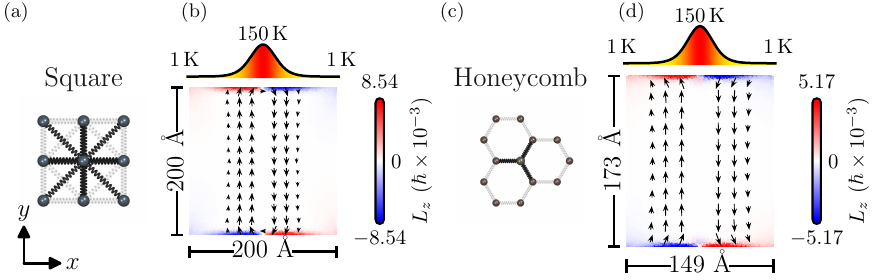}
  \caption{Phonon angular momentum Hall effect in minimal models of centrosymmetric lattices. Panels (a) and (c) show the lattice geometries and spring networks used for the square- and honeycomb-lattice models, respectively. Panels (b) and (d) show the corresponding real-space nonequilibrium response under a thermal bias applied along the horizontal direction. In the square lattice, both axial and diagonal bonds are included, whereas in the honeycomb lattice nearest-neighbor bonds already produce a finite effect. The blue--red colormap shows the local phonon angular momentum $L_z(s)$ from Eq.~\eqref{eq:L_def}, and the black arrows show the PAM current $\bm j^{(L_z)}(s)$ from Eq.~\eqref{eq:jsite_main}. For the real-space maps shown here, the square lattice uses axial and diagonal spring constants of 30~N/m and 15~N/m, respectively, while the honeycomb lattice uses nearest-neighbor springs only, with isotropic and anisotropic couplings of 80~N/m and 60~N/m, respectively. In both cases, we use a damping rate $\kappa=5\,\mathrm{ps}^{-1}$ at all sites.}
  \label{fig:square-honeycomb-overview}
\end{figure*}

To compute the response to an applied temperature gradient, we consider a weak temperature modulation around a uniform reference temperature,
\(
T_r=T^{(0)}+\delta T_r,
\)
with
\(
|\delta T_r|\ll T^{(0)}.
\)
Because the PAM current depends on $\mathbf{C}_{QQ}$, which is already nonzero for a uniform temperature background, a uniform shift of all bath temperatures is not automatically removed from the response. To isolate the response to an applied temperature gradient, we define the bath-site-centered kernel
$\tilde{\bm{\sigma}}^{(L_i)}(s|r)
=
\bm{\sigma}^{(L_i)}(s|r)
-
\frac{1}{N}\sum_{r'=1}^{N}\bm{\sigma}^{(L_i)}(s|r'),$
i.e., the response kernel with its average over the bath-site index $r$ subtracted. This definition ensures the neutrality property
$\sum_{r=1}^{N}\tilde{\bm{\sigma}}^{(L_i)}(s|r)=\bm 0$.

Using a linear temperature gradient,
\(
\delta T_r = \sum_{k=x,y,z}(r_{r,k}-r_{0,k})\,\partial_k T,
\)
where $\bm r_0$ is an arbitrary reference point, we can rewrite Eq.~\eqref{eq:j_sigma_decomp} into a thermal-gradient-induced PAM current
\begin{equation}
\delta j_j^{(L_i)}(s)
=
\sum_{k=x,y,z}
\sigma_{jk}^{(L_i)}(s)\,\partial_k T,
\label{eq:sigma_local_main}
\end{equation}
with the local PAM conductivity tensor given by
\begin{equation}
\sigma_{jk}^{(L_i)}(s)
=
\sum_{r=1}^{N}
(r_{r,k}-r_{0,k})\,\tilde{\sigma}_{j}^{(L_i)}(s|r).
\label{eq:sigma_local_explicit_main}
\end{equation}
Because of the neutrality property, $\sigma_{jk}^{(L_i)}(s)$ is independent of the choice of $\bm r_0$, as required for an intrinsic transport property. The bulk PAM conductivity is then obtained by averaging over the interior sites,
\begin{equation}
\overline{\sigma}_{jk}^{(L_i)}
=
\frac{1}{N_{\mathrm{bulk}}}
\sum_{s\in\mathrm{bulk}}
\sigma_{jk}^{(L_i)}(s).
\label{eq:sigma_bulk_main}
\end{equation}

In the following, we consider longitudinal thermal drives with a temperature gradient oriented along the $x$ direction. The relevant components of the PAM conductivity tensor are therefore the longitudinal component $\overline{\sigma}_{xx}^{(L_i)}$ and the transverse component $\overline{\sigma}_{yx}^{(L_i)}$. We define the PAM deflection angle as
\begin{equation}
\tan\theta_{L_i}
=
\frac{\overline{\sigma}_{yx}^{(L_i)}}{\overline{\sigma}_{xx}^{(L_i)}},
\label{eq:hall_angle_main}
\end{equation}
which measures the ratio of transverse to longitudinal PAM transport under an applied temperature gradient. In the presence of an external magnetic field, the conductivity components can acquire additional corrections, as derived in the Supplemental Material~\cite{supplementary}. The PAM deflection angle therefore characterizes the deflection, possibly induced by external fields, of the phonon-angular-momentum current relative to the longitudinal thermal drive. In analogy with the spin Hall angle, which compares a transverse spin current with a longitudinal charge current \cite{Obstbaum2016TuningSpinHallAngles}, we also define a Hall-like angle that compares the transverse phonon-angular-momentum current with the longitudinal transport of phonon kinetic energy,
\begin{equation}
\tan\theta_H
=
\kappa\,
\frac{\overline{\sigma}_{yx}^{(L_i)}}{\overline{\sigma}_{xx}^{(E)}},
\label{eq:mixed_angle_main}
\end{equation}
which relates the transverse PAM response to the longitudinal energy transport generated by the same thermal drive. Explicit formulas and related quantities for energy transport can be found in Supplemental Material~\cite{supplementary}.

\section{Phonon angular momentum Hall effect in fundamental geometries}
\label{sec:demo_geometries}

Now that we have established the formalism, we apply it to two minimal models of centrosymmetric two-dimensional lattices. We show the local phonon angular momentum $L_z(s)$ computed from Eq.~\eqref{eq:L_def} and the PAM current $\bm j^{(L_z)}(s)$ computed from Eq.~\eqref{eq:jsite_main} for a square lattice in Fig.~\ref{fig:square-honeycomb-overview}(a,b) and for a honeycomb lattice in Fig.~\ref{fig:square-honeycomb-overview}(c,d). In both cases, a longitudinal temperature gradient along the $x$ direction generates a transverse flow of phonon angular momentum along the $y$ direction, producing an edge accumulation of $L_z$. The accumulation appears near the sample boundaries, where the transverse current terminates, as expected from the continuity relation in Eq.~\eqref{eq:PAM_continuity_main}. It therefore follows from the finite geometry of the system rather than from polarization-dependent edge scattering or edge damping.

For the square lattice, a finite PAMHE requires diagonal springs connecting next-nearest neighbors. Without them, the lattice forces do not mix the $x$ and $y$ components of motion, and the current in Eq.~\eqref{eq:jL_def} vanishes. The diagonal bonds introduce this mixing and thereby generate a finite transverse PAM current. In the honeycomb lattice, the nearest-neighbor model already exhibits a clear PAMHE. These two examples therefore isolate the basic ingredients of the effect in the simplest possible settings. Details of the lattice constructions, together with the dependence of the edge accumulation on bond stiffness, damping, and temperature gradient, are given in Supplemental Material~\cite{supplementary}, along with the axial and diagonal spring network for the square lattice and the isotropic and anisotropic bond components of the honeycomb lattice.

These results demonstrate that the PAMHE does not require chirality or broken inversion symmetry. It already appears in minimal centrosymmetric lattices. We next consider the effect of an external magnetic field and the relation to the conventional phonon Hall effect.

\section{Magnetic-field dependence}
\label{sec:magnetic_field}

We now examine how an applied magnetic field modifies the transport currents. Within our framework, its effect is incorporated by adding a gyroscopic term $2\mathbf{J}\dot{\bm u}$ to the left-hand side of Eq.~\eqref{eq:EOM_realspace}. For a site $s$, this term reads
\begin{equation}
    2\left(\mathbf{J}\dot{\bm u}\right)_s
    =
    2 m_s \bm{\Omega}^{\mathrm g}_s \times \dot{\bm u}_s,
    \qquad
    \bm{\Omega}^{\mathrm g}_s = \gamma_s \bm B,
\end{equation}
where $\gamma_s$ is the gyromagnetic ratio assigned to site $s$. In the present two-dimensional geometry, $\bm{\Omega}^{\mathrm g}_s$ points out of the plane. This term couples the lattice dynamics to the applied field and produces field-dependent corrections to the steady-state correlations and, through them, to the local fields, currents, and conductivity tensors.
 
\begin{figure}[t]
  \centering
  \includegraphics[width=\columnwidth]{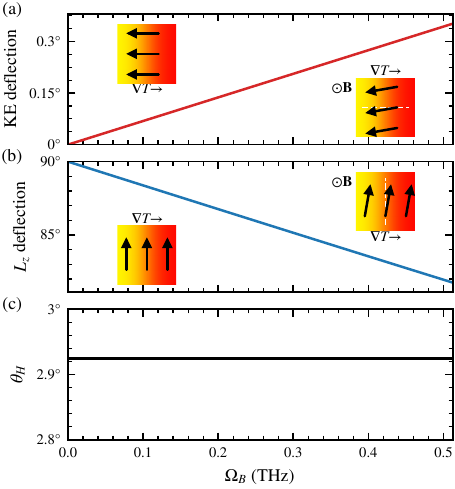}
\caption{Magnetic-field dependence of the kinetic-energy and PAM deflection angles, together with the Hall-like angle, for the square-lattice model of Fig.~\ref{fig:square-honeycomb-overview}. Panels (a) and (b) show the deflection angles of the kinetic-energy and PAM currents, respectively; the PAM deflection angle $\theta_{L_z}$ is defined in Eq.~\eqref{eq:hall_angle_main}, with an analogous definition for the energy current. Panel (c) shows the Hall-like angle $\theta_H$, defined in Eq.~\eqref{eq:mixed_angle_main}. At zero field, the kinetic-energy current is longitudinal, whereas the PAM current is transverse. An out-of-plane magnetic field deflects the kinetic-energy current and rotates the PAM current away from the purely transverse direction. By contrast, $\theta_H$ remains unchanged over the plotted field range, indicating that the field mainly affects the longitudinal PAM and transverse kinetic-energy components rather than the conductivity components entering Eq.~\eqref{eq:mixed_angle_main}. Here the field strength is expressed through the effective scale $\Omega_B=\gamma B$, where $B$ is the applied magnetic field and $\gamma$ is the gyromagnetic ratio assigned to the lattice.}
  \label{fig:deflection_angle}
\end{figure}

Figure~\ref{fig:deflection_angle} shows, for the square lattice, the magnetic-field dependence of the deflection angles of the PAM and kinetic-energy currents, defined in Eq.~\eqref{eq:hall_angle_main} and analogously for the kinetic-energy current, together with the Hall-like angle $\theta_H$ defined in Eq.~\eqref{eq:mixed_angle_main}. At zero field, the kinetic-energy current is purely longitudinal, corresponding to a deflection angle of $0^\circ$, since it follows the temperature gradient and carries energy from the hot central region toward the colder left and right edges. The PAM current, by contrast, is already purely transverse, corresponding to a deflection angle of $90^\circ$, as it transports angular momentum toward the transverse edges. The distinction is therefore already present at $B=0$: energy transport is longitudinal, whereas PAM transport is transverse.

Upon turning on the field, the kinetic-energy current acquires a transverse component, as expected for the conventional phonon Hall effect, while the PAM current is rotated away from the purely transverse direction and develops a longitudinal component along the temperature gradient. The applied field thus gives the kinetic-energy current its Hall character, whereas in the PAM channel it tilts a response that is already transverse at zero field. The contrasting field response of the two currents therefore highlights that the PAMHE is distinct from the conventional phonon Hall effect. By contrast, the Hall-like angle $\theta_H$ remains nearly unchanged over the plotted field range [Fig.~\ref{fig:deflection_angle}(c)]. This indicates that the magnetic field primarily modifies the longitudinal PAM and transverse kinetic-energy components, while leaving the combination entering $\theta_H$ in Eq.~\eqref{eq:mixed_angle_main} nearly unaffected.

\section{Phonon angular momentum Hall effect in example materials}

\begin{figure}[t]
  \centering
  \includegraphics[width=\columnwidth]{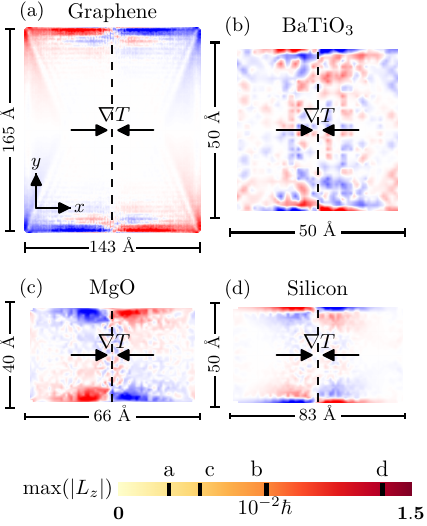}
\caption{Real-space edge accumulation of phonon angular momentum at zero magnetic field in representative materials: (a) graphene, (b) BaTiO$_3$, (c) MgO, and (d) silicon. The colormap shows the local phonon angular momentum $L_z(s)$ from Eq.~\eqref{eq:L_def}. The arrows indicate the applied thermal bias, and the dashed line marks the hot center of the sample. Opposite-sign accumulations appear near the upper and lower edges in each case, evidencing a transverse PAM response in finite samples. The black markers labeled a--d on the color bar indicate the corresponding values of $\max(|L_z|)$ for panels (a)--(d).}
  \label{fig:materials_p95_Lz}
\end{figure}

Finally, we apply the real-space nonequilibrium framework developed above to a range of representative materials, including graphene, silicon (Si), magnesium oxide (MgO), and barium titanate (BaTiO$_3$). For each case, the harmonic interatomic force constants are obtained from first-principles calculations and used to construct the finite-sample force-constant matrices entering Eq.~\eqref{eq:EOM_realspace}. Using these first-principles data, our finite-sample models appropriately reproduce the bulk phonon dispersions of the corresponding materials. The computational details are provided in Supplemental Material~\cite{supplementary}.

Figure~\ref{fig:materials_p95_Lz} shows the zero-field edge accumulation of phonon angular momentum in these atomistic systems, as characterized by the local field $L_z(s)$ defined in Eq.~\eqref{eq:L_def}. All calculations are performed with $T_{\mathrm{hot}}=150\,\mathrm{K}$, $T_{\mathrm{cold}}=1\,\mathrm{K}$, and $\kappa=5\,\mathrm{ps}^{-1}$. Panel (a) shows a graphene monolayer, whereas panels (b)--(d) show central slices through cuboidal Si, MgO, and BaTiO$_3$ samples with comparable transverse dimensions. A finite PAM accumulation appears in all four cases, demonstrating the existence of the PAMHE in real materials beyond the minimal lattice models discussed above. For the finite geometries and temperature gradients considered here, the accumulated PAM density is on the scale of $10^{-3}$ to $10^{-2}\hbar$ per atoms.

These values should be interpreted as finite-sample estimates rather than as direct predictions for experimentally relevant sample geometries. The present calculations are performed for samples with lateral dimensions on the scale of tens of angstroms, and the resulting accumulations depend on the sample size and crystal orientation, the magnitude and spatial profile of the applied temperature gradient, and the damping used to establish the nonequilibrium steady state. The main result is therefore not the numerical value obtained for a particular finite geometry, but the appearance of a nonzero transverse edge accumulation across all the materials considered here. In principle, the present framework can be extended to larger samples, but the computational bottleneck is the growth of the force-constant matrix and the cost of obtaining its eigenfrequencies and normal modes. A natural direction in that regime is to replace full diagonalization by sparse iterative eigensolvers for large symmetric matrices, such as Lanczos-type methods and their implicitly restarted implementations \cite{Lanczos1950,Lehoucq1998ARPACK}.


\section{Conclusion}
\label{sec:conclusion}

We have developed an atomistic real-space theory of the recently proposed PAMHE \cite{ParkPAMHE}, in which a longitudinal temperature gradient generates a transverse current of phonon angular momentum and an associated edge accumulation in a nonequilibrium steady state. We demonstrate that the PAM current arises universally from the polarization mixing of correlated atomic motion in a nonuniform temperature profile, a criterion that is realistically fulfilled in all crystalline materials. This criterion reduces the question of ``Does the PAMHE exist in this material?'' to ``How large is the PAMHE in this material?'' The answer to the latter question depends strongly on the material and sample geometry and calls for a systematic investigation across many materials to identify compounds that maximize the effect.

Our derivations of the phononic response functions show that phonon angular momentum can be considered a genuine transport quantity in nonequilibrium solids: it can flow through the bulk, accumulate at boundaries, and respond systematically to lattice geometry and vibrational polarization. The PAMHE provides a universal bulk source of thermally driven phonon angular momentum, complementary to, but more widely applicable than the recently predicted and discovered chirality-induced phonon angular momentum selectivity, which requires chiral or noncentrosymmetric crystal structures \cite{HamadaJfromTempGrad,Kim2023CPASS,Ohe2024CIPAM,Zhang2025PAM,Nabei2026OSE, feng2026vibrational}. 

Possible experimental routes to observe the effect include the following. Large temperature gradients can be induced by laser heating, as has been demonstrated in recent studies \cite{Zhang2025PAM,Kim2023CPASS,Nabei2026OSE}. A first route to detection involves the magnetic response of the circular atomic motion. The accumulated PAM density will simultaneously generate a phonon magnetic moment \cite{juraschek2:2017,Juraschek2019} at the sample edges, which could be detected using magneto-optic Kerr or Faraday measurements \cite{Luo2023,Basini2024,Davies2024,biggs2025ultrafastfaradayrotationprobe}. Furthermore, the circularly polarized phonons can be converted to magnons or spin currents at interfaces through magnetoelastic interactions \cite{Holanda2018,Sasaki2021SAW,Kim2023CPASS,Ohe2024CIPAM,Nabei2026OSE}. A second route involves direct measurements of the mechanical angular momentum. For example, X-ray scattering techniques have previously been successful in detecting circular motion of atoms \cite{Tauchert2022,Ueda2023,Ueda2025}. Furthermore, the induced angular momentum will generate a macroscopic torque on the sample that can be detected by cantilevers~\cite{Zhang2025PAM}.

The PAMHE complements the spin Hall and orbital Hall effects in the toolkit for controlling angular momentum in solids. Its universal occurrence in crystalline materials offers new possibilities for angular momentum transmutation between electronic and lattice degrees of freedom, and we anticipate applications in spintronic and orbitronic devices.

\begin{acknowledgments}
We thank Benedetta Flebus, Tobias Kampfrath, Alireza Qaiumzadeh, Rembert Duine, and Robin Neumann for useful discussions. This research was supported by the ERC Starting Grant CHIRALPHONONICS, no. 101166037. D.A.B.L. acknowledges support from the US Department of Energy, Office of Science, Office of Basic Energy Sciences Early Career Research Program under award number DE-SC-0021305. 
\end{acknowledgments}

\bibliography{refs}


\onecolumngrid
\clearpage

\setcounter{page}{1}

\begin{center}
\textbf{\large Supplemental Material:\\ Atomistic theory of the phonon angular momentum Hall effect}\\[0.4cm]
Daniel~A.~Bustamante~Lopez,$^{1,2}$ Verena Brehm,$^{2}$ Dominik~M.~Juraschek$^{2}$\\[0.15cm]
\affiliation{}
$^1${\itshape{\small Department of Physics, Boston University, Boston, Massachusetts 02215, USA}}\\
$^2${\itshape{\small Department of Applied Physics and Science Education,\\ Eindhoven University of Technology, 5612 AP Eindhoven, Netherlands}}\\
\end{center}

\date{\today}

\setcounter{section}{0}
\setcounter{equation}{0}
\setcounter{figure}{0}
\setcounter{table}{0}
\makeatletter
\renewcommand{\theequation}{S\arabic{equation}}
\renewcommand{\thefigure}{S\arabic{figure}}
\renewcommand{\thetable}{S\arabic{table}}

\section{Derivation of the nonequilibrium lattice framework}

We model a crystal lattice in the harmonic approximation coupled to local thermal reservoirs, which provides a microscopic setting for the study of nonequilibrium transport of lattice observables such as energy and angular momentum. Solving the stochastic equations of motion gives the steady-state correlations of atomic displacements and velocities. These correlations fully characterize the nonequilibrium state and form the basis for constructing the local densities and bond-resolved currents of phonon angular momentum and lattice energy. For weak temperature gradients, the same formalism yields the linear-response conductivity tensors that govern angular-momentum and energy transport. When the lattice is further coupled to magnetic order or to an external magnetic field, gyroscopic terms modify the dynamics and produce field-dependent corrections to these correlations, allowing the magnetothermal response of the lattice to be treated within the same framework. This provides a common microscopic description of the phonon angular momentum Hall effect (PAMHE), through the transverse current of phonon angular momentum generated by a longitudinal thermal drive, and of the conventional phonon Hall effect, through the corresponding transverse component of the energy current when time-reversal symmetry is broken.

\subsection{Harmonic lattice model}
\label{subsec:harmonic_lattice_model}

We consider a crystal lattice in the harmonic approximation, coupled to local thermal reservoirs. The lattice contains \(N\) sites in three spatial dimensions. Each site \(s\) has equilibrium position \(\bm{r}_s\), mass \(m_s\), and displacement \(\bm{u}_s(t)\in\mathbb{R}^3\). Collecting all displacements into the \(3N\)-component vector
\begin{equation}
\bm{u}(t)=\bigl(\bm{u}_1(t),\dots,\bm{u}_N(t)\bigr)^\top,
\end{equation}
the equations of motion read
\begin{equation}
\mathbf{M}\ddot{\bm{u}}+\kappa \mathbf{M}\dot{\bm{u}}+\mathbf{K}\bm{u}=\bm{\eta}(t).
\label{eq:S1_eom_real}
\end{equation}
Here
\begin{equation}
\mathbf{M}=\mathrm{diag}(m_1,\dots,m_N)\otimes \mathbf{I}_3
\end{equation}
is the mass matrix, \(\mathbf{K}\) is the harmonic force-constant matrix, \(\kappa\) is the damping rate, and \(\bm{\eta}(t)\) is the stochastic force exerted by the thermal reservoirs. The left-hand side of Eq.~\eqref{eq:S1_eom_real} describes the deterministic lattice dynamics: the first term is the inertial response of the atoms, the second term accounts for dissipation through the coupling to the reservoirs, and the third term contains the harmonic restoring forces generated by lattice distortions. The right-hand side represents the fluctuating thermal forces injected by the baths.

The reservoirs are taken to act locally and isotropically on the lattice. Their effect is described by Gaussian white noise with correlations
\begin{equation}
\langle \eta_{s,i}(t)\eta_{t,j}(t')\rangle
=
2\kappa m_s k_B T_s\,\delta_{st}\delta_{ij}\delta(t-t').
\label{eq:S1_noise_real}
\end{equation}
Here \(s\) and \(t\) label lattice sites, while \(i\) and \(j\) label Cartesian components of motion. This form means that each reservoir acts independently on its own site and couples equally to the three spatial directions. For a prescribed temperature profile \(\{T_s\}\), Eq.~\eqref{eq:S1_eom_real} defines a linear stochastic dynamics that relaxes to a nonequilibrium steady state.

The harmonic couplings are encoded in the matrix \(\mathbf{K}\), which is assembled from bond force constants. For each coupled pair of sites \((s,t)\) in the lattice, we introduce a symmetric \(3\times 3\) tensor \(\mathbf{\Phi}^{(st)}=\mathbf{\Phi}^{(ts)}\), which describes how the relative displacement of the two sites generates the restoring force associated with that bond. The contribution of bond \((s,t)\) to the force on site \(s\) is
$-\mathbf{\Phi}^{(st)}\bigl(\bm{u}_s-\bm{u}_t\bigr)$.
Summing over all bonds connected to site \(s\), the site-resolved blocks of \(\mathbf{K}\) take the form
\begin{equation}
\mathbf{K}_{ss}=\sum_{t\in \mathcal{N}(s)} \mathbf{\Phi}^{(st)},
\qquad
\mathbf{K}_{st}=-\mathbf{\Phi}^{(st)}
\quad (s\neq t),
\label{eq:S1_K_blocks}
\end{equation}
where \(\mathcal{N}(s)\) denotes the set of neighbors of site \(s\). In this way, the force on each atom depends only on relative displacements with respect to the atoms to which it is coupled.

For a central spring of stiffness \(k_{st}\) joining sites \(s\) and \(t\), the restoring force acts along the bond direction
\begin{equation}
\hat{\bm{d}}_{st}=\frac{\bm{r}_t-\bm{r}_s}{|\bm{r}_t-\bm{r}_s|},
\end{equation}
and the corresponding bond tensor is
\begin{equation}
\mathbf{\Phi}^{(st)}=k_{st}\,\hat{\bm{d}}_{st}\hat{\bm{d}}_{st}^{\top}.
\label{eq:S1_central_bond_tensor}
\end{equation}
More general force-constant networks can be treated in the same way by allowing \(\mathbf{\Phi}^{(st)}\) to contain noncentral or anisotropic couplings. This becomes important in lattices such as the honeycomb model discussed below.

The nonequilibrium steady state generated by Eq.~\eqref{eq:S1_eom_real} is fully characterized by the equal-time correlations of displacements and velocities. These correlations are the basic quantities from which the local lattice observables and their bond-resolved currents are constructed. To obtain them, it is convenient to transform the equations of motion to a basis in which the deterministic part of the lattice dynamics becomes decoupled.

\subsection{Mass-normalized coordinates and normal-mode decomposition}
\label{subsec:mass_normalized_modal_decomposition}

To obtain the steady-state correlations, it is convenient to rewrite the equations of motion in a basis in which the deterministic lattice dynamics becomes decoupled. This is done in two steps. We first introduce mass-normalized coordinates, which remove the mass matrix from the inertial term. We then diagonalize the resulting dynamical matrix, so that the deterministic motion is written in terms of independent normal modes.

We define the mass-normalized displacement and noise vectors as
\begin{equation}
\tilde{\bm{u}} \equiv \mathbf{M}^{1/2}\bm{u},
\qquad
\tilde{\bm{\eta}} \equiv \mathbf{M}^{-1/2}\bm{\eta},
\label{eq:S1_mass_normalized_defs}
\end{equation}
and the dynamical matrix as
\begin{equation}
\mathbf{D} \equiv \mathbf{M}^{-1/2}\mathbf{K}\mathbf{M}^{-1/2}.
\label{eq:S1_dynamical_matrix}
\end{equation}
In terms of these variables, Eq.~\eqref{eq:S1_eom_real} becomes
\begin{equation}
\ddot{\tilde{\bm{u}}}+\kappa \dot{\tilde{\bm{u}}}+\mathbf{D}\tilde{\bm{u}}=\tilde{\bm{\eta}}(t).
\label{eq:S1_eom_mass_normalized}
\end{equation}
In this form the explicit mass dependence has been absorbed into the coordinates, so that the deterministic motion is governed only by the dynamical matrix \(\mathbf{D}\).

The thermal noise also takes a simpler form in the mass-normalized basis. Its correlations are
\begin{equation}
\langle \tilde{\eta}_{s,i}(t)\tilde{\eta}_{t,j}(t')\rangle
=
2\kappa k_B T_s\,\delta_{st}\delta_{ij}\delta(t-t').
\label{eq:S1_noise_mass_normalized}
\end{equation}
It is convenient to collect the site temperatures into the matrix
\begin{equation}
\mathbf{T}=\mathrm{diag}(T_1,\dots,T_N)\otimes \mathbf{I}_3.
\label{eq:S1_temperature_matrix}
\end{equation}
The noise covariance can then be written as
\begin{equation}
\langle \tilde{\bm{\eta}}(t)\tilde{\bm{\eta}}(t')^{\top}\rangle
=
2\kappa k_B\,\mathbf{T}\,\delta(t-t').
\label{eq:S1_noise_mass_normalized_matrix}
\end{equation}
This shows that, in the mass-normalized basis, the local thermal driving is set directly by the temperature profile.

The matrix \(\mathbf{D}\) is real and symmetric, and therefore admits an orthonormal set of eigenvectors. We write
\begin{equation}
\mathbf{D}\mathbf{U}=\mathbf{U}\mathbf{\Omega}^2,
\qquad
\mathbf{U}^{\top}\mathbf{U}=\mathbf{I},
\label{eq:S1_D_diagonalization}
\end{equation}
where
\begin{equation}
\mathbf{\Omega}^2=\mathrm{diag}\!\left(\Omega_1^2,\dots,\Omega_{3N}^2\right).
\label{eq:S1_Omega_def}
\end{equation}
The columns \(\bm U_\mu\in\mathbb{R}^{3N}\) of \(\mathbf{U}\) are the mass-normalized normal-mode polarization vectors, and \(\Omega_\mu\) are the corresponding normal-mode frequencies. For each site \(s\), the vector \(\bm U_{s\mu}\in\mathbb{R}^{3}\) denotes the three-component block of \(\bm U_\mu\) associated with that site.

We now transform to modal coordinates,
\begin{equation}
\bm{Q}\equiv \mathbf{U}^{\top}\tilde{\bm{u}},
\qquad
\bm{\zeta}\equiv \mathbf{U}^{\top}\tilde{\bm{\eta}}.
\label{eq:S1_modal_defs}
\end{equation}
The equations of motion then reduce to
\begin{equation}
\ddot{Q}_{\mu}+\kappa \dot{Q}_{\mu}+\Omega_{\mu}^{2}Q_{\mu}=\zeta_{\mu}(t).
\label{eq:S1_modal_eom}
\end{equation}
The deterministic motion is now fully decoupled: each normal mode behaves as a damped harmonic oscillator with its own frequency \(\Omega_\mu\).

The nonequilibrium character of the problem appears through the modal noise, whose correlations are
\begin{equation}
\langle \zeta_{\mu}(t)\zeta_{\nu}(t')\rangle
=
W_{\mu\nu}\,\delta(t-t'),
\label{eq:S1_modal_noise_components}
\end{equation}
or, in matrix form,
\begin{equation}
\langle \bm{\zeta}(t)\bm{\zeta}(t')^{\top}\rangle
=
\mathbf{W}\,\delta(t-t'),
\qquad
\mathbf{W}=2\kappa k_B\,\mathbf{U}^{\top}\mathbf{T}\mathbf{U}.
\label{eq:S1_modal_noise_matrix}
\end{equation}
The matrix elements of \(\mathbf{W}\) can be written in site-resolved form as
\begin{equation}
W_{\mu\nu}
=
2\kappa k_B
\sum_{s=1}^{N}
T_s\,\bm U_{s\mu}\cdot\bm U_{s\nu}.
\label{eq:S1_W_components}
\end{equation}
This expression shows that \(W_{\mu\nu}\) is the temperature-weighted overlap of modes \(\mu\) and \(\nu\) across the lattice sites. Its role is to encode how the local thermal reservoirs populate the normal modes and, out of equilibrium, how they mix them. 

In the modal basis the deterministic lattice dynamics has been reduced to independent damped oscillators, while the nonequilibrium character of the problem is carried entirely by the matrix \(\mathbf{W}\). For a uniform temperature profile, \(\mathbf{W}\propto \mathbf{I}\), so each mode is driven independently. For a nonuniform temperature profile, off-diagonal elements \(W_{\mu\nu}\) appear and generate correlations between different normal modes in the steady state. When projected back to real space, these inter-mode correlations give the nonequilibrium contributions to the displacement and velocity correlations from which the local fields and bond-resolved currents are constructed.

\subsection{Steady-state solution in modal space}
\label{subsec:modal_steady_state}

At zero magnetic field, the equations of motion in the normal-mode basis take the form of independent damped oscillators,
\begin{equation}
\ddot{Q}_{\mu}+\kappa \dot{Q}_{\mu}+\Omega_{\mu}^{2}Q_{\mu}=\zeta_{\mu}(t).
\label{eq:S1_modal_eom_again}
\end{equation}
This makes the steady-state problem especially simple in modal space. The deterministic part of the motion is fully decoupled, and the effect of the nonequilibrium thermal driving is carried by the modal noise covariance \(\mathbf{W}\).

We solve Eq.~\eqref{eq:S1_modal_eom_again} in frequency space. For each mode,
\begin{equation}
Q_{\mu}(\omega)=G_{\mu}(\omega)\,\zeta_{\mu}(\omega),
\qquad
G_{\mu}(\omega)=\frac{1}{-\omega^{2}+i\kappa\omega+\Omega_{\mu}^{2}},
\label{eq:S1_modal_green}
\end{equation}
where \(G_\mu(\omega)\) is the response function of a damped harmonic oscillator with frequency \(\Omega_\mu\). The equal-time steady-state correlations follow from frequency integrals involving products of \(G_\mu(\omega)\) and \(G_\nu(-\omega)\). It is therefore convenient to define three scalar kernels,
\begin{subequations}
\begin{align}
C_{QQ}(\mu,\nu)
&\equiv
\int \frac{d\omega}{2\pi}\,
G_{\mu}(\omega)G_{\nu}(-\omega),
\label{eq:S1_CQQ_def}
\\
C_{Q\dot{Q}}(\mu,\nu)
&\equiv
\int \frac{d\omega}{2\pi}\,
(-i\omega)\,G_{\mu}(\omega)G_{\nu}(-\omega),
\label{eq:S1_CQdQ_def}
\\
C_{\dot{Q}\dot{Q}}(\mu,\nu)
&\equiv
\int \frac{d\omega}{2\pi}\,
\omega^{2}\,G_{\mu}(\omega)G_{\nu}(-\omega).
\label{eq:S1_CdQdQ_def}
\end{align}
\end{subequations}
These determine, respectively, the displacement covariances, the displacement--velocity covariances, and the velocity covariances in the modal basis.

Introducing the damped frequency
\begin{equation}
\widetilde{\Omega}_{\mu}=\sqrt{\Omega_{\mu}^{2}-\frac{\kappa^{2}}{4}},
\label{eq:S1_damped_frequency}
\end{equation}
the frequency integrals can be evaluated in closed form:
\begin{subequations}
\begin{align}
C_{QQ}(\mu,\nu)
&=
\frac{2\kappa}
{\left[\kappa^{2}+\left(\widetilde{\Omega}_{\mu}-\widetilde{\Omega}_{\nu}\right)^{2}\right]
 \left[\kappa^{2}+\left(\widetilde{\Omega}_{\mu}+\widetilde{\Omega}_{\nu}\right)^{2}\right]}
\notag
\\
&=
\frac{2\kappa}
{\left(\Omega_{\mu}^{2}-\Omega_{\nu}^{2}\right)^{2}
+2\kappa^{2}\left(\Omega_{\mu}^{2}+\Omega_{\nu}^{2}\right)},
\label{eq:S1_CQQ_closed}
\\[0.5em]
C_{Q\dot{Q}}(\mu,\nu)
&=
\frac{\widetilde{\Omega}_{\mu}^{2}-\widetilde{\Omega}_{\nu}^{2}}
{\left[\kappa^{2}+\left(\widetilde{\Omega}_{\mu}-\widetilde{\Omega}_{\nu}\right)^{2}\right]
 \left[\kappa^{2}+\left(\widetilde{\Omega}_{\mu}+\widetilde{\Omega}_{\nu}\right)^{2}\right]}
\notag
\\
&=
\frac{\Omega_{\mu}^{2}-\Omega_{\nu}^{2}}
{\left(\Omega_{\mu}^{2}-\Omega_{\nu}^{2}\right)^{2}
+2\kappa^{2}\left(\Omega_{\mu}^{2}+\Omega_{\nu}^{2}\right)},
\label{eq:S1_CQdQ_closed}
\\[0.5em]
C_{\dot{Q}\dot{Q}}(\mu,\nu)
&=
\frac{\kappa\left[\frac{\kappa^{2}}{2}+\widetilde{\Omega}_{\mu}^{2}+\widetilde{\Omega}_{\nu}^{2}\right]}
{\left[\kappa^{2}+\left(\widetilde{\Omega}_{\mu}-\widetilde{\Omega}_{\nu}\right)^{2}\right]
 \left[\kappa^{2}+\left(\widetilde{\Omega}_{\mu}+\widetilde{\Omega}_{\nu}\right)^{2}\right]}
\notag
\\
&=
\frac{\kappa\left(\Omega_{\mu}^{2}+\Omega_{\nu}^{2}\right)}
{\left(\Omega_{\mu}^{2}-\Omega_{\nu}^{2}\right)^{2}
+2\kappa^{2}\left(\Omega_{\mu}^{2}+\Omega_{\nu}^{2}\right)}.
\label{eq:S1_CdQdQ_closed}
\end{align}
\end{subequations}

The steady-state modal covariances then factorize into a thermal part, set by the matrix \(\mathbf{W}\), and a dynamical part, set by the kernels above:
\begin{subequations}
\begin{align}
\langle Q_{\mu}Q_{\nu}\rangle
&=
W_{\mu\nu}\,C_{QQ}(\mu,\nu),
\label{eq:S1_modal_cov_QQ}
\\
\langle Q_{\mu}\dot{Q}_{\nu}\rangle
&=
W_{\mu\nu}\,C_{Q\dot{Q}}(\mu,\nu),
\label{eq:S1_modal_cov_QdQ}
\\
\langle \dot{Q}_{\mu}\dot{Q}_{\nu}\rangle
&=
W_{\mu\nu}\,C_{\dot{Q}\dot{Q}}(\mu,\nu).
\label{eq:S1_modal_cov_dQdQ}
\end{align}
\end{subequations}
In matrix form,
\begin{equation}
\langle \bm{Q}\bm{Q}^{\top}\rangle=\mathbf{W}\circ \mathbf{C}_{QQ},
\qquad
\langle \bm{Q}\dot{\bm{Q}}^{\top}\rangle=\mathbf{W}\circ \mathbf{C}_{Q\dot{Q}},
\qquad
\langle \dot{\bm{Q}}\dot{\bm{Q}}^{\top}\rangle=\mathbf{W}\circ \mathbf{C}_{\dot{Q}\dot{Q}},
\label{eq:S1_modal_cov_matrix}
\end{equation}
where \(\circ\) denotes the Hadamard product, and \((\mathbf{C}_{QQ})_{\mu\nu}=C_{QQ}(\mu,\nu)\), with analogous definitions for \(\mathbf{C}_{Q\dot{Q}}\) and \(\mathbf{C}_{\dot{Q}\dot{Q}}\).

This factorized form separates the nonequilibrium driving from the vibrational dynamics of the lattice. The matrix \(\mathbf{W}\) contains the information about the temperature profile and the resulting mode mixing, while the kernels \(C_{QQ}\), \(C_{Q\dot{Q}}\), and \(C_{\dot{Q}\dot{Q}}\) depend only on the mode frequencies and the damping rate. In particular, \(C_{QQ}\) and \(C_{\dot{Q}\dot{Q}}\) are symmetric under \(\mu\leftrightarrow\nu\), whereas \(C_{Q\dot{Q}}\) is antisymmetric and vanishes on the diagonal. As a result, for a uniform temperature profile, where \(\mathbf{W}\propto \mathbf{I}\), one has
\begin{equation}
\langle \bm{Q}\dot{\bm{Q}}^{\top}\rangle = 0.
\label{eq:S1_uniform_no_QdQ}
\end{equation}
This means that displacement--velocity correlations require nonequilibrium mode mixing generated by a nonuniform temperature profile. After projection back to real space, these are the correlations that contribute to the local phonon angular momentum and, more generally, to the bond-resolved currents derived below.

As a consistency check, consider the equilibrium case of a uniform temperature profile, for which
\begin{equation}
W_{\mu\nu}=2\kappa k_B T\,\delta_{\mu\nu}.
\label{eq:S1_W_equilibrium}
\end{equation}
The diagonal modal covariances then reduce to
\begin{equation}
\langle Q_\mu^2\rangle
=
2\kappa k_B T\, C_{QQ}(\mu,\mu),
\qquad
\langle Q_\mu \dot Q_\mu\rangle
=
2\kappa k_B T\, C_{Q\dot Q}(\mu,\mu),
\qquad
\langle \dot Q_\mu^2\rangle
=
2\kappa k_B T\, C_{\dot Q\dot Q}(\mu,\mu).
\label{eq:S1_modal_diag_equilibrium}
\end{equation}
Using Eqs.~\eqref{eq:S1_CQQ_closed}--\eqref{eq:S1_CdQdQ_closed}, one finds
\begin{equation}
C_{QQ}(\mu,\mu)=\frac{1}{2\kappa \Omega_\mu^2},
\qquad
C_{Q\dot Q}(\mu,\mu)=0,
\qquad
C_{\dot Q\dot Q}(\mu,\mu)=\frac{1}{2\kappa}.
\label{eq:S1_kernel_diag_limits}
\end{equation}
Therefore,
\begin{equation}
\langle Q_\mu^2\rangle=\frac{k_B T}{\Omega_\mu^2},
\qquad
\langle Q_\mu \dot Q_\mu\rangle=0,
\qquad
\langle \dot Q_\mu^2\rangle=k_B T,
\label{eq:S1_modal_equipartition}
\end{equation}
which is the expected classical equilibrium result for an independent harmonic mode.

\subsection{Projection to real space}
\label{subsec:projection_real_space}

The steady state is most easily obtained in the normal-mode basis, but the physical observables of interest are defined in real space. We therefore now project the modal covariances back to the atomic displacements and velocities at each lattice site.

From the definitions of the mass-normalized and modal coordinates,
\begin{equation}
\tilde{\bm{u}}=\mathbf{M}^{1/2}\bm{u},
\qquad
\bm{Q}=\mathbf{U}^{\top}\tilde{\bm{u}},
\end{equation}
it follows that the real-space displacement vector can be written as
\begin{equation}
\bm{u}
=
\mathbf{M}^{-1/2}\mathbf{U}\bm{Q}
\equiv
\mathbf{R}\bm{Q},
\qquad
\mathbf{R}\equiv \mathbf{M}^{-1/2}\mathbf{U}.
\label{eq:S1_R_definition}
\end{equation}
The matrix \(\mathbf{R}\) therefore contains the normal-mode polarization vectors expressed in terms of the physical atomic displacements. Since \(\mathbf{R}\) is time independent, the velocities are obtained in the same way,
\begin{equation}
\dot{\bm{u}}=\mathbf{R}\dot{\bm{Q}}.
\label{eq:S1_velocity_projection}
\end{equation}

For each site \(s\), we denote by \(\mathbf{R}_s\) the \(3\times 3N\) block of \(\mathbf{R}\) associated with that site. The displacement and velocity of site \(s\) are then
\begin{equation}
\bm{u}_s=\mathbf{R}_s\bm{Q},
\qquad
\dot{\bm{u}}_s=\mathbf{R}_s\dot{\bm{Q}}.
\label{eq:S1_site_projection}
\end{equation}
This makes the modal solution directly connected to the lattice motion: the displacement of a given atom is obtained by superposing all normal modes with weights fixed by the corresponding block of \(\mathbf{R}\).

The equal-time real-space covariances follow directly from the modal covariances. For any pair of sites \(s\) and \(t\),
\begin{subequations}
\begin{align}
\langle \bm{u}_s \bm{u}_t^{\top}\rangle
&=
\mathbf{R}_s
\langle \bm{Q}\bm{Q}^{\top}\rangle
\mathbf{R}_t^{\top},
\label{eq:S1_realspace_cov_uu}
\\
\langle \bm{u}_s \dot{\bm{u}}_t^{\top}\rangle
&=
\mathbf{R}_s
\langle \bm{Q}\dot{\bm{Q}}^{\top}\rangle
\mathbf{R}_t^{\top},
\label{eq:S1_realspace_cov_udotu}
\\
\langle \dot{\bm{u}}_s \dot{\bm{u}}_t^{\top}\rangle
&=
\mathbf{R}_s
\langle \dot{\bm{Q}}\dot{\bm{Q}}^{\top}\rangle
\mathbf{R}_t^{\top}.
\label{eq:S1_realspace_cov_dotudotu}
\end{align}
\end{subequations}
Using the factorized modal solution obtained above, these expressions become
\begin{subequations}
\begin{align}
\langle \bm{u}_s \bm{u}_t^{\top}\rangle
&=
\mathbf{R}_s
\left(\mathbf{W}\circ \mathbf{C}_{QQ}\right)
\mathbf{R}_t^{\top},
\label{eq:S1_realspace_cov_uu_factorized}
\\
\langle \bm{u}_s \dot{\bm{u}}_t^{\top}\rangle
&=
\mathbf{R}_s
\left(\mathbf{W}\circ \mathbf{C}_{Q\dot{Q}}\right)
\mathbf{R}_t^{\top},
\label{eq:S1_realspace_cov_udotu_factorized}
\\
\langle \dot{\bm{u}}_s \dot{\bm{u}}_t^{\top}\rangle
&=
\mathbf{R}_s
\left(\mathbf{W}\circ \mathbf{C}_{\dot{Q}\dot{Q}}\right)
\mathbf{R}_t^{\top}.
\label{eq:S1_realspace_cov_dotudotu_factorized}
\end{align}
\end{subequations}

These matrices contain the full real-space information of the nonequilibrium steady state. The equal-site blocks, with \(s=t\), determine local quantities such as the phonon angular momentum density and the local energy density. The off-diagonal blocks, with \(s\neq t\), describe correlations between different lattice sites and provide the input needed to construct bond-resolved currents. In this way, the projection from modal space back to real space turns the modal solution into the real-space correlations from which the local transport observables are obtained.

\subsection{Local fields}
\label{subsec:local_fields}

The equal-site covariance blocks determine the local observables of the nonequilibrium steady state. Of particular interest here are the local phonon angular momentum and the local energy density, since their transport gives rise to the PAMHE and to heat flow. It is also useful to introduce the local vibrational amplitude, which measures the strength of the atomic motion at each lattice site.

The local displacement amplitude at site \(s\) is defined as
\begin{equation}
A(s)\equiv \Tr\langle \bm{u}_s \bm{u}_s^{\top}\rangle.
\label{eq:S1_local_amplitude}
\end{equation}
This quantity measures the mean-square vibrational amplitude of the atom around its equilibrium position.

The local kinetic-energy density is
\begin{equation}
E_{\mathrm{kin}}(s)\equiv \frac{m_s}{2}\Tr\langle \dot{\bm{u}}_s \dot{\bm{u}}_s^{\top}\rangle.
\label{eq:S1_local_kinetic_energy}
\end{equation}
It is determined by the local velocity fluctuations and gives the kinetic contribution to the lattice energy stored at site \(s\).

The local phonon angular momentum is obtained from the antisymmetric part of the displacement--velocity covariance. Let \(\mathbf{E}_i\), with \(i\in\{x,y,z\}\), denote the antisymmetric matrices defined by
\begin{equation}
(\mathbf{E}_i)_{jk}=\epsilon_{ijk},
\label{eq:S1_rotation_generators}
\end{equation}
with explicit form
\begin{equation}
\mathbf{E}_x=
\begin{pmatrix}
0 & 0 & 0\\
0 & 0 & 1\\
0 & -1 & 0
\end{pmatrix},
\qquad
\mathbf{E}_y=
\begin{pmatrix}
0 & 0 & -1\\
0 & 0 & 0\\
1 & 0 & 0
\end{pmatrix},
\qquad
\mathbf{E}_z=
\begin{pmatrix}
0 & 1 & 0\\
-1 & 0 & 0\\
0 & 0 & 0
\end{pmatrix}.
\label{eq:S1_rotation_generators_explicit}
\end{equation}
These satisfy \(\mathbf{E}_i^\top=-\mathbf{E}_i\). The local phonon-angular-momentum density is then
\begin{equation}
L_i(s)\equiv -m_s \Tr\!\left[\mathbf{E}_i \langle \bm{u}_s \dot{\bm{u}}_s^{\top}\rangle\right].
\label{eq:S1_local_PAM}
\end{equation}
Equivalently,
\begin{equation}
L_i(s)=m_s\left\langle \bigl(\bm{u}_s\times \dot{\bm{u}}_s\bigr)_i \right\rangle.
\label{eq:S1_local_PAM_cross}
\end{equation}
This field measures the local orbital angular momentum carried by the ionic motion at site \(s\). In the present context it is the basic local quantity associated with the phonon angular momentum Hall effect.

Using the real-space covariances derived above, these fields can be written directly in terms of the modal solution. The local displacement amplitude is
\begin{equation}
A(s)=\Tr\!\left[\mathbf{R}_s\left(\mathbf{W}\circ \mathbf{C}_{QQ}\right)\mathbf{R}_s^{\top}\right],
\label{eq:S1_local_amplitude_factorized}
\end{equation}
the local kinetic-energy density is
\begin{equation}
E_{\mathrm{kin}}(s)=
\frac{m_s}{2}
\Tr\!\left[\mathbf{R}_s\left(\mathbf{W}\circ \mathbf{C}_{\dot{Q}\dot{Q}}\right)\mathbf{R}_s^{\top}\right],
\label{eq:S1_local_kinetic_energy_factorized}
\end{equation}
and the local phonon-angular-momentum density is
\begin{equation}
L_i(s)=
-m_s \Tr\!\left[\mathbf{E}_i\,\mathbf{R}_s\left(\mathbf{W}\circ \mathbf{C}_{Q\dot{Q}}\right)\mathbf{R}_s^{\top}\right].
\label{eq:S1_local_PAM_factorized}
\end{equation}

These expressions show that different local fields probe different parts of the steady-state covariance structure. The displacement amplitude depends on the local displacement covariance, the kinetic-energy density depends on the local velocity covariance, and the phonon angular momentum depends on the local displacement--velocity covariance. This last point is especially important for the PAMHE. Since \(\mathbf{C}_{Q\dot{Q}}\) is antisymmetric and vanishes on the diagonal, a uniform temperature profile, for which \(\mathbf{W}\propto \mathbf{I}\), gives
\begin{equation}
L_i(s)=0.
\label{eq:S1_local_PAM_uniform_zero}
\end{equation}
A nonzero local phonon angular momentum therefore requires nonequilibrium mode mixing generated by a nonuniform temperature profile. When combined with polarization mixing in the force-constant network, these local correlations give rise to the phonon-angular-momentum currents derived next.

\subsection{Bond-resolved currents and continuity relations}
\label{subsec:S1_bond_balance}

The local fields introduced above are accompanied by bond-resolved currents obtained from their local balance equations. To derive them, it is useful to write the equation of motion at site \(s\) in bond form,
\begin{equation}
m_s \ddot{\bm{u}}_s
=
-\kappa m_s \dot{\bm{u}}_s
-
\sum_{t\in\mathcal{N}(s)}
\mathbf{\Phi}^{(st)}\bigl(\bm{u}_s-\bm{u}_t\bigr)
+
\bm{\eta}_s(t).
\label{eq:S1bc_site_eom}
\end{equation}
The first term on the right-hand side describes dissipation into the local reservoirs, the second term is the net elastic force exerted by the neighboring atoms, and the last term is the fluctuating force injected by the baths. The bond currents follow by differentiating the local fields and reorganizing the elastic contributions into sums over neighboring bonds.

We begin with the local kinetic-energy density,
\begin{equation}
E_{\mathrm{kin}}(s)
=
\frac{m_s}{2}\langle \dot{\bm{u}}_s\cdot\dot{\bm{u}}_s\rangle
=
\frac{m_s}{2}\Tr\langle \dot{\bm{u}}_s\dot{\bm{u}}_s^{\top}\rangle .
\label{eq:S1bc_Ekin_def}
\end{equation}
Differentiating with respect to time gives
\begin{equation}
\frac{d}{dt}E_{\mathrm{kin}}(s)
=
m_s\langle \dot{\bm{u}}_s\cdot\ddot{\bm{u}}_s\rangle .
\label{eq:S1bc_Ekin_dt_step1}
\end{equation}
Using Eq.~\eqref{eq:S1bc_site_eom}, we obtain
\begin{equation}
\frac{d}{dt}E_{\mathrm{kin}}(s)
=
-\kappa m_s \langle \dot{\bm{u}}_s\cdot\dot{\bm{u}}_s\rangle
-
\sum_{t\in\mathcal{N}(s)}
\left\langle
\dot{\bm{u}}_s^{\top}\mathbf{\Phi}^{(st)}(\bm{u}_s-\bm{u}_t)
\right\rangle
+
\langle \dot{\bm{u}}_s\cdot\bm{\eta}_s\rangle .
\label{eq:S1bc_Ekin_dt_step2}
\end{equation}
Since \(\mathbf{\Phi}^{(st)}\) is symmetric, the bond term can be written as
\begin{equation}
\left\langle
\dot{\bm{u}}_s^{\top}\mathbf{\Phi}^{(st)}(\bm{u}_s-\bm{u}_t)
\right\rangle
=
\Tr\!\left[
\mathbf{\Phi}^{(st)}
\left(
\langle \bm{u}_s\dot{\bm{u}}_s^{\top}\rangle
-
\langle \bm{u}_t\dot{\bm{u}}_s^{\top}\rangle
\right)
\right].
\label{eq:S1bc_E_bond_term}
\end{equation}
This motivates the bond-resolved energy current
\begin{equation}
j^{(E)}_{s\to t}
\equiv
\Tr\!\left[
\mathbf{\Phi}^{(st)}
\left(
\langle \bm{u}_s\dot{\bm{u}}_s^{\top}\rangle
-
\langle \bm{u}_t\dot{\bm{u}}_s^{\top}\rangle
\right)
\right].
\label{eq:S1bc_E_bond_current}
\end{equation}
The energy current is therefore governed by the correlation between the bond deformation \(\bm{u}_s-\bm{u}_t\) and the motion at site \(s\).

We next consider the local phonon angular momentum,
\begin{equation}
L_i(s)
=
m_s\left\langle \bigl(\bm{u}_s\times\dot{\bm{u}}_s\bigr)_i\right\rangle
=
-m_s \Tr\!\left[\mathbf{E}_i\langle \bm{u}_s\dot{\bm{u}}_s^{\top}\rangle\right].
\label{eq:S1bc_L_def}
\end{equation}
Differentiating with respect to time gives
\begin{equation}
\frac{d}{dt}L_i(s)
=
m_s\left\langle \dot{\bm{u}}_s^{\top}\mathbf{E}_i\dot{\bm{u}}_s\right\rangle
+
m_s\left\langle \bm{u}_s^{\top}\mathbf{E}_i\ddot{\bm{u}}_s\right\rangle .
\label{eq:S1bc_L_dt_step1}
\end{equation}
The first term vanishes identically because \(\mathbf{E}_i^\top=-\mathbf{E}_i\), so that
\begin{equation}
\bm{v}^{\top}\mathbf{E}_i\bm{v}=0
\qquad
\text{for any } \bm{v}\in\mathbb{R}^3.
\label{eq:S1bc_antisymmetry_identity}
\end{equation}
Using Eq.~\eqref{eq:S1bc_site_eom} in the second term, we obtain
\begin{equation}
\frac{d}{dt}L_i(s)
=
-\kappa L_i(s)
-
\sum_{t\in\mathcal{N}(s)}
\left\langle
\bm{u}_s^{\top}\mathbf{E}_i\mathbf{\Phi}^{(st)}(\bm{u}_s-\bm{u}_t)
\right\rangle
+
\left\langle
\bm{u}_s^{\top}\mathbf{E}_i\bm{\eta}_s
\right\rangle .
\label{eq:S1bc_L_dt_step2}
\end{equation}
The elastic term can be written as
\begin{equation}
\left\langle
\bm{u}_s^{\top}\mathbf{E}_i\mathbf{\Phi}^{(st)}(\bm{u}_s-\bm{u}_t)
\right\rangle
=
\Tr\!\left[
\mathbf{E}_i\mathbf{\Phi}^{(st)}
\left(
\langle \bm{u}_s\bm{u}_s^{\top}\rangle
-
\langle \bm{u}_t\bm{u}_s^{\top}\rangle
\right)
\right].
\label{eq:S1bc_L_bond_term}
\end{equation}
This identifies the bond-resolved phonon-angular-momentum current as
\begin{equation}
j^{(L_i)}_{s\to t}
\equiv
\Tr\!\left[
\mathbf{E}_i\mathbf{\Phi}^{(st)}
\left(
\langle \bm{u}_s\bm{u}_s^{\top}\rangle
-
\langle \bm{u}_t\bm{u}_s^{\top}\rangle
\right)
\right].
\label{eq:S1bc_L_bond_current}
\end{equation}
This shows that the PAM current is governed by nonlocal displacement correlations across a bond. The tensor \(\mathbf{\Phi}^{(st)}\) maps the bond stretch to the corresponding elastic force, and \(\mathbf{E}_i\) extracts the angular-momentum component about axis \(i\).

It is also useful to introduce a flux-like quantity associated with the displacement amplitude
\begin{equation}
A(s)=\Tr\langle \bm{u}_s\bm{u}_s^{\top}\rangle .
\label{eq:S1bc_A_def}
\end{equation}
Its first time derivative is
\begin{equation}
\frac{d}{dt}A(s)
=
2\Tr\langle \bm{u}_s\dot{\bm{u}}_s^{\top}\rangle ,
\label{eq:S1bc_A_dt}
\end{equation}
and differentiating once more gives
\begin{equation}
\frac{d^2}{dt^2}A(s)
=
2\Tr\langle \dot{\bm{u}}_s\dot{\bm{u}}_s^{\top}\rangle
+
2\langle \bm{u}_s\cdot\ddot{\bm{u}}_s\rangle .
\label{eq:S1bc_A_dt2_step1}
\end{equation}
Using Eq.~\eqref{eq:S1bc_site_eom}, we find
\begin{equation}
\frac{d^2}{dt^2}A(s)
=
2\Tr\langle \dot{\bm{u}}_s\dot{\bm{u}}_s^{\top}\rangle
-
2\kappa \Tr\langle \bm{u}_s\dot{\bm{u}}_s^{\top}\rangle
-
\frac{2}{m_s}
\sum_{t\in\mathcal{N}(s)}
\left\langle
\bm{u}_s^{\top}\mathbf{\Phi}^{(st)}(\bm{u}_s-\bm{u}_t)
\right\rangle
+
\frac{2}{m_s}\langle \bm{u}_s\cdot\bm{\eta}_s\rangle .
\label{eq:S1bc_A_dt2_step2}
\end{equation}
This motivates the bond quantity
\begin{equation}
j^{(A)}_{s\to t}
\equiv
\frac{2}{m_s}
\Tr\!\left[
\mathbf{\Phi}^{(st)}
\left(
\langle \bm{u}_s\bm{u}_s^{\top}\rangle
-
\langle \bm{u}_t\bm{u}_s^{\top}\rangle
\right)
\right].
\label{eq:S1bc_A_bond_flux}
\end{equation}
Although \(A(s)\) is not a conserved density, the quantity \(j^{(A)}_{s\to t}\) is useful as a bond-resolved measure of how vibrational amplitude is redistributed through the lattice.

At this stage, the bond contributions have been identified explicitly. The remaining terms describe local exchange with the thermal reservoirs,
\begin{equation}
\langle \dot{\bm{u}}_s\cdot\bm{\eta}_s\rangle,
\qquad
\langle \bm{u}_s^{\top}\mathbf{E}_i\bm{\eta}_s\rangle,
\qquad
\langle \bm{u}_s\cdot\bm{\eta}_s\rangle.
\label{eq:S1bc_bath_terms}
\end{equation}
Because the Langevin force is white noise, these terms are most conveniently evaluated by writing the stochastic dynamics in It\^o differential form \cite{Gardiner2004,VanKampen2007,Risken1996}. We write
\begin{equation}
\bm{\eta}_s(t)\,dt
=
\sqrt{2\kappa m_s k_B T_s}\,d\bm{W}_s(t),
\label{eq:S1bc_eta_ito}
\end{equation}
where \(d\bm{W}_s\) is a three-component Wiener increment satisfying
\begin{equation}
dW_{s,i}\,dW_{t,j}
=
\delta_{st}\delta_{ij}\,dt.
\label{eq:S1bc_wiener_rule}
\end{equation}
The stochastic equations of motion are then
\begin{equation}
d\bm{u}_s=\dot{\bm{u}}_s\,dt,
\label{eq:S1bc_u_ito}
\end{equation}
and
\begin{equation}
d\dot{\bm{u}}_s
=
\left[
-\kappa \dot{\bm{u}}_s
-
\frac{1}{m_s}
\sum_{t\in\mathcal{N}(s)}
\mathbf{\Phi}^{(st)}(\bm{u}_s-\bm{u}_t)
\right]dt
+
\sqrt{\frac{2\kappa k_B T_s}{m_s}}\,d\bm{W}_s .
\label{eq:S1bc_udot_ito}
\end{equation}

We first evaluate the bath terms entering the amplitude and phonon-angular-momentum balances. The displacement \(\bm{u}_s(t)\) depends on the noise history up to time \(t\), while \(d\bm{W}_s(t)\) is the new random kick during the interval \([t,t+dt]\). The two are therefore uncorrelated. Using Eq.~\eqref{eq:S1bc_eta_ito}, this gives
\begin{equation}
\langle \bm{u}_s\cdot\bm{\eta}_s\rangle\,dt
=
\sqrt{2\kappa m_s k_B T_s}\,
\langle \bm{u}_s\cdot d\bm{W}_s\rangle
=
0,
\label{eq:S1bc_u_eta_zero}
\end{equation}
and likewise
\begin{equation}
\langle \bm{u}_s^{\top}\mathbf{E}_i\bm{\eta}_s\rangle\,dt
=
\sqrt{2\kappa m_s k_B T_s}\,
\langle \bm{u}_s^{\top}\mathbf{E}_i\,d\bm{W}_s\rangle
=
0.
\label{eq:S1bc_uEeta_zero}
\end{equation}
These identities show that the baths do not inject a net displacement amplitude or a net local phonon angular momentum directly. In the present isotropic Langevin model, the local phonon angular momentum is transported through the lattice and relaxed by damping, but it is not sourced by an explicit stochastic term.

The kinetic-energy balance is different because the kinetic energy depends quadratically on the velocity, and therefore receives an It\^o correction. Applying It\^o's lemma to
\begin{equation}
E_{\mathrm{kin}}(s)=\frac{m_s}{2}\dot{\bm{u}}_s^2
\label{eq:S1bc_ito_Ekin_def}
\end{equation}
gives
\begin{equation}
dE_{\mathrm{kin}}(s)
=
m_s \dot{\bm{u}}_s\cdot d\dot{\bm{u}}_s
+
\frac{m_s}{2}\, d\dot{\bm{u}}_s\cdot d\dot{\bm{u}}_s .
\label{eq:S1bc_ito_Ekin}
\end{equation}
Using Eq.~\eqref{eq:S1bc_udot_ito}, the quadratic-variation term is
\begin{equation}
d\dot{\bm{u}}_s\cdot d\dot{\bm{u}}_s
=
\frac{2\kappa k_B T_s}{m_s}\,
d\bm{W}_s\cdot d\bm{W}_s .
\label{eq:S1bc_quadvar_step}
\end{equation}
Because
\begin{equation}
d\bm{W}_s\cdot d\bm{W}_s
=
\sum_{i=x,y,z} dW_{s,i}^2
=
3\,dt
\label{eq:S1bc_dW_square}
\end{equation}
in the present three-dimensional formulation, we obtain
\begin{equation}
d\dot{\bm{u}}_s\cdot d\dot{\bm{u}}_s
=
\frac{6\kappa k_B T_s}{m_s}\,dt .
\label{eq:S1bc_quadvar}
\end{equation}
Substituting this into Eq.~\eqref{eq:S1bc_ito_Ekin}, and taking the average, gives
\begin{equation}
\frac{d}{dt}E_{\mathrm{kin}}(s)
=
-\kappa m_s \langle \dot{\bm{u}}_s\cdot\dot{\bm{u}}_s\rangle
-
\sum_{t\in\mathcal{N}(s)}
\left\langle
\dot{\bm{u}}_s^{\top}\mathbf{\Phi}^{(st)}(\bm{u}_s-\bm{u}_t)
\right\rangle
+
3\kappa k_B T_s .
\label{eq:S1bc_Ekin_dt_ito}
\end{equation}
Comparing with Eq.~\eqref{eq:S1bc_Ekin_dt_step2}, we identify
\begin{equation}
\langle \dot{\bm{u}}_s\cdot\bm{\eta}_s\rangle
=
3\kappa k_B T_s .
\label{eq:S1bc_udot_eta}
\end{equation}
This is the local thermal power injected by the reservoir. Physically, the baths inject kinetic energy through stochastic velocity kicks, while the damping term removes energy continuously.

Collecting the results above, the local balance equations become
\begin{subequations}
\begin{align}
\frac{d}{dt}E_{\mathrm{kin}}(s)
+
\sum_{t\in\mathcal{N}(s)} j^{(E)}_{s\to t}
&=
-\kappa m_s \Tr\langle \dot{\bm{u}}_s\dot{\bm{u}}_s^{\top}\rangle
+
3\kappa k_B T_s,
\label{eq:S1bc_energy_balance}
\\
\frac{d}{dt}L_i(s)
+
\sum_{t\in\mathcal{N}(s)} j^{(L_i)}_{s\to t}
&=
-\kappa L_i(s),
\label{eq:S1bc_L_balance}
\\
\frac{d^2}{dt^2}A(s)
&=
2\Tr\langle \dot{\bm{u}}_s\dot{\bm{u}}_s^{\top}\rangle
-
2\kappa \Tr\langle \bm{u}_s\dot{\bm{u}}_s^{\top}\rangle
-
\sum_{t\in\mathcal{N}(s)} j^{(A)}_{s\to t}.
\label{eq:S1bc_A_balance}
\end{align}
\end{subequations}
The first equation describes the balance between elastic energy transport, local dissipation, and thermal injection. The second shows that phonon angular momentum is transported through the lattice and relaxed by damping, without a direct stochastic source term. The third is not a continuity equation in the usual sense, but it defines a useful bond quantity associated with the redistribution of vibrational amplitude.

For visualization, it is often convenient to convert the bond-resolved scalar currents into site current vectors by weighting each bond with its direction,
\begin{equation}
\bm{j}^{(\alpha)}(s)
\equiv
\sum_{t\in\mathcal{N}(s)}
j^{(\alpha)}_{s\to t}\,\hat{\bm{d}}_{st},
\qquad
\alpha\in\{E,A,L_x,L_y,L_z\}.
\label{eq:S1bc_site_current_vectors}
\end{equation}
These vectors provide a coarse-grained representation of the bond-resolved transport pattern in real space.

\subsection{Linear response to a weak temperature gradient}
\label{subsec:S1_linear_response_gradient}

The steady-state covariances derived above depend linearly on the bath temperatures. The same is therefore true for the local fields and bond-resolved currents. To describe the response to a weak thermal bias, it is enough to isolate the contribution of each bath site and then superpose these contributions for a general temperature profile.

We begin by decomposing the modal noise matrix \(\mathbf{W}\) into contributions from the individual bath sites. Let \(\mathbf{U}_r\) denote the \(3\times 3N\) block of \(\mathbf{U}\) associated with site \(r\), and define
\begin{equation}
\mathbf{\Pi}_r \equiv \mathbf{U}_r^{\top}\mathbf{U}_r .
\label{eq:S1lr_site_projector}
\end{equation}
The matrix \(\mathbf{\Pi}_r\) measures the overlap of normal modes on site \(r\). Because the blocks of \(\mathbf{U}\) span all lattice sites, they satisfy
\begin{equation}
\sum_{r=1}^{N}\mathbf{\Pi}_r=\mathbf{I}.
\label{eq:S1lr_projector_sum}
\end{equation}
Using the temperature matrix
\begin{equation}
\mathbf{T}=\mathrm{diag}(T_1,\dots,T_N)\otimes \mathbf{I}_3,
\end{equation}
the modal noise covariance
\begin{equation}
\mathbf{W}=2\kappa k_B\,\mathbf{U}^{\top}\mathbf{T}\mathbf{U}
\end{equation}
can be written as
\begin{equation}
\mathbf{W}
=
2\kappa k_B
\sum_{r=1}^{N}
T_r\,\mathbf{\Pi}_r .
\label{eq:S1lr_W_decomposition}
\end{equation}
Each bath site therefore contributes to \(\mathbf{W}\) through its local temperature \(T_r\) and its mode-overlap matrix \(\mathbf{\Pi}_r\).

Substituting Eq.~\eqref{eq:S1lr_W_decomposition} into the modal covariances gives
\begin{subequations}
\begin{align}
\langle \bm{Q}\bm{Q}^{\top}\rangle
&=
2\kappa k_B
\sum_{r=1}^{N}
T_r\,
\left(\mathbf{\Pi}_r\circ \mathbf{C}_{QQ}\right),
\label{eq:S1lr_modal_cov_QQ}
\\
\langle \bm{Q}\dot{\bm{Q}}^{\top}\rangle
&=
2\kappa k_B
\sum_{r=1}^{N}
T_r\,
\left(\mathbf{\Pi}_r\circ \mathbf{C}_{Q\dot{Q}}\right),
\label{eq:S1lr_modal_cov_QdQ}
\\
\langle \dot{\bm{Q}}\dot{\bm{Q}}^{\top}\rangle
&=
2\kappa k_B
\sum_{r=1}^{N}
T_r\,
\left(\mathbf{\Pi}_r\circ \mathbf{C}_{\dot{Q}\dot{Q}}\right).
\label{eq:S1lr_modal_cov_dQdQ}
\end{align}
\end{subequations}
Projecting these expressions back to real space gives
\begin{subequations}
\begin{align}
\langle \bm{u}_s \bm{u}_t^{\top}\rangle
&=
2\kappa k_B
\sum_{r=1}^{N}
T_r\,
\mathbf{R}_s
\left(\mathbf{\Pi}_r\circ \mathbf{C}_{QQ}\right)
\mathbf{R}_t^{\top},
\label{eq:S1lr_real_cov_uu}
\\
\langle \bm{u}_s \dot{\bm{u}}_t^{\top}\rangle
&=
2\kappa k_B
\sum_{r=1}^{N}
T_r\,
\mathbf{R}_s
\left(\mathbf{\Pi}_r\circ \mathbf{C}_{Q\dot{Q}}\right)
\mathbf{R}_t^{\top},
\label{eq:S1lr_real_cov_udotu}
\\
\langle \dot{\bm{u}}_s \dot{\bm{u}}_t^{\top}\rangle
&=
2\kappa k_B
\sum_{r=1}^{N}
T_r\,
\mathbf{R}_s
\left(\mathbf{\Pi}_r\circ \mathbf{C}_{\dot{Q}\dot{Q}}\right)
\mathbf{R}_t^{\top}.
\label{eq:S1lr_real_cov_dotudotu}
\end{align}
\end{subequations}
Every local field and every bond-resolved current can then be written as a sum of contributions coming from the individual bath sites.

For the coarse-grained site currents introduced in Eq.~\eqref{eq:S1bc_site_current_vectors}, we write
\begin{equation}
\bm{j}^{(\alpha)}(s)
=
\sum_{r=1}^{N}
\bm{\sigma}^{(\alpha)}(s|r)\,T_r,
\label{eq:S1lr_site_kernel_def}
\end{equation}
where \(\bm{\sigma}^{(\alpha)}(s|r)\in\mathbb{R}^{3}\) is the site-resolved response kernel. It gives the current of type \(\alpha\) flowing through observation site \(s\) per unit temperature bias applied at bath site \(r\). Here \(\alpha\in\{E,A,L_x,L_y,L_z\}\), with \(E\) denoting the kinetic-energy current, \(A\) the amplitude flux introduced above, and \(L_i\) the current of the \(i\)-th component of phonon angular momentum.

For compactness, we introduce the bond-difference matrix
\begin{equation}
\Delta \mathbf{R}_{st}\equiv \mathbf{R}_s-\mathbf{R}_t .
\label{eq:S1lr_bond_difference}
\end{equation}
The response kernels can then be written directly in terms of the relative displacement across the bond. For the kinetic-energy current,
\begin{equation}
\bm{\sigma}^{(E)}(s|r)
=
2\kappa k_B
\sum_{t\in\mathcal{N}(s)}
\hat{\bm{d}}_{st}\,
\Tr\!\left[
\mathbf{\Phi}^{(st)}
\,\Delta \mathbf{R}_{st}\,
\left(\mathbf{\Pi}_r\circ \mathbf{C}_{Q\dot{Q}}\right)
\mathbf{R}_s^{\top}
\right],
\label{eq:S1lr_sigma_E}
\end{equation}
for the amplitude flux,
\begin{equation}
\bm{\sigma}^{(A)}(s|r)
=
\frac{4\kappa k_B}{m_s}
\sum_{t\in\mathcal{N}(s)}
\hat{\bm{d}}_{st}\,
\Tr\!\left[
\mathbf{\Phi}^{(st)}
\,\Delta \mathbf{R}_{st}\,
\left(\mathbf{\Pi}_r\circ \mathbf{C}_{QQ}\right)
\mathbf{R}_s^{\top}
\right],
\label{eq:S1lr_sigma_A}
\end{equation}
and for the phonon-angular-momentum current,
\begin{equation}
\bm{\sigma}^{(L_i)}(s|r)
=
2\kappa k_B
\sum_{t\in\mathcal{N}(s)}
\hat{\bm{d}}_{st}\,
\Tr\!\left[
\mathbf{E}_i\mathbf{\Phi}^{(st)}
\,\Delta \mathbf{R}_{st}\,
\left(\mathbf{\Pi}_r\circ \mathbf{C}_{QQ}\right)
\mathbf{R}_s^{\top}
\right].
\label{eq:S1lr_sigma_L}
\end{equation}
The matrix \(\mathbf{\Pi}_r\) determines how bath site \(r\) populates and mixes the normal modes, the blocks of \(\mathbf{R}\) convert this modal covariance into real-space displacement and velocity correlations, and the bond tensor \(\mathbf{\Phi}^{(st)}\) turns those correlations into transport along bond \((s,t)\).

We now specialize to the regime of a weak temperature modulation around a uniform reference temperature,
\begin{equation}
T_r=T^{(0)}+\delta T_r,
\qquad
|\delta T_r|\ll T^{(0)}.
\label{eq:S1lr_small_temperature_modulation}
\end{equation}
The induced current is then linear in the local temperature offsets,
\begin{equation}
\delta \bm{j}^{(\alpha)}(s)
=
\sum_{r=1}^{N}
\bm{\sigma}^{(\alpha)}(s|r)\,\delta T_r .
\label{eq:S1lr_deltaj_linear}
\end{equation}

At this point it is useful to separate the response to a uniform temperature shift from the response to a temperature gradient. Summing Eq.~\eqref{eq:S1lr_sigma_E} over all bath sites and using Eq.~\eqref{eq:S1lr_projector_sum}, we obtain
\begin{equation}
\sum_{r=1}^{N}\bm{\sigma}^{(E)}(s|r)
=
2\kappa k_B
\sum_{t\in\mathcal{N}(s)}
\hat{\bm{d}}_{st}\,
\Tr\!\left[
\mathbf{\Phi}^{(st)}
\,\Delta \mathbf{R}_{st}\,
\left(\mathbf{I}\circ \mathbf{C}_{Q\dot{Q}}\right)
\mathbf{R}_s^{\top}
\right].
\label{eq:S1lr_sigma_E_sum}
\end{equation}
Since \(C_{Q\dot{Q}}(\mu,\mu)=0\), one has
\begin{equation}
\mathbf{I}\circ \mathbf{C}_{Q\dot{Q}}=0,
\label{eq:S1lr_CQdQ_diag_zero}
\end{equation}
and therefore
\begin{equation}
\sum_{r=1}^{N}\bm{\sigma}^{(E)}(s|r)=0.
\label{eq:S1lr_sigma_E_neutral}
\end{equation}
A uniform temperature shift thus produces no kinetic-energy current.

For the amplitude and phonon-angular-momentum channels, there is in general no analogous cancellation, because \(\mathbf{I}\circ \mathbf{C}_{QQ}\neq 0\). To isolate the response to a temperature gradient, we define
\begin{equation}
\tilde{\bm{\sigma}}^{(\alpha)}(s|r)
\equiv
\bm{\sigma}^{(\alpha)}(s|r)
-
\frac{1}{N}\sum_{r'=1}^{N}\bm{\sigma}^{(\alpha)}(s|r'),
\qquad
\alpha\in\{A,L_x,L_y,L_z\}.
\label{eq:S1lr_tilde_kernel}
\end{equation}
By construction,
\begin{equation}
\sum_{r=1}^{N}\tilde{\bm{\sigma}}^{(\alpha)}(s|r)=0.
\label{eq:S1lr_tilde_sum_zero}
\end{equation}
For the kinetic-energy channel we simply set
\begin{equation}
\tilde{\bm{\sigma}}^{(E)}(s|r)\equiv \bm{\sigma}^{(E)}(s|r),
\label{eq:S1lr_tilde_kernel_E}
\end{equation}
since Eq.~\eqref{eq:S1lr_sigma_E_neutral} already holds identically.

The gradient-induced part of the current can then be written as
\begin{equation}
\delta \bm{j}^{(\alpha)}(s)
=
\sum_{r=1}^{N}
\tilde{\bm{\sigma}}^{(\alpha)}(s|r)\,\delta T_r .
\label{eq:S1lr_deltaj_gradient_only}
\end{equation}
For a uniform imposed temperature gradient, we write
\begin{equation}
\delta T_r
=
\sum_{j=x,y,z}
(r_{r,j}-r_{0,j})\,\partial_j T,
\label{eq:S1lr_uniform_gradient}
\end{equation}
where \(\bm{r}_0\) is an arbitrary reference point. Substituting Eq.~\eqref{eq:S1lr_uniform_gradient} into Eq.~\eqref{eq:S1lr_deltaj_gradient_only} gives
\begin{equation}
\delta j_i^{(\alpha)}(s)
=
\sum_{j=x,y,z}
\sigma_{ij}^{(\alpha)}(s)\,\partial_j T,
\label{eq:S1lr_local_conductivity_law}
\end{equation}
with site-resolved conductivity tensor
\begin{equation}
\sigma_{ij}^{(\alpha)}(s)
\equiv
\sum_{r=1}^{N}
(r_{r,j}-r_{0,j})\,
\tilde{\sigma}_{i}^{(\alpha)}(s|r).
\label{eq:S1lr_local_conductivity_tensor}
\end{equation}
Thus \(\sigma_{ij}^{(\alpha)}(s)\) maps the \(j\)-component of the imposed temperature gradient to the \(i\)-component of the local current of type \(\alpha\).

Because Eq.~\eqref{eq:S1lr_tilde_sum_zero} holds, the conductivity tensor is independent of the choice of \(\bm{r}_0\). Indeed, shifting the reference point by a constant vector changes Eq.~\eqref{eq:S1lr_local_conductivity_tensor} by a term proportional to \(\sum_r \tilde{\sigma}_{i}^{(\alpha)}(s|r)\), which vanishes. A convenient explicit choice is
\begin{equation}
\bm{r}_0
=
\frac{1}{N}\sum_{r=1}^{N}\bm{r}_r,
\label{eq:S1lr_average_origin}
\end{equation}
the simple average of the lattice-site positions.

A bulk conductivity tensor is obtained by averaging the local conductivity over a set of interior sites,
\begin{equation}
\sigma^{(\alpha),\mathrm{bulk}}_{ij}
\equiv
\frac{1}{N_{\mathrm{bulk}}}
\sum_{s\in \mathrm{bulk}}
\sigma^{(\alpha)}_{ij}(s).
\label{eq:S1lr_bulk_conductivity}
\end{equation}
This bulk tensor gives the coarse-grained transport response of the lattice. In particular, for \(\alpha=L_i\) it describes the transport of the \(i\)-th component of phonon angular momentum under a thermal gradient, while for \(\alpha=E\) it describes the corresponding response of the kinetic-energy current.

\subsection{Gyroscopic dynamics and weak-magnetic-field expansion}
\label{subsec:S1_gyroscopic_dynamics}

We now extend the zero-field framework to the case in which the lattice is coupled to magnetic order or to an external magnetic field. In the classical lattice dynamics considered here, the field enters through a gyroscopic force that is linear in the ionic velocities and antisymmetric in Cartesian space.

For each site \(s\), we introduce an effective gyromagnetic ratio \(\gamma_s\), and define the local gyroscopic frequency
\begin{equation}
\bm{\Omega}^{\mathrm g}_s \equiv \gamma_s \bm{B}.
\label{eq:S1g_Omegag_def}
\end{equation}
The corresponding magnetic coupling can be written as
\begin{equation}
\mathcal{L}_B
=
\sum_{s=1}^{N}
\gamma_s \bm{B}\cdot \bm{L}^{\mathrm{ph}}_s,
\qquad
\bm{L}^{\mathrm{ph}}_s
=
m_s\,\bm{u}_s\times \dot{\bm{u}}_s ,
\label{eq:S1g_LB_def}
\end{equation}
so that the field couples directly to the local phonon angular momentum. Varying this term gives a gyroscopic force proportional to the velocity,
\begin{equation}
\bigl(2\mathbf{J}\dot{\bm{u}}\bigr)_s
=
2m_s\,\bm{\Omega}^{\mathrm g}_s\times \dot{\bm{u}}_s .
\label{eq:S1g_gyro_force}
\end{equation}

It is convenient to represent this force through a \(3\times 3\) antisymmetric matrix at each site. We define the site block \(\mathbf{J}_s\) by
\begin{equation}
\mathbf{J}_s \bm{v}
\equiv
m_s\,\bm{\Omega}^{\mathrm g}_s\times \bm{v},
\qquad
\mathbf{J}_s^{\top}=-\mathbf{J}_s ,
\label{eq:S1g_Js_def}
\end{equation}
for any \(\bm{v}\in\mathbb{R}^3\). In Cartesian form,
\begin{equation}
\mathbf{J}_s
=
m_s
\begin{pmatrix}
0 & -\Omega^{\mathrm g}_{s,z} & \Omega^{\mathrm g}_{s,y} \\
\Omega^{\mathrm g}_{s,z} & 0 & -\Omega^{\mathrm g}_{s,x} \\
-\Omega^{\mathrm g}_{s,y} & \Omega^{\mathrm g}_{s,x} & 0
\end{pmatrix}
=
m_s\gamma_s
\begin{pmatrix}
0 & -B_z & B_y \\
B_z & 0 & -B_x \\
-B_y & B_x & 0
\end{pmatrix}.
\label{eq:S1g_Js_matrix}
\end{equation}
Assembling these blocks over all sites gives the full gyroscopic matrix
\begin{equation}
\mathbf{J}=\mathrm{diag}\!\left(\mathbf{J}_1,\dots,\mathbf{J}_N\right),
\qquad
\mathbf{J}^{\top}=-\mathbf{J}.
\label{eq:S1g_J_def}
\end{equation}

The real-space equations of motion in the presence of the field become
\begin{equation}
\mathbf{M}\ddot{\bm{u}}
+
\kappa \mathbf{M}\dot{\bm{u}}
+
\mathbf{K}\bm{u}
+
2\mathbf{J}\dot{\bm{u}}
=
\bm{\eta}(t).
\label{eq:S1g_eom_real_B}
\end{equation}
The magnetic field therefore modifies only the deterministic part of the dynamics. The reservoir statistics remain unchanged at linear order in the field,
\begin{equation}
\langle \eta_{s,i}(t)\eta_{t,j}(t')\rangle
=
2\kappa m_s k_B T_s\,\delta_{st}\delta_{ij}\delta(t-t').
\label{eq:S1g_noise_real_B}
\end{equation}

Passing to the mass-normalized variables introduced earlier,
\begin{equation}
\tilde{\bm{u}}=\mathbf{M}^{1/2}\bm{u},
\qquad
\tilde{\bm{\eta}}=\mathbf{M}^{-1/2}\bm{\eta},
\qquad
\mathbf{D}=\mathbf{M}^{-1/2}\mathbf{K}\mathbf{M}^{-1/2},
\label{eq:S1g_mass_norm_defs}
\end{equation}
the equations of motion become
\begin{equation}
\ddot{\tilde{\bm{u}}}
+
\kappa \dot{\tilde{\bm{u}}}
+
\mathbf{D}\tilde{\bm{u}}
+
2\tilde{\mathbf{J}}\,\dot{\tilde{\bm{u}}}
=
\tilde{\bm{\eta}}(t),
\label{eq:S1g_eom_massnorm_B}
\end{equation}
with
\begin{equation}
\tilde{\mathbf{J}}
\equiv
\mathbf{M}^{-1/2}\mathbf{J}\mathbf{M}^{-1/2}.
\label{eq:S1g_J_tilde_def}
\end{equation}
Because \(\mathbf{M}\) is diagonal with scalar \(3\times 3\) blocks, \(\tilde{\mathbf{J}}\) remains antisymmetric,
\begin{equation}
\tilde{\mathbf{J}}^{\top}=-\tilde{\mathbf{J}}.
\label{eq:S1g_J_tilde_antisym}
\end{equation}

We now project to the normal-mode basis of \(\mathbf{D}\), defined by
\begin{equation}
\mathbf{D}\mathbf{U}=\mathbf{U}\mathbf{\Omega}^2,
\qquad
\mathbf{U}^{\top}\mathbf{U}=\mathbf{I},
\qquad
\bm{Q}=\mathbf{U}^{\top}\tilde{\bm{u}},
\qquad
\bm{\zeta}=\mathbf{U}^{\top}\tilde{\bm{\eta}}.
\label{eq:S1g_modal_defs}
\end{equation}
In this basis the equations of motion take the form
\begin{equation}
\ddot{\bm{Q}}
+
\kappa \dot{\bm{Q}}
+
\mathbf{\Omega}^2 \bm{Q}
+
2\mathbf{J}_{m}\dot{\bm{Q}}
=
\bm{\zeta}(t),
\label{eq:S1g_eom_modal_B}
\end{equation}
where
\begin{equation}
\mathbf{J}_{m}\equiv \mathbf{U}^{\top}\tilde{\mathbf{J}}\mathbf{U},
\qquad
\mathbf{J}_{m}^{\top}=-\mathbf{J}_{m}.
\label{eq:S1g_Jm_def}
\end{equation}
The matrix \(\mathbf{J}_{m}\) is linear in \(\bm{B}\) and mixes the normal modes through their velocities. The noise in modal space is unchanged,
\begin{equation}
\langle \bm{\zeta}(t)\bm{\zeta}(t')^{\top}\rangle
=
\mathbf{W}\,\delta(t-t'),
\label{eq:S1g_modal_noise_B}
\end{equation}
with the same zero-field matrix \(\mathbf{W}\) derived above.

We assume the gyroscopic coupling is weak and work only to first order in the magnetic field. The equal-time modal covariance matrices in the presence of the field are
\begin{subequations}
\begin{align}
\mathbf{C}_{QQ}^{(B)}
&\equiv
\langle \bm{Q}\bm{Q}^{\top}\rangle_{B},
\label{eq:S1g_covQQ_B}
\\
\mathbf{C}_{Q\dot{Q}}^{(B)}
&\equiv
\langle \bm{Q}\dot{\bm{Q}}^{\top}\rangle_{B},
\label{eq:S1g_covQdQ_B}
\\
\mathbf{C}_{\dot{Q}\dot{Q}}^{(B)}
&\equiv
\langle \dot{\bm{Q}}\dot{\bm{Q}}^{\top}\rangle_{B}.
\label{eq:S1g_covdQdQ_B}
\end{align}
\end{subequations}
We expand them around the zero-field steady state as
\begin{subequations}
\begin{align}
\mathbf{C}_{QQ}^{(B)}
&=
\mathbf{C}_{QQ}^{(0)}
+
\delta \mathbf{C}_{QQ},
\label{eq:S1g_covQQ_expand}
\\
\mathbf{C}_{Q\dot{Q}}^{(B)}
&=
\mathbf{C}_{Q\dot{Q}}^{(0)}
+
\delta \mathbf{C}_{Q\dot{Q}},
\label{eq:S1g_covQdQ_expand}
\\
\mathbf{C}_{\dot{Q}\dot{Q}}^{(B)}
&=
\mathbf{C}_{\dot{Q}\dot{Q}}^{(0)}
+
\delta \mathbf{C}_{\dot{Q}\dot{Q}},
\label{eq:S1g_covdQdQ_expand}
\end{align}
\end{subequations}
where
\begin{subequations}
\begin{align}
\mathbf{C}_{QQ}^{(0)}
&=
\langle \bm{Q}\bm{Q}^{\top}\rangle
=
\mathbf{W}\circ \mathbf{C}_{QQ},
\label{eq:S1g_covQQ_zero}
\\
\mathbf{C}_{Q\dot{Q}}^{(0)}
&=
\langle \bm{Q}\dot{\bm{Q}}^{\top}\rangle
=
\mathbf{W}\circ \mathbf{C}_{Q\dot{Q}},
\label{eq:S1g_covQdQ_zero}
\\
\mathbf{C}_{\dot{Q}\dot{Q}}^{(0)}
&=
\langle \dot{\bm{Q}}\dot{\bm{Q}}^{\top}\rangle
=
\mathbf{W}\circ \mathbf{C}_{\dot{Q}\dot{Q}},
\label{eq:S1g_covdQdQ_zero}
\end{align}
\end{subequations}
are the zero-field covariances obtained previously, and the corrections
\begin{equation}
\delta \mathbf{C}_{QQ},
\qquad
\delta \mathbf{C}_{Q\dot{Q}},
\qquad
\delta \mathbf{C}_{\dot{Q}\dot{Q}}
\label{eq:S1g_deltaC_list}
\end{equation}
are linear in the magnetic field.

These matrices contain the full weak-field correction to the nonequilibrium steady state. The magnetic field enters only through the antisymmetric mode-space matrix \(\mathbf{J}_{m}\), while the temperature profile and damping remain encoded in the zero-field quantities \(\mathbf{W}\), \(\mathbf{C}_{QQ}^{(0)}\), \(\mathbf{C}_{Q\dot{Q}}^{(0)}\), and \(\mathbf{C}_{\dot{Q}\dot{Q}}^{(0)}\). Next we derive the linear equations satisfied by \(\delta \mathbf{C}_{QQ}\), \(\delta \mathbf{C}_{Q\dot{Q}}\), and \(\delta \mathbf{C}_{\dot{Q}\dot{Q}}\), and solve them explicitly to first order in the field.

\subsection{Linear-in-\texorpdfstring{$B$}{B} correction to modal covariances}
\label{subsec:S1_linear_in_B_covariances}

Using the weak-field expansion introduced above, we now derive the equations satisfied by the first-order corrections
\begin{equation}
\delta \mathbf{C}_{QQ},
\qquad
\delta \mathbf{C}_{Q\dot{Q}},
\qquad
\delta \mathbf{C}_{\dot{Q}\dot{Q}}.
\label{eq:S1lb_deltaC_list}
\end{equation}
These matrices are linear in the magnetic field and contain the full \(B\)-odd correction to the modal steady state.

We start from the modal equation of motion in the presence of the gyroscopic coupling,
\begin{equation}
\ddot{\bm{Q}}
+
\kappa \dot{\bm{Q}}
+
\mathbf{\Omega}^{2}\bm{Q}
+
2\mathbf{J}_{m}\dot{\bm{Q}}
=
\bm{\zeta}(t),
\label{eq:S1lb_eom_modal_B}
\end{equation}
with modal noise correlations
\begin{equation}
\langle \bm{\zeta}(t)\bm{\zeta}(t')^{\top}\rangle
=
\mathbf{W}\,\delta(t-t').
\label{eq:S1lb_modal_noise}
\end{equation}
Here \(\mathbf{J}_{m}\) is antisymmetric and linear in \(\bm{B}\), while \(\mathbf{W}\) is unchanged at this order.

We first use the steady-state condition for the equal-time displacement covariance,
\begin{equation}
\frac{d}{dt}\langle \bm{Q}\bm{Q}^{\top}\rangle_{B}
=
\langle \dot{\bm{Q}}\bm{Q}^{\top}\rangle_{B}
+
\langle \bm{Q}\dot{\bm{Q}}^{\top}\rangle_{B}
=
0.
\label{eq:S1lb_dQQ_dt}
\end{equation}
Since
\begin{equation}
\langle \dot{\bm{Q}}\bm{Q}^{\top}\rangle_{B}
=
\left(\mathbf{C}_{Q\dot{Q}}^{(B)}\right)^{\top},
\label{eq:S1lb_dQ_Q_relation}
\end{equation}
this implies
\begin{equation}
\mathbf{C}_{Q\dot{Q}}^{(B)}
+
\left(\mathbf{C}_{Q\dot{Q}}^{(B)}\right)^{\top}
=
0.
\label{eq:S1lb_CQdQ_antisym_full}
\end{equation}
Using the weak-field expansion, we obtain
\begin{equation}
\delta \mathbf{C}_{Q\dot{Q}}
+
\delta \mathbf{C}_{Q\dot{Q}}^{\top}
=
0.
\label{eq:S1lb_deltaCQdQ_antisym}
\end{equation}
The linear-in-\(B\) correction to the mixed covariance is therefore antisymmetric.

Next we consider
\begin{equation}
\frac{d}{dt}\langle \bm{Q}\dot{\bm{Q}}^{\top}\rangle_{B}
=
\langle \dot{\bm{Q}}\dot{\bm{Q}}^{\top}\rangle_{B}
+
\langle \bm{Q}\ddot{\bm{Q}}^{\top}\rangle_{B}
=
0.
\label{eq:S1lb_dQdQ_dt}
\end{equation}
Using Eq.~\eqref{eq:S1lb_eom_modal_B}, we find
\begin{equation}
\mathbf{C}_{\dot{Q}\dot{Q}}^{(B)}
-
\mathbf{C}_{QQ}^{(B)}\mathbf{\Omega}^{2}
-
\kappa \mathbf{C}_{Q\dot{Q}}^{(B)}
+
2\mathbf{C}_{Q\dot{Q}}^{(B)}\mathbf{J}_{m}
+
\langle \bm{Q}\bm{\zeta}^{\top}\rangle_{B}
=
0.
\label{eq:S1lb_cov_eq_1_pre}
\end{equation}
The last term vanishes in the It\^o convention. The vector \(\bm{Q}(t)\) depends on the noise history up to time \(t\), while \(\bm{\zeta}(t)\,dt\) is the new Wiener increment during the interval \([t,t+dt]\). The two are therefore uncorrelated, so
\begin{equation}
\langle \bm{Q}\bm{\zeta}^{\top}\rangle_{B}=0.
\label{eq:S1lb_Qzeta_zero}
\end{equation}
Equation~\eqref{eq:S1lb_cov_eq_1_pre} then becomes
\begin{equation}
\mathbf{C}_{\dot{Q}\dot{Q}}^{(B)}
-
\mathbf{C}_{QQ}^{(B)}\mathbf{\Omega}^{2}
-
\kappa \mathbf{C}_{Q\dot{Q}}^{(B)}
+
2\mathbf{C}_{Q\dot{Q}}^{(B)}\mathbf{J}_{m}
=
0.
\label{eq:S1lb_cov_eq_1_full}
\end{equation}
Keeping only terms linear in the field gives
\begin{equation}
\delta \mathbf{C}_{\dot{Q}\dot{Q}}
-
\delta \mathbf{C}_{QQ}\mathbf{\Omega}^{2}
-
\kappa \delta \mathbf{C}_{Q\dot{Q}}
+
2\mathbf{C}_{Q\dot{Q}}^{(0)}\mathbf{J}_{m}
=
0.
\label{eq:S1lb_cov_eq_1_linear}
\end{equation}

The velocity covariance requires an It\^o correction, exactly as in the real-space derivation above. Writing the stochastic differential equation corresponding to Eq.~\eqref{eq:S1lb_eom_modal_B},
\begin{equation}
d\dot{\bm{Q}}
=
\left(
-\kappa \dot{\bm{Q}}
-
\mathbf{\Omega}^{2}\bm{Q}
-
2\mathbf{J}_{m}\dot{\bm{Q}}
\right)dt
+
d\bm{\mathcal W},
\label{eq:S1lb_dQdot_ito}
\end{equation}
with
\begin{equation}
\langle d\bm{\mathcal W}\,d\bm{\mathcal W}^{\top}\rangle
=
\mathbf{W}\,dt,
\label{eq:S1lb_modal_wiener}
\end{equation}
we obtain
\begin{equation}
\frac{d}{dt}\langle \dot{\bm{Q}}\dot{\bm{Q}}^{\top}\rangle_{B}
=
\langle \ddot{\bm{Q}}\dot{\bm{Q}}^{\top}\rangle_{B}
+
\langle \dot{\bm{Q}}\ddot{\bm{Q}}^{\top}\rangle_{B}
+
\mathbf{W}
=
0.
\label{eq:S1lb_ddQdQ_dt}
\end{equation}
Using Eq.~\eqref{eq:S1lb_eom_modal_B}, this gives
\begin{equation}
2\kappa \mathbf{C}_{\dot{Q}\dot{Q}}^{(B)}
+
\left[\mathbf{\Omega}^{2},\mathbf{C}_{Q\dot{Q}}^{(B)}\right]
-
2\left[\mathbf{C}_{\dot{Q}\dot{Q}}^{(B)},\mathbf{J}_{m}\right]
=
\mathbf{W},
\label{eq:S1lb_cov_eq_2_full}
\end{equation}
where \([\,\cdot,\cdot\,]\) denotes the commutator. Subtracting the zero-field relation and keeping only terms linear in the field gives
\begin{equation}
2\kappa \delta \mathbf{C}_{\dot{Q}\dot{Q}}
+
\left[\mathbf{\Omega}^{2},\delta \mathbf{C}_{Q\dot{Q}}\right]
-
2\left[\mathbf{C}_{\dot{Q}\dot{Q}}^{(0)},\mathbf{J}_{m}\right]
=
0.
\label{eq:S1lb_cov_eq_2_linear}
\end{equation}

It is convenient to define the three driver matrices
\begin{subequations}
\begin{align}
\mathbf{A}
&\equiv
\left[\mathbf{J}_{m},\mathbf{C}_{Q\dot{Q}}^{(0)}\right],
\label{eq:S1lb_A_def}
\\
\mathbf{B}
&\equiv
\left\{\mathbf{C}_{Q\dot{Q}}^{(0)},\mathbf{J}_{m}\right\},
\label{eq:S1lb_B_def}
\\
\mathbf{C}
&\equiv
\left[\mathbf{C}_{\dot{Q}\dot{Q}}^{(0)},\mathbf{J}_{m}\right],
\label{eq:S1lb_C_def}
\end{align}
\end{subequations}
where \(\{\,\cdot,\cdot\,\}\) denotes the anticommutator. Because \(\mathbf{C}_{Q\dot{Q}}^{(0)}\) and \(\mathbf{J}_{m}\) are antisymmetric, while \(\mathbf{C}_{\dot{Q}\dot{Q}}^{(0)}\) is symmetric, these matrices satisfy
\begin{equation}
\mathbf{A}^{\top}=-\mathbf{A},
\qquad
\mathbf{B}^{\top}=\mathbf{B},
\qquad
\mathbf{C}^{\top}=\mathbf{C}.
\label{eq:S1lb_ABC_symmetry}
\end{equation}

Taking the transpose of Eq.~\eqref{eq:S1lb_cov_eq_1_linear} and combining it with Eq.~\eqref{eq:S1lb_deltaCQdQ_antisym}, we obtain the coupled matrix equations
\begin{subequations}
\begin{align}
2\,\delta \mathbf{C}_{\dot{Q}\dot{Q}}
-
\left\{\mathbf{\Omega}^{2},\delta \mathbf{C}_{QQ}\right\}
+
2\mathbf{B}
&=
0,
\label{eq:S1lb_matrix_eq_1}
\\
\left[\mathbf{\Omega}^{2},\delta \mathbf{C}_{QQ}\right]
-
2\kappa\,\delta \mathbf{C}_{Q\dot{Q}}
-
2\mathbf{A}
&=
0,
\label{eq:S1lb_matrix_eq_2}
\\
\left[\mathbf{\Omega}^{2},\delta \mathbf{C}_{Q\dot{Q}}\right]
+
2\kappa\,\delta \mathbf{C}_{\dot{Q}\dot{Q}}
-
2\mathbf{C}
&=
0.
\label{eq:S1lb_matrix_eq_3}
\end{align}
\end{subequations}
These equations decouple element by element in the normal-mode basis, because \(\mathbf{\Omega}^{2}\) is diagonal.

For each pair of modes \((\mu,\nu)\), Eqs.~\eqref{eq:S1lb_matrix_eq_1}--\eqref{eq:S1lb_matrix_eq_3} become
\begin{subequations}
\begin{align}
2\,\delta C_{\dot{Q}\dot{Q},\mu\nu}
-
\left(\Omega_\mu^{2}+\Omega_\nu^{2}\right)\delta C_{QQ,\mu\nu}
+
2B_{\mu\nu}
&=
0,
\label{eq:S1lb_component_eq_1}
\\
\left(\Omega_\mu^{2}-\Omega_\nu^{2}\right)\delta C_{QQ,\mu\nu}
-
2\kappa\,\delta C_{Q\dot{Q},\mu\nu}
-
2A_{\mu\nu}
&=
0,
\label{eq:S1lb_component_eq_2}
\\
\left(\Omega_\mu^{2}-\Omega_\nu^{2}\right)\delta C_{Q\dot{Q},\mu\nu}
+
2\kappa\,\delta C_{\dot{Q}\dot{Q},\mu\nu}
-
2C_{\mu\nu}
&=
0.
\label{eq:S1lb_component_eq_3}
\end{align}
\end{subequations}
Solving this linear system gives
\begin{subequations}
\begin{align}
\delta C_{QQ,\mu\nu}
&=
\frac{
2\left(\Omega_\mu^{2}-\Omega_\nu^{2}\right)A_{\mu\nu}
+
4\kappa^{2} B_{\mu\nu}
+
4\kappa C_{\mu\nu}}
{
\left(\Omega_\mu^{2}-\Omega_\nu^{2}\right)^{2}
+
2\kappa^{2}\left(\Omega_\mu^{2}+\Omega_\nu^{2}\right)
},
\label{eq:S1lb_deltaCQQ_solution}
\\
\delta C_{Q\dot{Q},\mu\nu}
&=
\frac{
-2\kappa\left(\Omega_\mu^{2}+\Omega_\nu^{2}\right)A_{\mu\nu}
+
2\kappa\left(\Omega_\mu^{2}-\Omega_\nu^{2}\right)B_{\mu\nu}
+
2\left(\Omega_\mu^{2}-\Omega_\nu^{2}\right)C_{\mu\nu}}
{
\left(\Omega_\mu^{2}-\Omega_\nu^{2}\right)^{2}
+
2\kappa^{2}\left(\Omega_\mu^{2}+\Omega_\nu^{2}\right)
},
\label{eq:S1lb_deltaCQdQ_solution}
\\
\delta C_{\dot{Q}\dot{Q},\mu\nu}
&=
\frac{
\left(\Omega_\mu^{2}+\Omega_\nu^{2}\right)\left(\Omega_\mu^{2}-\Omega_\nu^{2}\right)A_{\mu\nu}
-
\left(\Omega_\mu^{2}-\Omega_\nu^{2}\right)^{2}B_{\mu\nu}
+
2\kappa\left(\Omega_\mu^{2}+\Omega_\nu^{2}\right)C_{\mu\nu}}
{
\left(\Omega_\mu^{2}-\Omega_\nu^{2}\right)^{2}
+
2\kappa^{2}\left(\Omega_\mu^{2}+\Omega_\nu^{2}\right)
}.
\label{eq:S1lb_deltaCdQdQ_solution}
\end{align}
\end{subequations}
These formulas remain valid on the diagonal, \(\mu=\nu\), in which case \(A_{\mu\mu}=0\) because \(\mathbf{A}\) is antisymmetric. One then obtains
\begin{equation}
\delta C_{Q\dot{Q},\mu\mu}=0,
\qquad
\delta C_{\dot{Q}\dot{Q},\mu\mu}=\frac{C_{\mu\mu}}{\kappa},
\qquad
\delta C_{QQ,\mu\mu}=\frac{B_{\mu\mu}+C_{\mu\mu}/\kappa}{\Omega_\mu^{2}}.
\label{eq:S1lb_diagonal_sector}
\end{equation}

These expressions are real and satisfy the expected symmetry properties,
\begin{equation}
\delta \mathbf{C}_{QQ}^{\top}=\delta \mathbf{C}_{QQ},
\qquad
\delta \mathbf{C}_{Q\dot{Q}}^{\top}=-\delta \mathbf{C}_{Q\dot{Q}},
\qquad
\delta \mathbf{C}_{\dot{Q}\dot{Q}}^{\top}=\delta \mathbf{C}_{\dot{Q}\dot{Q}}.
\label{eq:S1lb_deltaC_symmetry}
\end{equation}
The first and third follow because \(A_{\nu\mu}=-A_{\mu\nu}\), while \(B_{\nu\mu}=B_{\mu\nu}\) and \(C_{\nu\mu}=C_{\mu\nu}\). The second then follows directly from Eq.~\eqref{eq:S1lb_deltaCQdQ_solution}.

It is also useful to note the equilibrium limit. If the temperature profile is uniform, then \(\mathbf{W}\propto \mathbf{I}\), so that \(\mathbf{C}_{Q\dot{Q}}^{(0)}=0\) and \(\mathbf{C}_{\dot{Q}\dot{Q}}^{(0)}\propto \mathbf{I}\). In that case
\begin{equation}
\mathbf{A}=0,
\qquad
\mathbf{B}=0,
\qquad
\mathbf{C}=0,
\label{eq:S1lb_ABC_equilibrium_zero}
\end{equation}
and therefore
\begin{equation}
\delta \mathbf{C}_{QQ}
=
\delta \mathbf{C}_{Q\dot{Q}}
=
\delta \mathbf{C}_{\dot{Q}\dot{Q}}
=
0.
\label{eq:S1lb_deltaC_equilibrium_zero}
\end{equation}
A uniform magnetic field therefore produces no steady \(B\)-odd correction in classical thermal equilibrium. The weak-field corrections become nonzero only out of equilibrium, when the temperature profile generates mode mixing through \(\mathbf{W}\).

Projecting these matrices back to real space then gives the field-odd corrections to the local phonon angular momentum, the bond-resolved PAM current, and the kinetic-energy current.

\subsection{\texorpdfstring{$B$}{B}-odd fields and currents}
\label{subsec:S1_Bodd_fields_currents}

The linear-in-\(B\) corrections derived above are projected back to real space in the same way as the zero-field covariances. Since the matrices \(\mathbf{A}\), \(\mathbf{B}\), and \(\mathbf{C}\) are linear in the zero-field covariances \(\mathbf{C}_{Q\dot{Q}}^{(0)}\) and \(\mathbf{C}_{\dot{Q}\dot{Q}}^{(0)}\), they are also linear in the bath temperatures. It is therefore convenient to resolve explicitly the contribution of each bath site \(r\).

Using the decomposition of the modal noise matrix,
\begin{equation}
\mathbf{W}
=
2\kappa k_B
\sum_{r=1}^{N}
T_r\,\mathbf{\Pi}_r,
\label{eq:S1Bo_W_decomposition}
\end{equation}
the zero-field modal covariances can be written as
\begin{subequations}
\begin{align}
\mathbf{C}_{QQ}^{(0)}
&=
\sum_{r=1}^{N}
T_r\,\mathbf{C}_{QQ}^{(0|r)},
\label{eq:S1Bo_CQQ0_decomp}
\\
\mathbf{C}_{Q\dot{Q}}^{(0)}
&=
\sum_{r=1}^{N}
T_r\,\mathbf{C}_{Q\dot{Q}}^{(0|r)},
\label{eq:S1Bo_CQdQ0_decomp}
\\
\mathbf{C}_{\dot{Q}\dot{Q}}^{(0)}
&=
\sum_{r=1}^{N}
T_r\,\mathbf{C}_{\dot{Q}\dot{Q}}^{(0|r)},
\label{eq:S1Bo_CdQdQ0_decomp}
\end{align}
\end{subequations}
with
\begin{subequations}
\begin{align}
\mathbf{C}_{QQ}^{(0|r)}
&\equiv
2\kappa k_B\,
\left(\mathbf{\Pi}_r\circ \mathbf{C}_{QQ}\right),
\label{eq:S1Bo_CQQ0r_def}
\\
\mathbf{C}_{Q\dot{Q}}^{(0|r)}
&\equiv
2\kappa k_B\,
\left(\mathbf{\Pi}_r\circ \mathbf{C}_{Q\dot{Q}}\right),
\label{eq:S1Bo_CQdQ0r_def}
\\
\mathbf{C}_{\dot{Q}\dot{Q}}^{(0|r)}
&\equiv
2\kappa k_B\,
\left(\mathbf{\Pi}_r\circ \mathbf{C}_{\dot{Q}\dot{Q}}\right).
\label{eq:S1Bo_CdQdQ0r_def}
\end{align}
\end{subequations}
Each of these matrices gives the zero-field modal covariance generated per unit temperature bias applied at bath site \(r\).

For each bath site \(r\), we define the corresponding driver matrices
\begin{subequations}
\begin{align}
\mathbf{A}^{(r)}
&\equiv
\left[\mathbf{J}_{m},\mathbf{C}_{Q\dot{Q}}^{(0|r)}\right],
\label{eq:S1Bo_Ar_def}
\\
\mathbf{B}^{(r)}
&\equiv
\left\{\mathbf{C}_{Q\dot{Q}}^{(0|r)},\mathbf{J}_{m}\right\},
\label{eq:S1Bo_Br_def}
\\
\mathbf{C}^{(r)}
&\equiv
\left[\mathbf{C}_{\dot{Q}\dot{Q}}^{(0|r)},\mathbf{J}_{m}\right].
\label{eq:S1Bo_Cr_def}
\end{align}
\end{subequations}
The corresponding linear-in-\(B\) modal corrections,
\begin{equation}
\delta \mathbf{C}_{QQ}^{(r)},
\qquad
\delta \mathbf{C}_{Q\dot{Q}}^{(r)},
\qquad
\delta \mathbf{C}_{\dot{Q}\dot{Q}}^{(r)},
\label{eq:S1Bo_deltaCr_list}
\end{equation}
follow from the mode-resolved solutions derived above. For each pair of modes \((\mu,\nu)\),
\begin{subequations}
\begin{align}
\delta C_{QQ,\mu\nu}^{(r)}
&=
\frac{
2\left(\Omega_\mu^{2}-\Omega_\nu^{2}\right)A_{\mu\nu}^{(r)}
+
4\kappa^{2} B_{\mu\nu}^{(r)}
+
4\kappa C_{\mu\nu}^{(r)}
}
{
\left(\Omega_\mu^{2}-\Omega_\nu^{2}\right)^{2}
+
2\kappa^{2}\left(\Omega_\mu^{2}+\Omega_\nu^{2}\right)
},
\label{eq:S1Bo_deltaCQQr_solution}
\\
\delta C_{Q\dot{Q},\mu\nu}^{(r)}
&=
\frac{
-2\kappa\left(\Omega_\mu^{2}+\Omega_\nu^{2}\right)A_{\mu\nu}^{(r)}
+
2\kappa\left(\Omega_\mu^{2}-\Omega_\nu^{2}\right)B_{\mu\nu}^{(r)}
+
2\left(\Omega_\mu^{2}-\Omega_\nu^{2}\right)C_{\mu\nu}^{(r)}
}
{
\left(\Omega_\mu^{2}-\Omega_\nu^{2}\right)^{2}
+
2\kappa^{2}\left(\Omega_\mu^{2}+\Omega_\nu^{2}\right)
},
\label{eq:S1Bo_deltaCQdQr_solution}
\\
\delta C_{\dot{Q}\dot{Q},\mu\nu}^{(r)}
&=
\frac{
\left(\Omega_\mu^{2}+\Omega_\nu^{2}\right)\left(\Omega_\mu^{2}-\Omega_\nu^{2}\right)A_{\mu\nu}^{(r)}
-
\left(\Omega_\mu^{2}-\Omega_\nu^{2}\right)^{2}B_{\mu\nu}^{(r)}
+
2\kappa\left(\Omega_\mu^{2}+\Omega_\nu^{2}\right)C_{\mu\nu}^{(r)}
}
{
\left(\Omega_\mu^{2}-\Omega_\nu^{2}\right)^{2}
+
2\kappa^{2}\left(\Omega_\mu^{2}+\Omega_\nu^{2}\right)
}.
\label{eq:S1Bo_deltaCdQdQr_solution}
\end{align}
\end{subequations}
By linearity,
\begin{subequations}
\begin{align}
\delta \mathbf{C}_{QQ}
&=
\sum_{r=1}^{N}
T_r\,\delta \mathbf{C}_{QQ}^{(r)},
\label{eq:S1Bo_deltaCQQ_decomp}
\\
\delta \mathbf{C}_{Q\dot{Q}}
&=
\sum_{r=1}^{N}
T_r\,\delta \mathbf{C}_{Q\dot{Q}}^{(r)},
\label{eq:S1Bo_deltaCQdQ_decomp}
\\
\delta \mathbf{C}_{\dot{Q}\dot{Q}}
&=
\sum_{r=1}^{N}
T_r\,\delta \mathbf{C}_{\dot{Q}\dot{Q}}^{(r)}.
\label{eq:S1Bo_deltaCdQdQ_decomp}
\end{align}
\end{subequations}

Projecting back to real space with the site blocks \(\mathbf{R}_s\) gives the field-odd corrections to the real-space covariance blocks:
\begin{subequations}
\begin{align}
\delta \langle \bm{u}_s\bm{u}_{t}^{\top}\rangle
&=
\sum_{r=1}^{N}
T_r\,
\mathbf{R}_s\,
\delta \mathbf{C}_{QQ}^{(r)}\,
\mathbf{R}_{t}^{\top},
\label{eq:S1Bo_real_cov_uu}
\\
\delta \langle \bm{u}_s\dot{\bm{u}}_{t}^{\top}\rangle
&=
\sum_{r=1}^{N}
T_r\,
\mathbf{R}_s\,
\delta \mathbf{C}_{Q\dot{Q}}^{(r)}\,
\mathbf{R}_{t}^{\top},
\label{eq:S1Bo_real_cov_udotu}
\\
\delta \langle \dot{\bm{u}}_s\dot{\bm{u}}_{t}^{\top}\rangle
&=
\sum_{r=1}^{N}
T_r\,
\mathbf{R}_s\,
\delta \mathbf{C}_{\dot{Q}\dot{Q}}^{(r)}\,
\mathbf{R}_{t}^{\top}.
\label{eq:S1Bo_real_cov_dotudotu}
\end{align}
\end{subequations}

The field-odd correction to the local displacement amplitude is
\begin{equation}
\delta A(s)
=
\Tr\!\left[
\mathbf{R}_s\,
\delta \mathbf{C}_{QQ}\,
\mathbf{R}_s^{\top}
\right]
=
\sum_{r=1}^{N}
T_r\,
\Tr\!\left[
\mathbf{R}_s\,
\delta \mathbf{C}_{QQ}^{(r)}\,
\mathbf{R}_s^{\top}
\right].
\label{eq:S1Bo_deltaA}
\end{equation}
The field-odd correction to the local kinetic-energy density is
\begin{equation}
\delta E_{\mathrm{kin}}(s)
=
\frac{m_s}{2}
\Tr\!\left[
\mathbf{R}_s\,
\delta \mathbf{C}_{\dot{Q}\dot{Q}}\,
\mathbf{R}_s^{\top}
\right]
=
\frac{m_s}{2}
\sum_{r=1}^{N}
T_r\,
\Tr\!\left[
\mathbf{R}_s\,
\delta \mathbf{C}_{\dot{Q}\dot{Q}}^{(r)}\,
\mathbf{R}_s^{\top}
\right].
\label{eq:S1Bo_deltaEkin}
\end{equation}
The field-odd correction to the local phonon angular momentum is
\begin{equation}
\delta L_i(s)
=
-m_s
\Tr\!\left[
\mathbf{E}_i\,
\mathbf{R}_s\,
\delta \mathbf{C}_{Q\dot{Q}}\,
\mathbf{R}_s^{\top}
\right]
=
-m_s
\sum_{r=1}^{N}
T_r\,
\Tr\!\left[
\mathbf{E}_i\,
\mathbf{R}_s\,
\delta \mathbf{C}_{Q\dot{Q}}^{(r)}\,
\mathbf{R}_s^{\top}
\right].
\label{eq:S1Bo_deltaLi}
\end{equation}

Using the same notation as in the zero-field case,
\begin{equation}
\Delta \mathbf{R}_{st}\equiv \mathbf{R}_s-\mathbf{R}_t,
\label{eq:S1Bo_DeltaR}
\end{equation}
the field-odd correction to the bond kinetic-energy current is
\begin{equation}
\delta j_{s\to t}^{(E)}
=
\Tr\!\left[
\mathbf{\Phi}^{(st)}
\,\Delta \mathbf{R}_{st}\,
\delta \mathbf{C}_{Q\dot{Q}}\,
\mathbf{R}_s^{\top}
\right].
\label{eq:S1Bo_deltajE_bond}
\end{equation}
The corresponding correction to the amplitude flux is
\begin{equation}
\delta j_{s\to t}^{(A)}
=
\frac{2}{m_s}
\Tr\!\left[
\mathbf{\Phi}^{(st)}
\,\Delta \mathbf{R}_{st}\,
\delta \mathbf{C}_{QQ}\,
\mathbf{R}_s^{\top}
\right].
\label{eq:S1Bo_deltajA_bond}
\end{equation}
The field-odd correction to the bond phonon-angular-momentum current is
\begin{equation}
\delta j_{s\to t}^{(L_i)}
=
\Tr\!\left[
\mathbf{E}_i\mathbf{\Phi}^{(st)}
\,\Delta \mathbf{R}_{st}\,
\delta \mathbf{C}_{QQ}\,
\mathbf{R}_s^{\top}
\right].
\label{eq:S1Bo_deltajL_bond}
\end{equation}

As before, the corresponding site current vectors are obtained by weighting each bond with its direction,
\begin{equation}
\delta \bm{j}^{(\alpha)}(s)
=
\sum_{t\in\mathcal{N}(s)}
\delta j_{s\to t}^{(\alpha)}\,\hat{\bm{d}}_{st},
\qquad
\alpha\in\{E,A,L_x,L_y,L_z\}.
\label{eq:S1Bo_delta_site_current}
\end{equation}
Because the modal corrections are linear in the bath temperatures, these site currents can be written as
\begin{equation}
\delta \bm{j}^{(\alpha)}(s)
=
\sum_{r=1}^{N}
T_r\,\bm{\sigma}^{(\alpha,B)}(s|r),
\label{eq:S1Bo_site_kernel_def}
\end{equation}
which defines the \(B\)-odd site-resolved response kernels.

Reading off the coefficient of \(T_r\) gives, for the kinetic-energy current,
\begin{equation}
\bm{\sigma}^{(E,B)}(s|r)
=
\sum_{t\in\mathcal{N}(s)}
\hat{\bm{d}}_{st}\,
\Tr\!\left[
\mathbf{\Phi}^{(st)}
\,\Delta \mathbf{R}_{st}\,
\delta \mathbf{C}_{Q\dot{Q}}^{(r)}\,
\mathbf{R}_s^{\top}
\right],
\label{eq:S1Bo_sigmaE_B}
\end{equation}
for the amplitude flux,
\begin{equation}
\bm{\sigma}^{(A,B)}(s|r)
=
\frac{2}{m_s}
\sum_{t\in\mathcal{N}(s)}
\hat{\bm{d}}_{st}\,
\Tr\!\left[
\mathbf{\Phi}^{(st)}
\,\Delta \mathbf{R}_{st}\,
\delta \mathbf{C}_{QQ}^{(r)}\,
\mathbf{R}_s^{\top}
\right],
\label{eq:S1Bo_sigmaA_B}
\end{equation}
and for the phonon-angular-momentum current,
\begin{equation}
\bm{\sigma}^{(L_i,B)}(s|r)
=
\sum_{t\in\mathcal{N}(s)}
\hat{\bm{d}}_{st}\,
\Tr\!\left[
\mathbf{E}_i\mathbf{\Phi}^{(st)}
\,\Delta \mathbf{R}_{st}\,
\delta \mathbf{C}_{QQ}^{(r)}\,
\mathbf{R}_s^{\top}
\right].
\label{eq:S1Bo_sigmaL_B}
\end{equation}
These kernels give the field-odd current of type \(\alpha\) flowing through observation site \(s\) per unit temperature bias applied at bath site \(r\).

The local \(B\)-odd fields and currents are thus obtained from the same geometric tensors \(\mathbf{E}_i\) and \(\mathbf{\Phi}^{(st)}\) as in the zero-field theory, with the zero-field modal covariances replaced by their linear-in-\(B\) corrections.

\subsection{\texorpdfstring{$B$}{B}-odd conductivity tensors}
\label{subsec:S1_Bodd_conductivity}

The linear-in-\(B\) correction to the current at site \(s\) is a linear functional of the temperature profile. For each transport channel \(\alpha\in\{E,A,L_x,L_y,L_z\}\), we write the current in the presence of the field as
\begin{equation}
\bm{j}^{(\alpha)}(s;B)
=
\bm{j}^{(\alpha,0)}(s)
+
\bm{j}^{(\alpha,B)}(s)
+
O(B^{2}),
\label{eq:S1Bc_current_expansion}
\end{equation}
where \(\bm{j}^{(\alpha,0)}(s)\) is the zero-field current and \(\bm{j}^{(\alpha,B)}(s)\) is the part linear in the magnetic field. The dependence of \(\bm{j}^{(\alpha,B)}(s)\) on the bath temperatures can be written as
\begin{equation}
\bm{j}^{(\alpha,B)}(s)
=
\sum_{r=1}^{N}
T_r\,\bm{\sigma}^{(\alpha,B)}(s|r),
\label{eq:S1Bc_site_kernel_def}
\end{equation}
where \(\bm{\sigma}^{(\alpha,B)}(s|r)\) gives the linear-in-\(B\) current correction of type \(\alpha\) flowing through observation site \(s\) per unit temperature bias applied at bath site \(r\).

We consider again a weak temperature modulation around a uniform reference temperature,
\begin{equation}
T_r=T^{(0)}+\delta T_r,
\qquad
|\delta T_r|\ll T^{(0)}.
\label{eq:S1Bc_small_temperature_modulation}
\end{equation}
Substituting Eq.~\eqref{eq:S1Bc_small_temperature_modulation} into Eq.~\eqref{eq:S1Bc_site_kernel_def} gives
\begin{equation}
\bm{j}^{(\alpha,B)}(s)
=
T^{(0)}\sum_{r=1}^{N}\bm{\sigma}^{(\alpha,B)}(s|r)
+
\sum_{r=1}^{N}\bm{\sigma}^{(\alpha,B)}(s|r)\,\delta T_r.
\label{eq:S1Bc_current_split}
\end{equation}
For a spatially uniform temperature profile, the linear-in-\(B\) modal corrections vanish, and the linear-in-\(B\) currents vanish with them. Therefore the site-resolved kernels satisfy
\begin{equation}
\sum_{r=1}^{N}\bm{\sigma}^{(\alpha,B)}(s|r)=0.
\label{eq:S1Bc_sigma_sum_zero}
\end{equation}
The linear-in-\(B\) sector does not require any additional neutralization: the response to a uniform temperature offset is already absent, and only the temperature variation contributes. Equation~\eqref{eq:S1Bc_current_split} then reduces to
\begin{equation}
\bm{j}^{(\alpha,B)}(s)
=
\sum_{r=1}^{N}
\bm{\sigma}^{(\alpha,B)}(s|r)\,\delta T_r .
\label{eq:S1Bc_jB_gradient_only}
\end{equation}

For a uniform imposed temperature gradient, we write
\begin{equation}
\delta T_r
=
\sum_{j=x,y,z}
(r_{r,j}-r_{0,j})\,\partial_j T,
\label{eq:S1Bc_uniform_gradient}
\end{equation}
where \(\bm{r}_0\) is an arbitrary reference point. Substituting Eq.~\eqref{eq:S1Bc_uniform_gradient} into Eq.~\eqref{eq:S1Bc_jB_gradient_only} gives
\begin{equation}
j_i^{(\alpha,B)}(s)
=
\sum_{j=x,y,z}
\sigma_{ij}^{(\alpha,B)}(s)\,\partial_j T,
\label{eq:S1Bc_local_conductivity_law}
\end{equation}
with site-resolved conductivity tensor
\begin{equation}
\sigma_{ij}^{(\alpha,B)}(s)
\equiv
\sum_{r=1}^{N}
(r_{r,j}-r_{0,j})\,\sigma_{i}^{(\alpha,B)}(s|r).
\label{eq:S1Bc_local_conductivity_tensor}
\end{equation}
Thus \(\sigma_{ij}^{(\alpha,B)}(s)\) maps the \(j\)-component of the imposed temperature gradient to the \(i\)-component of the linear-in-\(B\) current correction of type \(\alpha\).

This definition does not depend on the choice of the reference point \(\bm{r}_0\). A shift \(\bm{r}_0\to \bm{r}_0+\bm{a}\) changes \(\sigma_{ij}^{(\alpha,B)}(s)\) by
$-\;a_j\sum_{r=1}^{N}\sigma_i^{(\alpha,B)}(s|r)$,
and this vanishes because of Eq.~\eqref{eq:S1Bc_sigma_sum_zero}. One may therefore choose \(\bm{r}_0\) for convenience, for instance as the average lattice position,
\begin{equation}
\bm{r}_0
=
\frac{1}{N}\sum_{r=1}^{N}\bm{r}_r.
\label{eq:S1Bc_average_origin}
\end{equation}

A bulk conductivity tensor is obtained by averaging the local conductivity over a set of interior sites,
\begin{equation}
\sigma_{ij}^{(\alpha,B),\mathrm{bulk}}
\equiv
\frac{1}{N_{\mathrm{bulk}}}
\sum_{s\in\mathrm{bulk}}
\sigma_{ij}^{(\alpha,B)}(s).
\label{eq:S1Bc_bulk_conductivity}
\end{equation}
This bulk tensor gives the coarse-grained linear-in-\(B\) transport correction. In particular, for \(\alpha=E\) it describes the magnetic-field correction to the kinetic-energy current response, while for \(\alpha=L_i\) it describes the corresponding correction to the transport of the \(i\)-th component of phonon angular momentum.

It is also useful to separate longitudinal and transverse parts of this correction. We therefore define
\begin{subequations}
\begin{align}
\sigma_{ij}^{(\alpha,B),\mathrm{S}}
&\equiv
\frac{1}{2}
\left(
\sigma_{ij}^{(\alpha,B),\mathrm{bulk}}
+
\sigma_{ji}^{(\alpha,B),\mathrm{bulk}}
\right),
\label{eq:S1Bc_bulk_conductivity_sym}
\\
\sigma_{ij}^{(\alpha,B),\mathrm{A}}
&\equiv
\frac{1}{2}
\left(
\sigma_{ij}^{(\alpha,B),\mathrm{bulk}}
-
\sigma_{ji}^{(\alpha,B),\mathrm{bulk}}
\right).
\label{eq:S1Bc_bulk_conductivity_antisym}
\end{align}
\end{subequations}
The symmetric part gives the magnetic-field correction to the longitudinal response, while the antisymmetric part gives the magnetic-field correction to the transverse deflection of the current.

For a longitudinal thermal drive along \(x\), the deflection angle in the presence of the field is
\begin{equation}
\tan \theta_{\alpha}(B)
\equiv
\frac{
\sigma_{yx}^{(\alpha),\mathrm{bulk}}
+
\sigma_{yx}^{(\alpha,B),\mathrm{bulk}}
}{
\sigma_{xx}^{(\alpha),\mathrm{bulk}}
+
\sigma_{xx}^{(\alpha,B),\mathrm{bulk}}
},
\label{eq:S1Bc_Hall_angle_total}
\end{equation}
where \(\sigma_{ij}^{(\alpha),\mathrm{bulk}}\) denotes the corresponding zero-field bulk conductivity tensor. This definition includes both the zero-field transverse response and the linear-in-\(B\) correction. For \(\alpha=L_i\), it gives the deflection angle associated with the transport of the \(i\)-th component of phonon angular momentum, while for \(\alpha=E\) it gives the corresponding deflection angle of the kinetic-energy current.

\subsection{Specialization to two-dimensional lattices}
\label{subsec:S1_two_dimensional_specialization}

We now specialize the general framework to two-dimensional lattices. Each site \(s\) carries only in-plane displacements,
\begin{equation}
\bm{u}_s=
\begin{pmatrix}
u_{s,x}\\
u_{s,y}
\end{pmatrix}
\in \mathbb{R}^{2},
\qquad
\dot{\bm{u}}_s\in\mathbb{R}^{2},
\label{eq:S12D_us_def}
\end{equation}
and the global displacement vector has length \(2N\),
\begin{equation}
\bm{u}=
\bigl(\bm{u}_1,\dots,\bm{u}_N\bigr)^{\top}\in\mathbb{R}^{2N}.
\label{eq:S12D_u_def}
\end{equation}
The mass matrix is
\begin{equation}
\mathbf{M}
=
\mathrm{diag}(m_1,\dots,m_N)\otimes \mathbf{I}_2,
\label{eq:S12D_M_def}
\end{equation}
and the harmonic force-constant matrix is assembled from symmetric \(2\times 2\) bond tensors \(\mathbf{\Phi}^{(st)}\) exactly as in three dimensions,
\begin{equation}
\mathbf{K}_{ss}=\sum_{t\in\mathcal{N}(s)}\mathbf{\Phi}^{(st)},
\qquad
\mathbf{K}_{st}=-\mathbf{\Phi}^{(st)}
\quad (s\neq t).
\label{eq:S12D_K_blocks}
\end{equation}

The equations of motion keep the same form,
\begin{equation}
\mathbf{M}\ddot{\bm{u}}
+
\kappa \mathbf{M}\dot{\bm{u}}
+
\mathbf{K}\bm{u}
=
\bm{\eta}(t),
\label{eq:S12D_eom_real}
\end{equation}
with local white-noise baths
\begin{equation}
\langle \eta_{s,i}(t)\eta_{t,j}(t')\rangle
=
2\kappa m_s k_B T_s\,\delta_{st}\delta_{ij}\delta(t-t'),
\qquad
i,j\in\{x,y\}.
\label{eq:S12D_noise_real}
\end{equation}
The mass-normalized variables are defined in the same way,
\begin{equation}
\tilde{\bm{u}}=\mathbf{M}^{1/2}\bm{u},
\qquad
\mathbf{D}=\mathbf{M}^{-1/2}\mathbf{K}\mathbf{M}^{-1/2},
\qquad
\bm{Q}=\mathbf{U}^{\top}\tilde{\bm{u}},
\label{eq:S12D_massnorm_defs}
\end{equation}
so the modal equations, the kernels \(C_{QQ}\), \(C_{Q\dot{Q}}\), \(C_{\dot{Q}\dot{Q}}\), and the factorized modal covariances retain exactly the same form as in the three-dimensional derivation, with the only change that all vectors and matrices now live in \(\mathbb{R}^{2N}\).

Projecting back to real space also proceeds without change,
\begin{equation}
\bm{u}=\mathbf{R}\bm{Q},
\qquad
\mathbf{R}=\mathbf{M}^{-1/2}\mathbf{U},
\label{eq:S12D_R_def}
\end{equation}
with site blocks \(\mathbf{R}_s\in\mathbb{R}^{2\times 2N}\). The real-space covariance blocks are therefore
\begin{subequations}
\begin{align}
\langle \bm{u}_s\bm{u}_t^{\top}\rangle
&=
\mathbf{R}_s
\langle \bm{Q}\bm{Q}^{\top}\rangle
\mathbf{R}_t^{\top},
\label{eq:S12D_real_cov_uu}
\\
\langle \bm{u}_s\dot{\bm{u}}_t^{\top}\rangle
&=
\mathbf{R}_s
\langle \bm{Q}\dot{\bm{Q}}^{\top}\rangle
\mathbf{R}_t^{\top},
\label{eq:S12D_real_cov_udotu}
\\
\langle \dot{\bm{u}}_s\dot{\bm{u}}_t^{\top}\rangle
&=
\mathbf{R}_s
\langle \dot{\bm{Q}}\dot{\bm{Q}}^{\top}\rangle
\mathbf{R}_t^{\top}.
\label{eq:S12D_real_cov_dotudotu}
\end{align}
\end{subequations}
All local fields, bond currents, and conductivity tensors follow from these blocks exactly as before, with traces now taken over the two in-plane coordinates only.

In strictly two-dimensional motion, the only nonzero component of the phonon angular momentum is the out-of-plane one. The antisymmetric generator reduces to
\begin{equation}
\mathbf{E}_z=
\begin{pmatrix}
0 & 1\\
-1 & 0
\end{pmatrix},
\label{eq:S12D_Ez}
\end{equation}
and the local phonon angular momentum becomes
\begin{equation}
L_z(s)
=
-m_s\Tr\!\left[
\mathbf{E}_z
\langle \bm{u}_s\dot{\bm{u}}_s^{\top}\rangle
\right]
=
m_s\left\langle
u_{s,x}\dot{u}_{s,y}-u_{s,y}\dot{u}_{s,x}
\right\rangle .
\label{eq:S12D_Lz}
\end{equation}
The corresponding bond current is
\begin{equation}
j^{(L_z)}_{s\to t}
=
\Tr\!\left[
\mathbf{E}_z\mathbf{\Phi}^{(st)}
\left(
\langle \bm{u}_s\bm{u}_s^{\top}\rangle
-
\langle \bm{u}_t\bm{u}_s^{\top}\rangle
\right)
\right].
\label{eq:S12D_jLz_bond}
\end{equation}
The amplitude and kinetic-energy fields and currents keep the same form as in three dimensions, with all traces understood in the two-dimensional subspace.

The bath injection into the kinetic-energy balance is correspondingly reduced. Repeating the It\^o calculation with two active Cartesian components gives
\begin{equation}
\langle \dot{\bm{u}}_s\cdot\bm{\eta}_s\rangle
=
2\kappa k_B T_s,
\label{eq:S12D_udot_eta}
\end{equation}
so the local kinetic-energy balance becomes
\begin{equation}
\frac{d}{dt}E_{\mathrm{kin}}(s)
+
\sum_{t\in\mathcal{N}(s)}j^{(E)}_{s\to t}
=
-\kappa m_s\Tr\langle \dot{\bm{u}}_s\dot{\bm{u}}_s^{\top}\rangle
+
2\kappa k_B T_s.
\label{eq:S12D_energy_balance}
\end{equation}

For a weak magnetic field, the physically relevant two-dimensional case is an out-of-plane field,
\begin{equation}
\bm{B}=B_z \hat{\bm z}.
\label{eq:S12D_Bz}
\end{equation}
The local gyroscopic frequency is then
\begin{equation}
\bm{\Omega}^{\mathrm g}_s
=
\gamma_s B_z \hat{\bm z},
\label{eq:S12D_Omegag}
\end{equation}
and the gyroscopic block at site \(s\) reduces to
\begin{equation}
\mathbf{J}_s
=
m_s\gamma_s B_z
\begin{pmatrix}
0 & -1\\
1 & 0
\end{pmatrix}
=
-\,m_s\gamma_s B_z\,\mathbf{E}_z.
\label{eq:S12D_Js}
\end{equation}
The full weak-field expansion derived above then applies with this reduced \(2\times 2\) structure. In two dimensions the only angular-momentum transport channel is therefore the \(L_z\) channel, while the linear-in-\(B\) correction modifies both the kinetic-energy current and the transverse phonon-angular-momentum current in the same formal way as in three dimensions.

\subsection{Square-lattice specialization}
\label{subsec:S1_square_lattice}

To illustrate the two-dimensional framework in the simplest setting, we consider a monatomic square lattice with lattice constant \(a\) and two in-plane displacement components per site. The equilibrium positions may be written as
\begin{equation}
\bm{r}_s = a\,(n_{s,x},n_{s,y}),
\qquad
n_{s,x},n_{s,y}\in\mathbb{Z},
\label{eq:S1sq_positions}
\end{equation}
for the sites belonging to the finite sample.

Each site is coupled to two bond families: axial nearest neighbors and diagonal next-nearest neighbors. The axial bond directions are
\begin{equation}
\hat{\bm d}\in\{\pm \hat{\bm x},\,\pm \hat{\bm y}\},
\label{eq:S1sq_axial_directions}
\end{equation}
while the diagonal bond directions are
\begin{equation}
\hat{\bm d}\in
\left\{
\frac{\pm \hat{\bm x}\pm \hat{\bm y}}{\sqrt{2}}
\right\}.
\label{eq:S1sq_diagonal_directions}
\end{equation}
We denote by \(\mathcal N_{\mathrm{ax}}(s)\) and \(\mathcal N_{\mathrm{diag}}(s)\) the corresponding neighbor sets, so that
\begin{equation}
\mathcal N(s)=\mathcal N_{\mathrm{ax}}(s)\cup \mathcal N_{\mathrm{diag}}(s).
\label{eq:S1sq_neighbor_sets}
\end{equation}

The bond tensors are taken to be central. For an axial nearest-neighbor bond \((s,t)\),
\begin{equation}
\mathbf{\Phi}^{(st)}
=
K_{\mathrm{ax}}\,
\hat{\bm d}_{st}\hat{\bm d}_{st}^{\top},
\qquad
t\in\mathcal N_{\mathrm{ax}}(s),
\label{eq:S1sq_axial_bond_tensor}
\end{equation}
while for a diagonal next-nearest-neighbor bond,
\begin{equation}
\mathbf{\Phi}^{(st)}
=
K_{\mathrm{diag}}\,
\hat{\bm d}_{st}\hat{\bm d}_{st}^{\top},
\qquad
t\in\mathcal N_{\mathrm{diag}}(s).
\label{eq:S1sq_diagonal_bond_tensor}
\end{equation}
The real-space force-constant blocks are therefore
\begin{equation}
\mathbf{K}_{ss}
=
\sum_{t\in\mathcal N(s)}
\mathbf{\Phi}^{(st)},
\qquad
\mathbf{K}_{st}
=
-\mathbf{\Phi}^{(st)}
\quad (s\neq t).
\label{eq:S1sq_K_blocks}
\end{equation}

For axial bonds, the bond tensors are diagonal in the Cartesian basis. For example,
\begin{equation}
\hat{\bm x}\hat{\bm x}^{\top}
=
\begin{pmatrix}
1 & 0\\
0 & 0
\end{pmatrix},
\qquad
\hat{\bm y}\hat{\bm y}^{\top}
=
\begin{pmatrix}
0 & 0\\
0 & 1
\end{pmatrix}.
\label{eq:S1sq_axial_examples}
\end{equation}
The axial network therefore couples horizontal bonds only to \(x\)-motion and vertical bonds only to \(y\)-motion. If the diagonal springs are removed, \(K_{\mathrm{diag}}=0\), the force network is separable into Cartesian components. In that limit the stiffness matrix can be arranged in block-diagonal form in polarization space, and the in-plane motions remain unmixed.

This immediately suppresses the out-of-plane phonon-angular-momentum current. In two dimensions the bond-resolved PAM current is
\begin{equation}
j_{s\to t}^{(L_z)}
=
\Tr\!\left[
\mathbf{E}_z\mathbf{\Phi}^{(st)}
\left(
\langle \bm{u}_s\bm{u}_s^{\top}\rangle
-
\langle \bm{u}_t\bm{u}_s^{\top}\rangle
\right)
\right],
\label{eq:S1sq_jLz_general}
\end{equation}
with
\begin{equation}
\mathbf{E}_z=
\begin{pmatrix}
0 & 1\\
-1 & 0
\end{pmatrix}.
\label{eq:S1sq_Ez}
\end{equation}
When \(K_{\mathrm{diag}}=0\), both the bond tensors and the covariance blocks remain diagonal in the \(x\)-\(y\) basis, so the trace in Eq.~\eqref{eq:S1sq_jLz_general} vanishes identically. A finite PAMHE in the square lattice therefore requires
\begin{equation}
K_{\mathrm{diag}}\neq 0.
\label{eq:S1sq_Kdiag_nonzero}
\end{equation}

The diagonal bonds provide the required polarization mixing. For a bond along
\begin{equation}
\hat{\bm d}=\frac{\hat{\bm x}+\hat{\bm y}}{\sqrt{2}},
\label{eq:S1sq_diag_plus_dir}
\end{equation}
one has
\begin{equation}
\hat{\bm d}\hat{\bm d}^{\top}
=
\frac{1}{2}
\begin{pmatrix}
1 & 1\\
1 & 1
\end{pmatrix},
\label{eq:S1sq_diag_plus_tensor}
\end{equation}
whereas for a bond along
\begin{equation}
\hat{\bm d}=\frac{\hat{\bm x}-\hat{\bm y}}{\sqrt{2}},
\label{eq:S1sq_diag_minus_dir}
\end{equation}
one finds
\begin{equation}
\hat{\bm d}\hat{\bm d}^{\top}
=
\frac{1}{2}
\begin{pmatrix}
1 & -1\\
-1 & 1
\end{pmatrix}.
\label{eq:S1sq_diag_minus_tensor}
\end{equation}
These tensors contain off-diagonal matrix elements and therefore mix the two in-plane polarizations. Once a nonuniform temperature profile generates nonequilibrium mode mixing, the diagonal force network converts those correlations into a transverse phonon-angular-momentum current.

\subsection{Honeycomb-lattice specialization}
\label{subsec:S1_honeycomb_lattice}

We now specialize the two-dimensional framework to the honeycomb lattice. In contrast to the square lattice, the honeycomb geometry contains a two-site basis and three inequivalent nearest-neighbor bond directions already at the minimal level. A finite transverse phonon-angular-momentum response can therefore arise with first-neighbor couplings alone, provided the bond tensors mix the two in-plane polarizations. This is the minimal honeycomb model used in the lattice calculations below. 

We consider a two-dimensional honeycomb lattice with two sites per unit cell, labeled \(A\) and \(B\), and two in-plane displacement components per site. Only first-neighbor \(A\)–\(B\) bonds are retained. Each site is therefore connected to three nearest neighbors on the opposite sublattice. Denoting the nearest-neighbor distance by \(a_{\mathrm{nn}}\), the three bond directions may be chosen as
\begin{equation}
\hat{\bm d}_{1}=
\begin{pmatrix}
1\\
0
\end{pmatrix},
\qquad
\hat{\bm d}_{2}=
\begin{pmatrix}
-\frac{1}{2}\\[2pt]
\frac{\sqrt{3}}{2}
\end{pmatrix},
\qquad
\hat{\bm d}_{3}=
\begin{pmatrix}
-\frac{1}{2}\\[2pt]
-\frac{\sqrt{3}}{2}
\end{pmatrix},
\label{eq:S1hc_bond_directions}
\end{equation}
so that the three bonds are separated by angles of \(2\pi/3\). For a bond \((s,t)\), we denote its in-plane angle by \(\theta_{st}\).

The first-neighbor force network is described by a symmetric \(2\times 2\) bond tensor
\begin{equation}
\mathbf{\Phi}^{(st)}
=
A_{1}\mathbf{I}_{2}
+
B_{1}\mathbf{Q}_{2}(\theta_{st}),
\label{eq:S1hc_bond_tensor}
\end{equation}
with
\begin{equation}
\mathbf{Q}_{2}(\theta)
=
\begin{pmatrix}
\cos 2\theta & \sin 2\theta\\
\sin 2\theta & -\cos 2\theta
\end{pmatrix}.
\label{eq:S1hc_Q2_def}
\end{equation}
The parameter \(A_{1}\) sets the isotropic part of the nearest-neighbor coupling. It contributes the same restoring force to every in-plane direction and does not distinguish one bond orientation from another. The parameter \(B_{1}\) sets the bond-direction-dependent part. It is this term that makes the restoring force depend on the bond angle and therefore mixes the two in-plane polarizations. 

Using the bond direction \(\hat{\bm d}_{st}=(\cos\theta_{st},\sin\theta_{st})^{\top}\), one may also write
\begin{equation}
\mathbf{Q}_{2}(\theta_{st})
=
2\hat{\bm d}_{st}\hat{\bm d}_{st}^{\top}-\mathbf{I}_{2},
\label{eq:S1hc_Q2_dyadic}
\end{equation}
so that
\begin{equation}
\mathbf{\Phi}^{(st)}
=
(A_{1}-B_{1})\mathbf{I}_{2}
+
2B_{1}\hat{\bm d}_{st}\hat{\bm d}_{st}^{\top}.
\label{eq:S1hc_bond_tensor_rewritten}
\end{equation}
This separates the part that is the same in every in-plane direction from the part that follows the bond orientation.

It is useful to decompose the bond tensor into motion along the bond and motion perpendicular to it. Since
\begin{equation}
\mathbf{I}_{2}
=
\hat{\bm d}_{st}\hat{\bm d}_{st}^{\top}
+
\left(\mathbf{I}_{2}-\hat{\bm d}_{st}\hat{\bm d}_{st}^{\top}\right),
\label{eq:S1hc_projector_decomp}
\end{equation}
Eq.~\eqref{eq:S1hc_bond_tensor_rewritten} becomes
\begin{equation}
\mathbf{\Phi}^{(st)}
=
(A_{1}+B_{1})\,\hat{\bm d}_{st}\hat{\bm d}_{st}^{\top}
+
(A_{1}-B_{1})\,
\left(\mathbf{I}_{2}-\hat{\bm d}_{st}\hat{\bm d}_{st}^{\top}\right).
\label{eq:S1hc_long_trans_decomp}
\end{equation}
The coefficient \(A_{1}+B_{1}\) is therefore the longitudinal stiffness for relative motion along the bond, while \(A_{1}-B_{1}\) is the transverse stiffness for relative motion perpendicular to the bond. If
\begin{equation}
A_{1}=B_{1},
\label{eq:S1hc_central_limit_condition}
\end{equation}
the transverse stiffness vanishes and the bond reduces to a purely central spring,
\begin{equation}
\mathbf{\Phi}^{(st)}
=
2A_{1}\,\hat{\bm d}_{st}\hat{\bm d}_{st}^{\top}.
\label{eq:S1hc_central_limit}
\end{equation}
When \(A_{1}\neq B_{1}\), the bond also responds to relative displacements perpendicular to its axis. In the honeycomb lattice this additional directional structure strengthens the mixing between the two in-plane polarizations. 

The real-space force-constant blocks follow from the same construction used throughout,
\begin{equation}
\mathbf{K}_{ss}
=
\sum_{t\in\mathcal{N}(s)}
\mathbf{\Phi}^{(st)},
\qquad
\mathbf{K}_{st}
=
-\mathbf{\Phi}^{(st)}
\quad (s\neq t),
\label{eq:S1hc_K_blocks}
\end{equation}
where \(\mathcal{N}(s)\) denotes the set of first neighbors of site \(s\).

For the three nearest-neighbor directions, the bond tensors are
\begin{subequations}
\begin{align}
\mathbf{\Phi}_{1}
&=
A_{1}\mathbf{I}_{2}
+
B_{1}
\begin{pmatrix}
1 & 0\\
0 & -1
\end{pmatrix},
\label{eq:S1hc_Phi1}
\\
\mathbf{\Phi}_{2}
&=
A_{1}\mathbf{I}_{2}
+
B_{1}
\begin{pmatrix}
-\frac{1}{2} & -\frac{\sqrt{3}}{2}\\[2pt]
-\frac{\sqrt{3}}{2} & \frac{1}{2}
\end{pmatrix},
\label{eq:S1hc_Phi2}
\\
\mathbf{\Phi}_{3}
&=
A_{1}\mathbf{I}_{2}
+
B_{1}
\begin{pmatrix}
-\frac{1}{2} & \frac{\sqrt{3}}{2}\\[2pt]
\frac{\sqrt{3}}{2} & \frac{1}{2}
\end{pmatrix}.
\label{eq:S1hc_Phi3}
\end{align}
\end{subequations}
The tensors \(\mathbf{\Phi}_{2}\) and \(\mathbf{\Phi}_{3}\) contain off-diagonal elements, so the two in-plane polarizations are already mixed at first-neighbor level. This is the main difference with the square lattice, where an additional diagonal bond family was needed to generate the same effect. 

The honeycomb model therefore contains the ingredients needed for a finite PAMHE: a two-site basis, three inequivalent bond directions, and bond tensors that mix the in-plane polarizations.

The mechanism is the same as in the square lattice, but it appears here in a different geometrical setting. In the square case, polarization mixing is introduced by diagonal next-nearest-neighbor springs. In the honeycomb case, the two-site basis together with the three bond directions already gives a nearest-neighbor network that can redirect phonon angular momentum into the transverse direction under a longitudinal thermal drive.

\section{Square- and honeycomb-lattice analysis}
\label{sec:SM_square_honeycomb_analysis}

We now apply the two-dimensional real-space formalism to the square and honeycomb lattices introduced above. In both cases, we evaluate the local phonon-angular-momentum field $L_z(s)$ and the bond-resolved PAM current $j_{s\to t}^{(L_z)}$ from Eqs.~\eqref{eq:S12D_Lz} and \eqref{eq:S12D_jLz_bond}, and we visualize the corresponding site current vectors $\bm j^{(L_z)}(s)$ using Eq.~\eqref{eq:S1bc_site_current_vectors}. The thermal bias is applied along $\hat{\bm x}$, so the $x$ direction is the longitudinal direction and the $y$ direction is the transverse direction. The orientation of the PAM flow is therefore characterized by the deflection angle
\[
\tan\theta_{L_z}
=
\frac{\sigma_{yx}^{(L_z),\mathrm{bulk}}}{\sigma_{xx}^{(L_z),\mathrm{bulk}}},
\]
using the bulk conductivity tensor in Eq.~\eqref{eq:S1lr_bulk_conductivity}. The same construction can also be used for the kinetic-energy current when discussing its field-induced deflection. When an external magnetic field is included, we use the full field-dependent deflection angle in Eq.~\eqref{eq:S1Bc_Hall_angle_total}, which includes both the zero-field conductivities and their linear-in-$B$ corrections.

\begin{figure}[t]
  \centering
  \includegraphics[width=\textwidth]{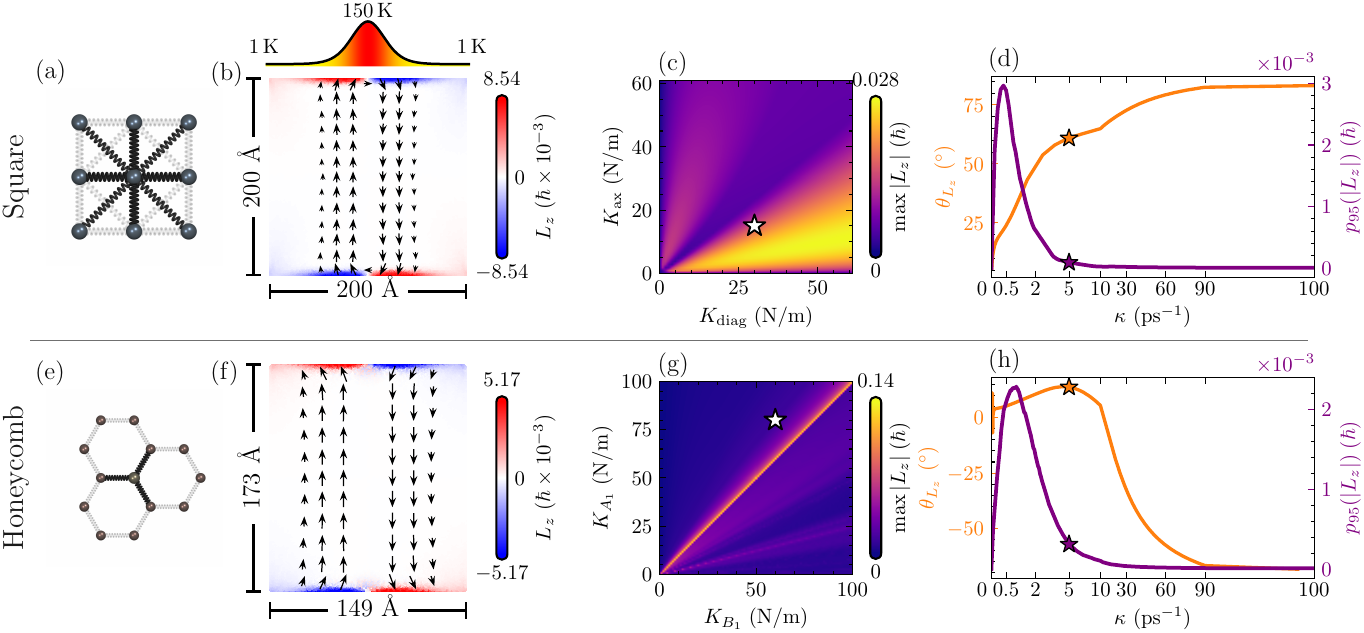}
  \caption{Phonon angular-momentum Hall effect in minimal square and honeycomb lattices. \textbf{(a--d) Square lattice:} (a) lattice geometry and spring network with axial bonds of stiffness $K_{\rm ax}$ and diagonal bonds of stiffness $K_{\rm diag}$. (b) Real-space distribution of $L_z(s)$ and site current vectors $\bm j^{(L_z)}(s)$. (c) Edge accumulation of phonon angular momentum as a function of $K_{\rm ax}$ and $K_{\rm diag}$, measured by $\max |L_z|$, i.e. the maximum site-resolved magnitude of the local phonon angular momentum in the sample for each pair of stiffness parameters. (d) Deflection angle $\theta_{L_z}$ and $p_{95}(|L_z|)$ as functions of the damping rate $\kappa$, where $p_{95}(|L_z|)$ is the 95th percentile of the site-resolved magnitude $|L_z(s)|$ and is used to characterize the accumulation while reducing the influence of isolated outliers. \textbf{(e--h) Honeycomb lattice:} (e) lattice geometry and nearest-neighbor bond model with isotropic and anisotropic couplings $A_1$ and $B_1$. (f) Real-space distribution of $L_z(s)$ and site current vectors $\bm j^{(L_z)}(s)$. (g) Edge accumulation as a function of $A_1$ and $B_1$, again measured by $\max |L_z|$. (h) Deflection angle $\theta_{L_z}$ and $p_{95}(|L_z|)$ as functions of $\kappa$. The stars in panels (c,d,g,h) mark the parameter values used for the real-space maps in panels (b,f).}
  \label{fig:SM_square_honeycomb}
\end{figure}

Figure~\ref{fig:SM_square_honeycomb} summarizes the square- and honeycomb-lattice results. For each lattice, the first column shows the force network, the second column shows the real-space distribution of $L_z(s)$ together with the site current vectors $\bm j^{(L_z)}(s)$, the third column shows how the edge accumulation changes as the bond parameters are varied, and the fourth column shows how the deflection angle and the edge accumulation depend on the damping rate. In both lattices, the same physical pattern appears: the thermal bias drives a PAM current with opposite transverse components on the left and right halves of the sample, and this produces accumulation of opposite signs of $L_z$ at the upper and lower edges.

For the real-space maps, the spring-parameter scans, and the damping-dependent accumulation shown in Fig.~\ref{fig:SM_square_honeycomb}, we use open-boundary finite samples, since the edge buildup of $L_z$ is itself a boundary manifestation of the transverse PAM flow. By contrast, the deflection angle is extracted from periodic-boundary conductivity calculations, which suppress the boundary accumulation and therefore give a cleaner characterization of the bulk orientation of the PAM current. In the damping-dependent curves shown here, we additionally include an out-of-plane magnetic field and evaluate the full field-dependent deflection angle. Physically, this allows us to follow how the magnetic field tilts the PAM transport away from the purely transverse direction by inducing a longitudinal component of phonon-angular-momentum flow and thereby changing the balance between the transverse and longitudinal conductivity components.

\subsection{Square lattice}

The square lattice gives the clearest minimal realization of the polarization-mixing requirement. Its force network is built from axial nearest-neighbor bonds of stiffness $K_{\rm ax}$ and diagonal next-nearest-neighbor bonds of stiffness $K_{\rm diag}$, with bond tensors given in Eqs.~\eqref{eq:S1sq_axial_bond_tensor} and \eqref{eq:S1sq_diagonal_bond_tensor}. The axial bonds alone do not produce a PAMHE. If $K_{\rm diag}=0$, the force network is separable in the two in-plane polarizations: horizontal bonds couple only $x$ motion and vertical bonds couple only $y$ motion. In that limit the covariance blocks remain diagonal in polarization space, and the trace in the bond-current expression vanishes, so the PAM current is zero [Eq.~\eqref{eq:S1sq_jLz_general}]. A finite PAMHE therefore requires the diagonal bond family, as summarized in Eq.~\eqref{eq:S1sq_Kdiag_nonzero}. Once $K_{\rm diag}\neq 0$, the diagonal bond tensors contain off-diagonal matrix elements, mix the $x$ and $y$ polarizations, and convert the nonequilibrium correlations generated by the thermal bias into a transverse PAM current.

This is seen directly in Fig.~\ref{fig:SM_square_honeycomb}(c). The edge accumulation is strongly suppressed near the axial limit $K_{\rm diag}=0$ and grows once the diagonal bonds become appreciable. It is largest when the diagonal bonds are comparable to, or somewhat stronger than, the axial bonds. In that regime the diagonal springs provide the strongest mixing between the two in-plane polarizations, so the nonequilibrium correlations generated by the thermal bias are converted most efficiently into a transverse current of phonon angular momentum. When $K_{\rm ax}\gg K_{\rm diag}$, the motion becomes increasingly axial and the transverse response weakens.

The real-space map in Fig.~\ref{fig:SM_square_honeycomb}(b) shows how this response appears in the sample. The current vectors develop opposite $\hat{\bm y}$ components on the left and right halves, and the corresponding upper and lower edges accumulate opposite signs of $L_z$. The edge accumulation therefore follows directly from the transverse deflection of the PAM current. Figure~\ref{fig:SM_square_honeycomb}(d) shows that the accumulation is largest at intermediate damping. For small $\kappa$, the coupling to the baths is weak and the nonequilibrium correlations remain small. For large $\kappa$, overdamping suppresses the displacement--velocity correlations that enter the local PAM density, so the accumulation is reduced again in the strongly damped regime. The deflection angle remains large over a broad range of damping, which shows that the current stays predominantly transverse even though the thermal bias is longitudinal. Once the out-of-plane magnetic field is included in the conductivity analysis, however, the PAM current acquires a finite longitudinal component as well, so the deflection angle measures how the field reshapes the balance between longitudinal and transverse PAM transport rather than simply whether the current is present.

\subsection{Honeycomb lattice}

The honeycomb lattice realizes the same mechanism in a different geometric setting. Its nearest-neighbor bond tensor is given by Eq.~\eqref{eq:S1hc_bond_tensor}, with isotropic and anisotropic couplings $A_1$ and $B_1$. The equivalent forms in Eqs.~\eqref{eq:S1hc_bond_tensor_rewritten} and \eqref{eq:S1hc_long_trans_decomp} show the same interaction in terms of bond-direction-dependent, longitudinal, and transverse components. Because the honeycomb lattice has two sublattices and three inequivalent bond directions, the in-plane polarizations are already mixed at the nearest-neighbor level. A finite PAMHE therefore appears without the need for an additional bond family.

Figure~\ref{fig:SM_square_honeycomb}(g) shows how the edge accumulation changes across the $(A_1,B_1)$ plane. The accumulation is largest when $A_1$ and $B_1$ are of comparable magnitude. In that region the isotropic and anisotropic parts of the bond tensor both contribute substantially, and the three bond directions mix the in-plane motion most strongly. Away from this region one part of the force network dominates, the polarization mixing is reduced, and the accumulation decreases.

The real-space response in Fig.~\ref{fig:SM_square_honeycomb}(f) again shows opposite transverse current components on the two sides of the sample and opposite accumulation at the upper and lower edges. The nearest-neighbor honeycomb model therefore already supports the same qualitative PAMHE pattern as the square lattice: a longitudinal thermal drive produces a transverse current of phonon angular momentum and an edge accumulation of opposite signs at opposite transverse boundaries. The damping dependence in Fig.~\ref{fig:SM_square_honeycomb}(h) shows that the accumulation is again largest at intermediate damping. As in the square lattice, the low-$\kappa$ limit reflects the weak coupling of the lattice to the reservoirs, so the nonequilibrium PAM response remains small, whereas the high-$\kappa$ limit reflects overdamping, which suppresses the correlations required to sustain the local angular-momentum buildup. The deflection angle is nonmonotonic and can change sign at large $\kappa$. This reflects the competition between different parts of the honeycomb spectrum, including acoustic and optical branches associated with the two-sublattice structure, which contribute with different signs and become dominant in different damping windows. In the field-dependent conductivity analysis, the magnetic field again induces a longitudinal component of PAM transport, and the sign change indicates that the relative balance between longitudinal and transverse response can be reshuffled strongly as different mode families take over. By the time this occurs, however, the overall accumulation is already small.

Taken together, the square and honeycomb results isolate the two ingredients required for a finite PAMHE in harmonic lattices. The nonuniform temperature profile generates the nonequilibrium correlations, and the elastic network must mix the in-plane polarizations strongly enough to redirect the resulting phonon angular momentum into the transverse direction. Once both ingredients are present, the response has the same qualitative form in both lattices: a longitudinal thermal drive produces a transverse PAM current and a corresponding edge accumulation of $L_z$.

\section{Graphene force-constant model}
\label{sec:SM_graphene_model}

For the graphene calculations, we extend the minimal first-neighbor honeycomb model to a shell-resolved force-constant model that reproduces the lattice dynamics more accurately. We include the first three neighbor shells: first-neighbor A--B bonds, second-neighbor A--A/B--B bonds, and third-neighbor A--B bonds. In this way, the model retains the same physical ingredients as the honeycomb analysis, while providing a more realistic description of graphene.

The in-plane part of the force-constant block is written in the same form as the honeycomb bond tensor in Eq.~\eqref{eq:S1hc_bond_tensor}. For a bond $(s,t)$ in shell $k\in\{1,2,3\}$ with in-plane angle $\theta_{st}$, we write
\begin{equation}
\mathbf B^{\parallel}_{st}(k,\theta_{st}) = A_k \mathbf I_2 + B_k \mathbf Q_2(\theta_{st}),
\label{eq:SM_graphene_Bparallel}
\end{equation}
with the same angular matrix $\mathbf Q_2(\theta)$ used in Eqs.~\eqref{eq:S1hc_Q2_def} and \eqref{eq:S1hc_Q2_dyadic}. The parameters $A_k$ and $B_k$ describe the isotropic and anisotropic in-plane couplings for shell $k$.

To describe graphene as a three-dimensional lattice with out-of-plane motion, we supplement the in-plane block by a flexural coupling $Z_k$ and write the full bond block as
\begin{equation}
\mathbf B_{st}(k,\theta_{st}) = \mathrm{diag}\!\left(\mathbf B^{\parallel}_{st}(k,\theta_{st}),\, Z_k\right),
\label{eq:SM_graphene_Bfull}
\end{equation}
so that $A_k$ and $B_k$ control the in-plane restoring forces, while $Z_k$ controls the $z$--$z$ coupling associated with the flexural motion.

The finite-sample force-constant matrix is assembled from these shell-resolved bond blocks in the same way as for the other lattice models,
\begin{equation}
\mathbf K_{ss} = \sum_{t\in\mathcal N(s)} \mathbf B_{st}, \qquad
\mathbf K_{st} = -\mathbf B_{st} \quad (s\neq t),
\label{eq:SM_graphene_Kblocks}
\end{equation}
with the neighbor shell determining which parameter set $(A_k,B_k,Z_k)$ is used for the bond $(s,t)$. The difference from the minimal honeycomb model is therefore the inclusion of additional neighbor shells, which are needed to reproduce the phonon dispersion of graphene more faithfully.

The shell-resolved parameters used in the calculations are listed in Table~\ref{tab:SM_graphene_params}. The first-neighbor model is sufficient to demonstrate the PAMHE mechanism, while the three-shell parametrization is used for the graphene calculations.

\begin{table}[t]
  \caption{Shell-resolved graphene force constants used in the calculations. Shell 1 corresponds to first-neighbor A--B bonds, shell 2 to second-neighbor A--A/B--B bonds, and shell 3 to third-neighbor A--B bonds. Units are N/m. The values are taken from Ref.~\cite{FALKOVSKY20085189}.}
  \label{tab:SM_graphene_params}
  \begin{ruledtabular}
  \begin{tabular}{c c c c c}
    Shell $k$ & Pair type & $A_k$ & $B_k$ & $Z_k$ \\
    \hline
    1 & AB    & 289.525 & 116.305  & 100.043 \\
    2 & AA/BB & 14.777  & 48.784   & -12.090 \\
    3 & AB    & 5.091   & -26.513  & -6.010
  \end{tabular}
  \end{ruledtabular}
\end{table}

\section{Real-material calculations from first-principles force constants}
\label{sec:SM_real_materials}

We obtain the harmonic interatomic force constants of Si, MgO, and BaTiO$_3$ from first principles using the Vienna \textit{ab initio} simulation package (\textsc{VASP}) together with the frozen-phonon method as implemented in the \textsc{Phonopy} package. We use the \textsc{VASP} projector augmented-wave (PAW) pseudopotentials with valence electron configurations Si $(3s^2 3p^2)$, Mg $(3s^2)$, O $(2s^2 2p^4)$, Ba $(5s^2 5p^6 6s^2)$, and Ti $(3d^3 4s^1)$, and choose the Perdew--Burke--Ernzerhof (PBE) form of the generalized gradient approximation for the exchange-correlation functional. 

For Si, we relax the two-atom diamond primitive cell using a plane-wave energy cutoff of 600~eV and a $12 \times 12 \times 12$ $k$-point mesh, converging the Hellmann--Feynman forces to $10^{-4}$~eV/\AA. The relaxed structure has a conventional cubic lattice constant of $a = 5.47$~\AA\ and a primitive-cell volume of $V_c = 40.89$~\AA$^3$. We then calculate the force constants in a $3 \times 3 \times 3$ supercell containing 54 atoms using finite displacements of 0.01~\AA\ and a $4 \times 4 \times 4$ $\Gamma$-centered $k$-point mesh. 

For MgO, we relax the two-atom rocksalt primitive cell using a plane-wave energy cutoff of 600~eV and an $8 \times 8 \times 8$ $\Gamma$-centered $k$-point mesh, converging the Hellmann--Feynman forces to $10^{-4}$~eV/\AA. The relaxed structure has a conventional cubic lattice constant of $a = 4.24$~\AA\ and a primitive-cell volume of $V_c = 19.12$~\AA$^3$. We then calculate the force constants in a $4 \times 4 \times 4$ supercell containing 128 atoms using finite displacements of 0.01~\AA\ and a $3 \times 3 \times 3$ $\Gamma$-centered $k$-point mesh. 

For cubic BaTiO$_3$, we relax the five-atom primitive cell using a plane-wave energy cutoff of 700~eV and an $8 \times 8 \times 8$ $\Gamma$-centered $k$-point mesh, converging the Hellmann--Feynman forces to $10^{-3}$~eV/\AA. The relaxed structure has a lattice constant of $a = 4.03$~\AA\ and a unit-cell volume of $V_c = 65.68$~\AA$^3$. We then calculate the force constants in a $2 \times 2 \times 2$ supercell containing 40 atoms using finite displacements of 0.01~\AA\ and an $8 \times 8 \times 8$ $\Gamma$-centered $k$-point mesh. The force constants are used to assemble the block force-constant matrix $\mathbf{K}$ of the finite crystal according to the bond construction in Eq.~\eqref{eq:S1_K_blocks}.

Before evaluating nonequilibrium observables, we verify that the first-principles force constants reproduce the phonon dispersions of the corresponding bulk materials along standard high-symmetry paths \cite{Wang2006MgOPhonon,Tinte1999BaTiO3Atomistic,Kazan2010SiThermalConductivity}. We then use these validated real-space force constants to construct finite cuboidal samples for the nonequilibrium calculations. For each material, the crystal is oriented so that the first primitive lattice axis $\bm{a}_1$ points along the positive $x$ direction. The thermal bias is applied along $x$, with a hot central region and colder faces normal to $x$. The two halves of the sample therefore correspond to opposite directions of the temperature gradient along $x$.

In this geometry, the relevant angular-momentum components are $L_y$ and $L_z$. These are the components whose transport acquires a Hall-like transverse character under the longitudinal thermal drive along $x$. We evaluate the corresponding local phonon-angular-momentum fields from Eq.~\eqref{eq:S1_local_PAM}. Their transport response is described by the conductivity tensor introduced in Eq.~\eqref{eq:S1lr_bulk_conductivity}, while the finite-sample maps show the associated accumulation pattern in real space. Opposite-sign accumulation on opposite transverse faces indicates that the transport of $L_y$ or $L_z$ has acquired a transverse component under the longitudinal drive.

\begin{figure}[t]
  \centering
  \includegraphics[width=\textwidth]{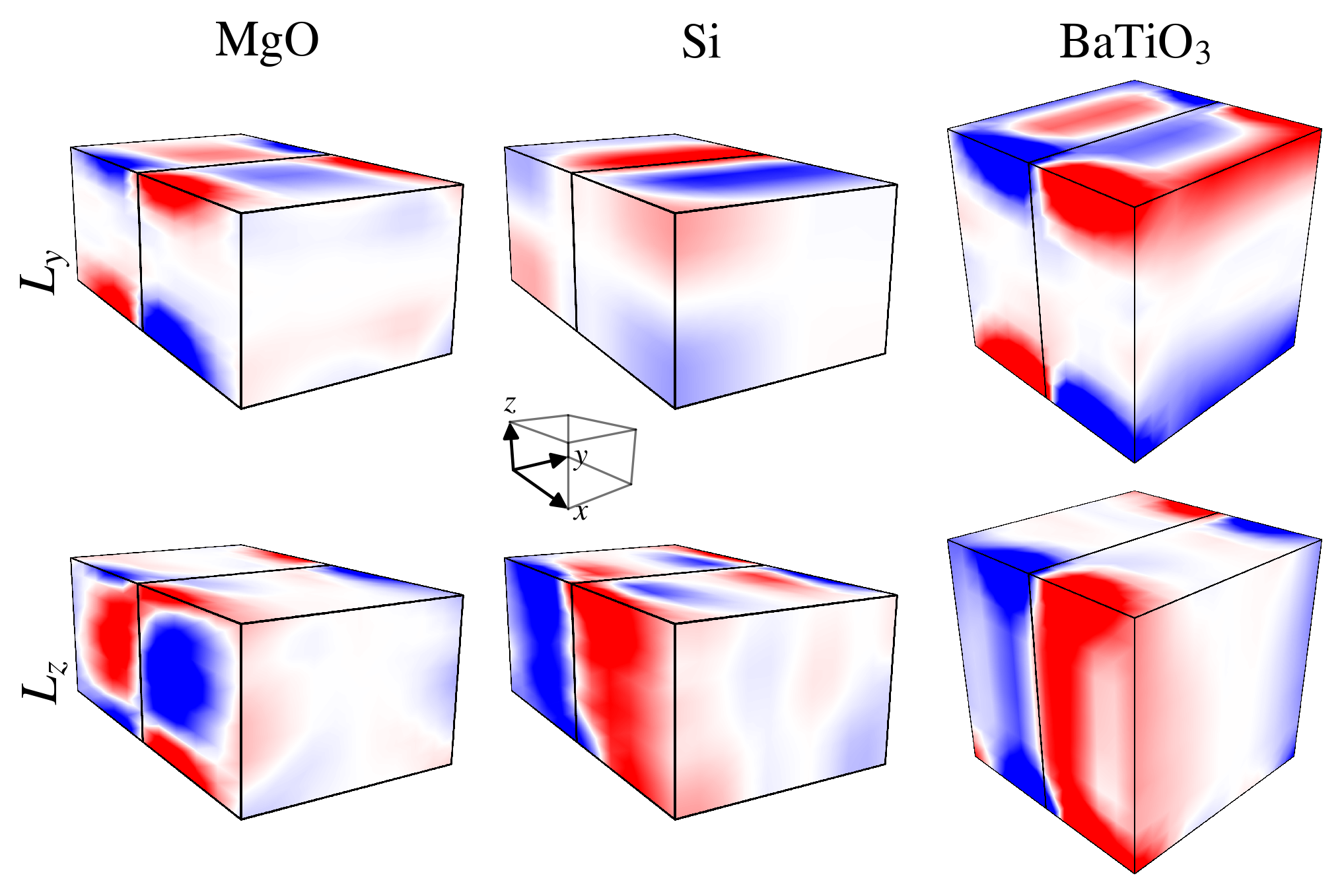}
  \caption{Real-space phonon-angular-momentum fields in finite samples of MgO, Si, and BaTiO$_3$ in the absence of an external magnetic field. For each material, the first primitive lattice axis $\bm{a}_1$ is aligned with the positive $x$ direction, and the sample is hottest at its center and colder at the two faces perpendicular to $x$. The top row shows the local phonon angular momentum $L_y$, and the bottom row shows $L_z$. In all three materials, opposite-sign accumulation appears on opposite transverse faces of the cuboid, showing the Hall-like redistribution of these angular-momentum components under a longitudinal thermal drive.}
  \label{fig:SM_real_materials_LyLz}
\end{figure}

Figure~\ref{fig:SM_real_materials_LyLz} shows the resulting zero-field finite-sample fields for MgO, Si, and BaTiO$_3$. The top row shows $L_y$, and the bottom row shows $L_z$. In all three materials, the temperature gradient along $x$ produces opposite-sign accumulations of these components on opposite transverse faces of the cuboid. The sign structure follows the same Hall-like pattern discussed for the model lattices: the drive is longitudinal, while the phonon angular momentum is redistributed toward the transverse boundaries and accumulates there with opposite signs on opposite faces.

\section{Analogous spin Hall angle}

In the spin Hall effect, one commonly characterizes the conversion between a longitudinal charge current and a transverse spin current by a Hall-angle-like quantity. This suggests an analogous construction for the present problem, where a longitudinal thermal drive generates transport of both phonon kinetic energy and phonon angular momentum. In our two-dimensional geometry, the temperature gradient is applied along \(x\), the transverse direction is \(y\), and the relevant bulk conductivity components are therefore the longitudinal components \(\sigma^{(\alpha),\mathrm{bulk}}_{xx}\) and the transverse components \(\sigma^{(\alpha),\mathrm{bulk}}_{yx}\), with \(\alpha\in\{E,L_z\}\).

So far, we have characterized the orientation of each transport channel through its deflection angle. For the phonon-angular-momentum current this is
\begin{equation}
\tan\theta_{L_z}
=
\frac{\sigma^{(L_z),\mathrm{bulk}}_{yx}}
{\sigma^{(L_z),\mathrm{bulk}}_{xx}},
\end{equation}
and for the kinetic-energy current it is
\begin{equation}
\tan\theta_{E}
=
\frac{\sigma^{(E),\mathrm{bulk}}_{yx}}
{\sigma^{(E),\mathrm{bulk}}_{xx}}.
\end{equation}
These quantities compare the transverse and longitudinal components of the \emph{same} transport channel and therefore measure how strongly that current is deflected away from the imposed thermal drive.

To compare instead the transverse transport of phonon angular momentum with the longitudinal transport of phonon kinetic energy, it is useful to introduce a mixed Hall-angle-like quantity. Since \(\sigma^{(L_z)}_{yx}/\sigma^{(E)}_{xx}\) has units of time in our convention, we multiply this ratio by a rate in order to form a dimensionless angle variable. Using the damping rate \(\kappa\), we define the mixed angle
\begin{equation}
\tan\theta_H
=
\,\kappa\,
\frac{\sigma^{(L_z),\mathrm{bulk}}_{yx}}
{\sigma^{(E),\mathrm{bulk}}_{xx}}.
\label{eq:mixed_angle}
\end{equation}
With this definition, \(\theta_H\) compares the transverse phonon-angular-momentum transport directly to the longitudinal kinetic-energy transport generated by the same thermal drive.

It is also useful to remove the explicit \(\kappa\)-dependence from this comparison and define a reference-rate version, which we call the conversion angle:
\begin{equation}
\tan\tilde{\theta}_H
=
\,\kappa_0\,
\frac{\sigma^{(L_z),\mathrm{bulk}}_{yx}}
{\sigma^{(E),\mathrm{bulk}}_{xx}},
\label{eq:conversion_angle}
\end{equation}
where \(\kappa_0\) is a fixed reference rate.  The angle \(\tilde{\theta}_H\) isolates the relative strength of transverse phonon-angular-momentum transport and longitudinal kinetic-energy transport without tying that comparison to the actual damping value used in a given run.

In the presence of an out-of-plane magnetic field, the same constructions are extended by replacing the zero-field conductivities by the corresponding full field-dependent conductivities, including the linear-in-\(B\) corrections. For the mixed angle this gives
\begin{equation}
\tan\theta_H(B)
=
\,\kappa\,
\frac{\sigma^{(L_z),\mathrm{bulk}}_{yx}+\sigma^{(L_z,B),\mathrm{bulk}}_{yx}}
{\sigma^{(E),\mathrm{bulk}}_{xx}+\sigma^{(E,B),\mathrm{bulk}}_{xx}},
\end{equation}
and for the reference-rate form
\begin{equation}
\tan\tilde{\theta}_H(B)
=
\,\kappa_0\,
\frac{\sigma^{(L_z),\mathrm{bulk}}_{yx}+\sigma^{(L_z,B),\mathrm{bulk}}_{yx}}
{\sigma^{(E),\mathrm{bulk}}_{xx}+\sigma^{(E,B),\mathrm{bulk}}_{xx}}.
\end{equation}
These quantities complement the usual deflection angles \(\theta_{L_z}\) and \(\theta_E\): the deflection angles quantify the orientation of each current separately, whereas \(\theta_H\) and \(\tilde{\theta}_H\) compare the transverse PAM response directly with the longitudinal kinetic-energy response.

\begin{table}[h]
\centering
\caption{Mixed angle \(\theta_H\) and conversion angle \(\tilde{\theta}_H\) for damping \(\kappa=5~\mathrm{ps}^{-1}\) and reference damping \(\kappa_0=1~\mathrm{ps}^{-1}\) for the square and honeycomb lattices.}
\label{tab:mixed_conversion_angles}
\begin{tabular}{l@{\hspace{1.2cm}}c@{\hspace{1.2cm}}c}
\hline\hline
Lattice & \(\theta_H\) & \(\tilde{\theta}_H\) \\
\hline
Square
& \(2.92^\circ = 0.051\)
& \(0.585^\circ =0.0102\) \\
Honeycomb
& \(0.28^\circ = 0.0049\)
& \(0.057^\circ = 0.00099\) \\
\hline\hline
\end{tabular}
\end{table}

\begin{figure*}[t]
    \centering
    \includegraphics[width=\linewidth]{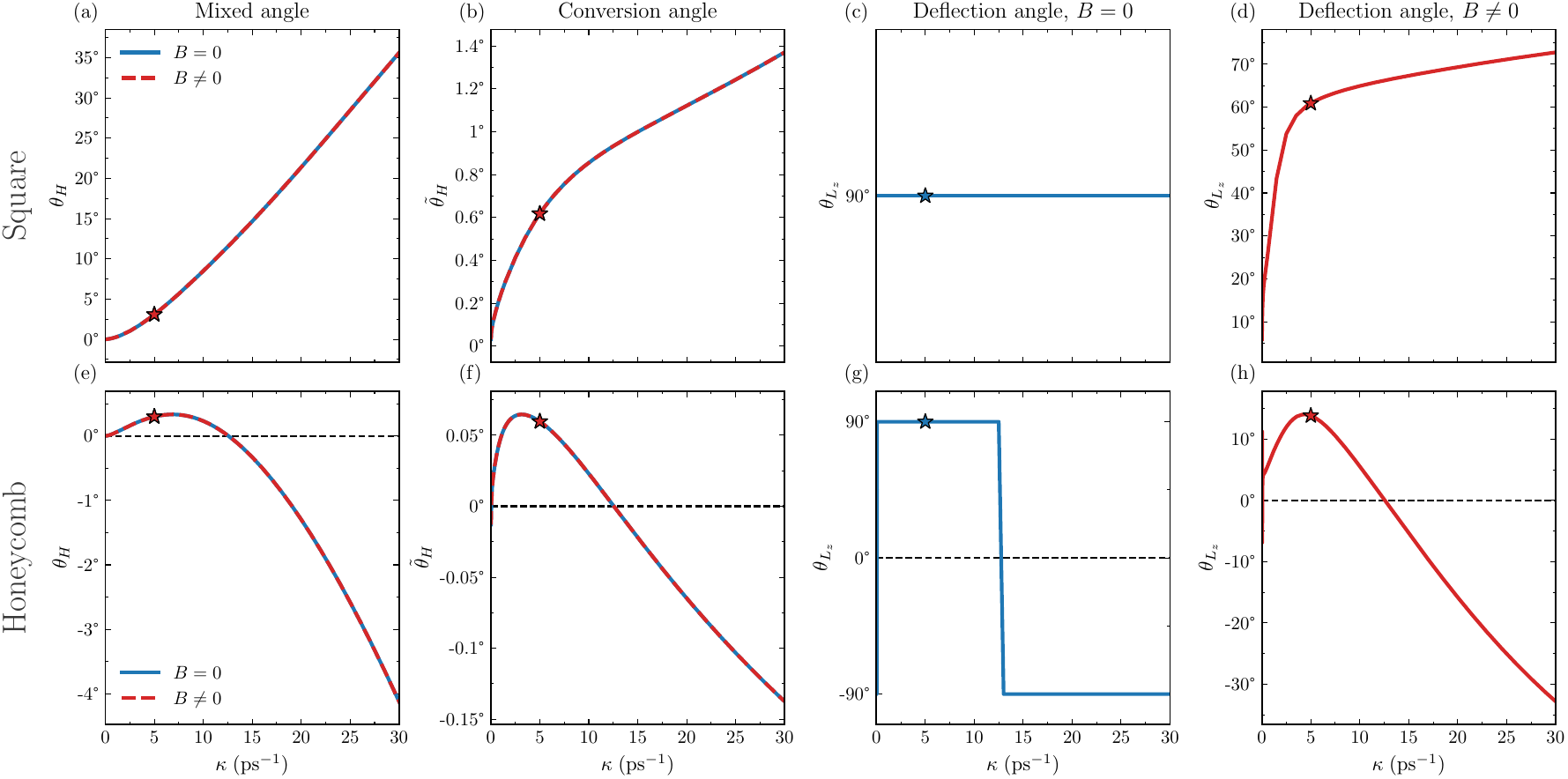}
    \caption{
    Comparison of Hall-angle-like measures for the square and honeycomb lattices as functions of the damping rate \(\kappa\). The first column shows the mixed angle \(\theta_H\) defined in Eq.~\eqref{eq:mixed_angle}; the second column shows the reference-rate angle \(\tilde{\theta}_H\) defined in Eq.~\eqref{eq:conversion_angle}; the third column shows the zero-field PAM deflection angle \(\theta_{L_z}\); and the fourth column shows the corresponding field-dependent PAM deflection angle including the linear-in-\(B\) conductivity corrections. In the second column we use the reference rate \(\kappa_0=1~\mathrm{ps}^{-1}\). Blue solid curves denote the zero-field conductivities and red dashed curves the full field-dependent conductivities. The stars mark the parameter set used for the real-space current maps discussed in the main text and in Fig.~\ref{fig:SM_square_honeycomb}. The zero-field deflection angle shows that the direction of transverse phonon-angular-momentum transport can reverse as the damping is varied; in particular, for the honeycomb lattice the sign change of \(\theta_{L_z}\) indicates a damping-controlled reversal of the transverse PAM flow while the response remains nearly purely transverse. For the field-dependent curves, we use comparatively large effective magnetic-field scales, defined by $\Omega_B \equiv \gamma B_z$, with $\Omega_B \approx 2\,\mathrm{THz}$ for the square lattice and $\Omega_B \approx 4\,\mathrm{THz}$ for the honeycomb lattice, so that the magnetic-field-induced changes are more clearly visible.}
    \label{fig:analogous_spin_angle}
\end{figure*}

Figure~\ref{fig:analogous_spin_angle} compares these quantities for the square and honeycomb lattices. The first column shows the mixed angle \(\theta_H\), the second column shows the reference-rate angle \(\tilde{\theta}_H\), the third column shows the zero-field PAM deflection angle \(\theta_{L_z}\), and the fourth column shows the corresponding field-dependent PAM deflection angle including the linear-in-\(B\) corrections. This comparison separates three distinct questions: how strongly the PAM current is deflected, how strongly the kinetic-energy current remains longitudinal, and how efficiently a longitudinal thermal drive is converted into transverse transport of phonon angular momentum relative to longitudinal transport of phonon kinetic energy.

\section{Molecular-dynamics simulations}
\label{sec:SM_MD}

To complement the covariance-based framework, we also solve the lattice dynamics directly in real time. These simulations reproduce the same steady-state pattern as the analytical treatment: a longitudinal thermal drive redistributes phonon angular momentum transversely and produces accumulation of opposite signs of $L_z$ at opposite sample edges. They therefore provide a direct numerical check of the steady-state response obtained from the covariance formalism.

The dynamical variables are the site displacements and velocities. For a two-dimensional lattice, the equations of motion are
\begin{equation}
m_s \ddot{\bm u}_s
=
\bm F^{\rm spr}_s(\{\bm u\})
-\kappa_s m_s \dot{\bm u}_s
-2m_s\Omega_s^{g}\,\hat{\bm z}\times \dot{\bm u}_s
+\bm \eta_s(t),
\label{eq:SM_MD_eom}
\end{equation}
where $\bm F^{\rm spr}_s$ is the total spring force on site $s$, $\kappa_s$ is the local damping, and $\bm \eta_s(t)$ is Gaussian thermal noise. The magnetic field is optional. We parameterize its coupling by the site-resolved gyroscopic frequency
\begin{equation}
\Omega_s^{g}=\gamma_s B_z,
\label{eq:SM_MD_Omegag}
\end{equation}
where $\gamma_s$ is the effective gyromagnetic ratio assigned to site $s$. Equivalently, one may write
\begin{equation}
\gamma_s=\frac{q_s^{\mathrm{eff}}}{2m_s},
\qquad
\Omega_s^{g}=\frac{q_s^{\mathrm{eff}} B_z}{2m_s},
\label{eq:SM_MD_gamma_qeff}
\end{equation}
in terms of an effective charge-like parameter $q_s^{\mathrm{eff}}$. When the gyromagnetic ratio is uniform, $\gamma_s=\gamma$, the magnetic-field strength can be expressed through the effective scale
\begin{equation}
\Omega_B \equiv \gamma B_z,
\label{eq:SM_MD_OmegaB}
\end{equation}
so that $\Omega_s^{g}=\Omega_B$ for all sites. This is the same gyroscopic frequency scale that enters the analytical weak-field formulation.

The local baths satisfy the fluctuation--dissipation relation
\begin{equation}
\langle \eta_{s\mu}(t)\eta_{s'\nu}(t')\rangle
=
2\kappa_s m_s k_B T_s\,
\delta_{ss'}\delta_{\mu\nu}\delta(t-t'),
\label{eq:SM_MD_noise}
\end{equation}
with a prescribed site-dependent temperature profile $T_s$. The nonequilibrium drive is imposed through a hot band centered in the sample and colder regions toward the left and right edges. In practice, the temperature varies smoothly along the transport direction $x$ and is close to $T_{\rm hot}$ in the central region and to $T_{\rm cold}$ outside it. A convenient form is
\begin{equation}
T(x)
=
T_{\rm cold}
+
\frac{T_{\rm hot}-T_{\rm cold}}{2}
\left[
\tanh\!\left(\frac{x-x_{\rm L}}{\ell_T}\right)
-
\tanh\!\left(\frac{x-x_{\rm R}}{\ell_T}\right)
\right],
\label{eq:SM_MD_Tprofile}
\end{equation}
where $x_{\rm L}$ and $x_{\rm R}$ define the hot region and $\ell_T$ controls the smoothness of its edges.

The finite square lattice is the same axial-plus-diagonal network analyzed above, and the finite honeycomb lattice is the same open honeycomb graph with sites on the two sublattices A and B. In both cases, the spring force is evaluated bond by bond from the instantaneous extension of each harmonic bond. For a bond $(s,t)$ with spring constant $K_{st}$, rest length $\ell_{st}$, and instantaneous bond vector
\begin{equation}
\bm \rho_{st}(t)=\bm r^{\,0}_{t}-\bm r^{\,0}_{s}+\bm u_t(t)-\bm u_s(t),
\qquad
\rho_{st}=|\bm \rho_{st}|,
\qquad
\hat{\bm \rho}_{st}=\frac{\bm \rho_{st}}{\rho_{st}},
\label{eq:SM_MD_rhost}
\end{equation}
the force exerted on site $s$ by site $t$ is
\begin{equation}
\bm F_{s\leftarrow t}
=
-K_{st}\bigl(\rho_{st}-\ell_{st}\bigr)\hat{\bm \rho}_{st},
\label{eq:SM_MD_bondforce}
\end{equation}
and the total spring force is
\begin{equation}
\bm F^{\rm spr}_s
=
\sum_{t\in\mathcal N(s)} \bm F_{s\leftarrow t}.
\label{eq:SM_MD_totforce}
\end{equation}
To prevent rigid drifting of the entire sample and to define the sample edges sharply, selected boundary sites are pinned. After each full time step, the displacements and velocities of the pinned sites are reset to zero,
\begin{equation}
\bm u_s=0,
\qquad
\dot{\bm u}_s=0,
\qquad
s\in\partial\Omega.
\label{eq:SM_MD_pinned}
\end{equation}

\subsection{Square lattice: BAOAB integration with Boris rotation}

For the square lattice, the stochastic dynamics is integrated with a BAOAB splitting scheme~\cite{LeimkuhlerMatthews2013}, while the gyroscopic term is treated with a Boris rotation~\cite{Boris1970,ZenitaniUmeda2018}. This combination is well suited to Langevin dynamics with a velocity-dependent gyroscopic force: the thermostat is handled through an exact Ornstein--Uhlenbeck step, while the field-induced rotation is advanced by a stable update of the velocity.

Let $(\bm u_n,\bm v_n)$ denote the full vectors of lattice displacements and velocities at time $t_n$, and let $\Delta t$ be the time step. One time step consists of a half kick from the spring force,
\begin{equation}
\bm v^{(1)}
=
\bm v_n
+
\frac{\Delta t}{2}\,\mathbf M^{-1}\bm F^{\rm spr}(\bm u_n),
\label{eq:SM_MD_B1}
\end{equation}
followed by a half Boris rotation. For an out-of-plane field, the velocity of each site rotates in the $(x,y)$ plane. Defining
\begin{equation}
t_s=\Omega_s^{g}\Delta t,
\qquad
s_s=\frac{2t_s}{1+t_s^2},
\label{eq:SM_MD_Boris_ts}
\end{equation}
which is equivalently $t_s=(q_s^{\mathrm{eff}}B_z/m_s)(\Delta t/2)$, the rotation is
\begin{align}
v'_{x,s} &= v^{(1)}_{x,s}+t_s v^{(1)}_{y,s}, \\
v'_{y,s} &= v^{(1)}_{y,s}-t_s v^{(1)}_{x,s}, \\
v^{(2)}_{x,s} &= v^{(1)}_{x,s}+s_s v'_{y,s}, \\
v^{(2)}_{y,s} &= v^{(1)}_{y,s}-s_s v'_{x,s}.
\label{eq:SM_MD_Boris_update}
\end{align}
For a uniform gyromagnetic ratio, $\gamma_s=\gamma$, one simply has $t_s=\Omega_B\Delta t$ for all sites.

The positions are then advanced by a half drift,
\begin{equation}
\bm u^{(1)}
=
\bm u_n+\frac{\Delta t}{2}\,\bm v^{(2)},
\label{eq:SM_MD_A1}
\end{equation}
after which damping and noise are applied through the exact Ornstein--Uhlenbeck solution of the local Langevin equation,
\begin{equation}
\bm v^{(3)}
=
a_s\,\bm v^{(2)}+b_s\,\bm \xi_s,
\label{eq:SM_MD_OUupdate}
\end{equation}
with
\begin{equation}
a_s=e^{-\kappa_s\Delta t},
\qquad
b_s=\sqrt{\frac{k_B T_s}{m_s}\bigl(1-a_s^2\bigr)},
\label{eq:SM_MD_OUcoeff}
\end{equation}
and $\bm\xi_s$ a two-component standard Gaussian random vector. A second half drift gives
\begin{equation}
\bm u^{(2)}
=
\bm u^{(1)}+\frac{\Delta t}{2}\,\bm v^{(3)},
\label{eq:SM_MD_A2}
\end{equation}
and the Boris rotation is then applied once more over half a step. Finally, the spring force is evaluated at the updated positions and applied for the last half kick,
\begin{equation}
\bm v_{n+1}
=
\bm v^{(4)}
+
\frac{\Delta t}{2}\,\mathbf M^{-1}\bm F^{\rm spr}(\bm u^{(2)}),
\qquad
\bm u_{n+1}=\bm u^{(2)}.
\label{eq:SM_MD_B2}
\end{equation}
After the full update, the pinned boundary sites are reset according to Eq.~\eqref{eq:SM_MD_pinned}.

\subsection{Honeycomb lattice: fourth-order Runge--Kutta with thermal kicks}

For the honeycomb lattice, the deterministic part of the dynamics is integrated with an explicit fourth-order Runge--Kutta step, and the local baths are incorporated afterwards through Gaussian velocity kicks. This provides an independent numerical route to the same nonequilibrium steady state and makes it possible to check that the transverse phonon-angular-momentum accumulation is not tied to the BAOAB/Boris implementation used for the square lattice.

The deterministic equations for each site are
\begin{equation}
\dot{\bm u}_s=\bm v_s,
\qquad
\dot{\bm v}_s
=
\frac{1}{m_s}\bm F^{\rm spr}_s(\{\bm u\})
-\kappa_s \bm v_s
-2\Omega_s^{g}\,\hat{\bm z}\times \bm v_s.
\label{eq:SM_MD_honey_rhs_site}
\end{equation}
In the calculations discussed here, we take a uniform gyromagnetic ratio, so that $\Omega_s^{g}=\Omega_B$ for all sites and the field strength is set by the single parameter $\Omega_B$.

One Runge--Kutta step advances the positions and velocities according to
\begin{align}
\bm u_{n+1}
&=
\bm u_n+\frac{\Delta t}{6}
\left(
\bm k_1^{u}+2\bm k_2^{u}+2\bm k_3^{u}+\bm k_4^{u}
\right), \\
\bm v_{n+1}^{\rm det}
&=
\bm v_n+\frac{\Delta t}{6}
\left(
\bm k_1^{v}+2\bm k_2^{v}+2\bm k_3^{v}+\bm k_4^{v}
\right),
\label{eq:SM_MD_RK4}
\end{align}
with the standard four-stage evaluation of the right-hand side of Eq.~\eqref{eq:SM_MD_honey_rhs_site}. After this deterministic update, the thermal bath is applied through Gaussian velocity kicks,
\begin{equation}
v_{x,s}\leftarrow v_{x,s}^{\rm det}+\sqrt{\Delta t}\,\sigma_{v,s}\,\xi_{x,s},
\qquad
v_{y,s}\leftarrow v_{y,s}^{\rm det}+\sqrt{\Delta t}\,\sigma_{v,s}\,\xi_{y,s},
\label{eq:SM_MD_honey_kick}
\end{equation}
where
\begin{equation}
\sigma_{v,s}
=
\sqrt{\frac{2\kappa_s k_B T_s}{m_s}},
\label{eq:SM_MD_sigma_v}
\end{equation}
and $\xi_{x,s}$ and $\xi_{y,s}$ are independent standard Gaussian variables. The pinned boundary sites are then reset according to Eq.~\eqref{eq:SM_MD_pinned}.

\subsection{Steady-state averaging and observables}

The local phonon angular momentum is evaluated directly from the trajectories as
\begin{equation}
L_z(s,t)
=
m_s\bigl[u_{s,x}(t)\dot u_{s,y}(t)-u_{s,y}(t)\dot u_{s,x}(t)\bigr],
\label{eq:SM_MD_Lz}
\end{equation}
and we also record the local displacement amplitude and kinetic energy,
\begin{equation}
u_{\rm amp}(s,t)=\sqrt{u_{s,x}^2(t)+u_{s,y}^2(t)},
\qquad
E_{\mathrm{kin}}(s,t)=\frac{m_s}{2}\bigl[\dot u_{s,x}^2(t)+\dot u_{s,y}^2(t)\bigr].
\label{eq:SM_MD_obs}
\end{equation}
Each trajectory is first evolved through a warmup interval to remove transients. The remaining steady-state part of the run is sampled at a fixed stride and accumulated into time-averaged spatial maps and edge-resolved signals. To reduce residual stochastic fluctuations, the observables are averaged over long trajectories and over independent realizations with different random seeds.

The use of two independent numerical approaches---BAOAB/Boris for the square lattice and RK4 with thermal kicks for the honeycomb lattice---provides a direct check on the steady-state patterns. Both implementations give the same transverse redistribution of phonon angular momentum and the same accumulation of opposite signs of $L_z$ at opposite edges, supporting the conclusion that the observed PAMHE patterns are properties of the driven lattice rather than artifacts of a particular solver.


\end{document}